%% file: main.tex
\documentclass[12pt]{report}

\usepackage[
a4paper,
vmargin=2.7cm,
hmargin=2.7cm,
]{geometry}
\usepackage{tikz-cd}

\usepackage{xcolor} 
\definecolor{azuluc3m}{RGB}{0,0,102}
\usepackage{amsfonts,amssymb,amsthm,mathtools}  
\usepackage{tensor}
\usepackage{tikz}
\usepackage{pdfpages}
\usepackage{lmodern}
\newlength\longest
\usepackage{amssymb}
\usetikzlibrary{matrix}
\usetikzlibrary{calc}
\usepackage{cancel}
\usepackage[T1]{fontenc}
\usepackage{lmodern} 
 \usepackage{hyperref}
\hypersetup{colorlinks=true,
	linkcolor=black, citecolor = azuluc3m, urlcolor = azuluc3m} % links to resources outside the document in blue
\usepackage{titling}

\usepackage{url}

\newcommand{\volg}{\mathrm{vol}_{g}}
\newcommand{\volgg}{\mathrm{vol}_{\overline{g}}}
\newcommand{\R}{\mathrm{R}}
\newcommand{\K}{\mathrm{K}}
\newcommand{\N}{\mathrm{N}}
\newcommand{\E}{\mathrm{E}}

\newcommand{\e}{\mathrm{e}}
\newcommand{\F}{\mathrm{F}}
\newcommand{\A}{\mathrm{A}}
\newcommand{\de}{\mathrm{d}}

\usepackage[listings]{tcolorbox}
\newtcblisting{EvalBox}[2][]{%
  colback=white,
  arc=0pt,
  boxrule=0.5pt,
  text only,
  title=#2,#1}

\theoremstyle{definition}
\newtheorem*{definition}{Definition}
\newtheorem*{proposition}{Proposition}

\usepackage{fancyhdr}
\fancypagestyle{plain}{
  
  \fancyhf{}
  \fancyfoot[C]{\thepage}
}

\begin{document}

\includepdf[pages=-]{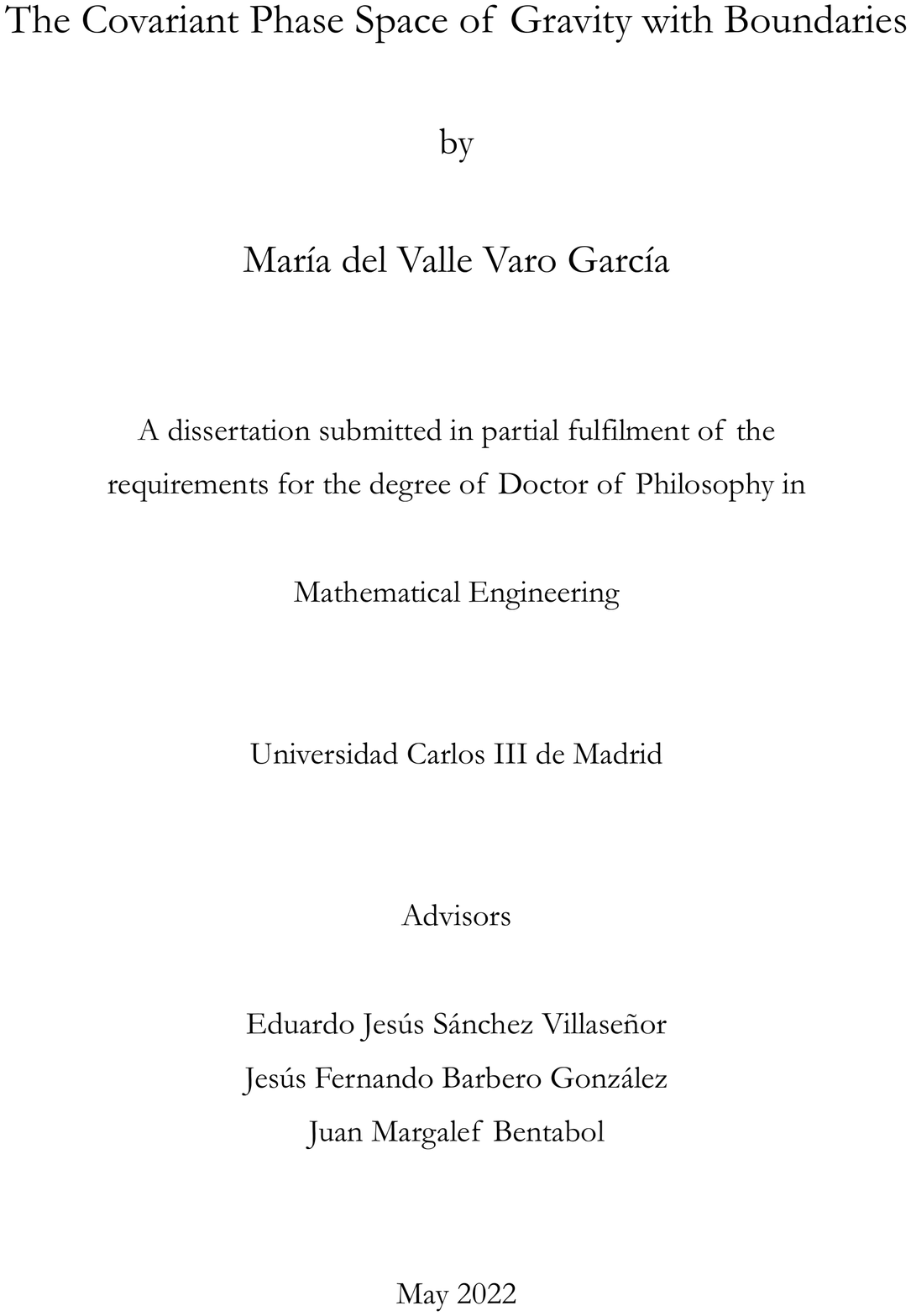}

\newpage
\thispagestyle{empty}
\mbox{}

%%%%%%%%%% COPYRIGHT %%%%%%%%%
\clearpage
\newpage
\thispagestyle{empty}
	\vspace*{\fill}
	\begin{center}
	This thesis is distributed under license ``\textbf{Creative Commons Atributtion - Non Commercial - Non Derivatives}''.\\
	\vspace*{0.5cm}
	\includegraphics[width=4.2cm]{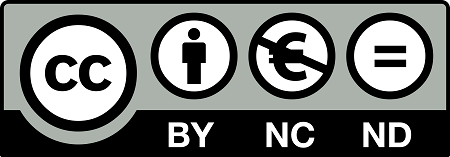} % Creative Commons logo
	\end{center}
\clearpage

\newpage
\thispagestyle{empty}
\mbox{}

%%%%%%%%%%DEDICATORIA%%%%%%%%%
\newpage
\thispagestyle{empty}
\null\vfill
\begin{flushright}
\textit{A mis padres,\\
Cristóbal y Margarita.}
\end{flushright}
\vfill\vfill
\mbox{}
\newpage
\thispagestyle{empty}
\mbox{}

%%%%%%%%%% AGRADECIMIENTOS %%%%%%%%%
\include{acknowledgements}
\newpage
\newpage
\thispagestyle{empty}
\mbox{}

%%%%%%%%%%%%%%%%%%% ABSTRACT %%%%%%%%%%%%%%%%%%
\include{abstract}

\newpage
\thispagestyle{empty}
\mbox{}

%%%%%%%%%%%%%%%%%%%RESEARCH OVERVIEW %%%%%%%%%%%%%%%%%%
\include{researchoverview}
\newpage
\thispagestyle{empty}
\mbox{}

\include{conferences}
\newpage
\thispagestyle{empty}
\mbox{}

%%%%%%%%%%%%%%%%%% TABLE OF CONTENTS %%%%%%%%%%%%%%%%%%
\tableofcontents
%%%%%%%%QUOTE%%%%%%

\clearpage
\thispagestyle{empty}
\null\vfill

\textit{``Ich kann es nun einmal nicht lassen, in diesem Drama von Mathematik und Physik $-$ die sich im Dunkeln befruchten, aber von Angesicht zu Angesicht so gerne einander verkennen und verleugnen $-$ die Rolle des (wie ich genugsam erfuhr, oft unerwünschten) Boten zu spielen.''}

  \begin{flushright}
\small{Philosophie der Mathematik und Naturwissenschaft $-$ Hermann Weyl}
\end{flushright}

\vfill\vfill
\clearpage

\newpage
\thispagestyle{empty}
\mbox{}

%%%%%%%%%%CHAPTERS%%%%%%%%%
\include{introduction}
\include{historicaloverview}

\newpage
\thispagestyle{empty}
\mbox{}

\include{CPS}
\newpage
\thispagestyle{empty}
\mbox{}

\include{EH}

\include{PALATINI}
\newpage
\thispagestyle{empty}
\mbox{}

\include{HMS}
\newpage
\thispagestyle{empty}
\mbox{}

\include{equivalence}

\newpage
\thispagestyle{empty}
\mbox{}

\include{conclusions}

\newpage
\thispagestyle{empty}
\mbox{}

\include{appendix}
\newpage
\thispagestyle{empty}
\mbox{}

%\nocite*{}
\bibliographystyle{plainnat}
\bibliography{refs}
%%%%%%%%%%%%%%%%%%
\newpage
\thispagestyle{empty}
\mbox{}

\newpage
\thispagestyle{empty}
\mbox{}

\end{document}

%% file: acknowledgements.tex
\chapter*{Acknowledgements}

I owe a great deal to the people who have helped and advised me over the last three years, without whom I could not have written this thesis. 

First and foremost, I am tremendously grateful to my supervisors, who gave me such an absorbing topic to work on and offered their guidance, Eduardo J.S. Villaseñor, J. Fernando Barbero G. and Juan Margalef Bentabol. I want to thank your incredible patience, continuous encouragement and your priceless time answering all of my questions. It has been a privilege to learn from you and work under your supervision.

I thank Tomás Ortín for making me fall in love with general relativity in my master studies and his patience with my endless and possibly pointless questions about spacetime and black holes. I also inevitably need to thank Mercedes Martín Benito for encouraging me in the first place to pursue a PhD in this particular field of research. 

Special thanks also to Laura Hiscott for correcting my English, and to Pablo H. Ufarte for helping me with the German language. Thank you to María José and Juan Manuel for their constant hospitality, particularly during the lockdown. I also thank my best friends, Marta and Evelyn, for always being there and listening to all my rambling. Without a doubt, I need to thank my volleyball team for taking my mind out of the seriousness of my studies when I most needed it. 

Last but not least, I could not have done any of this without the constant support and love from my family, Asier, Laura, Myriam and my parents. Thank you for believing in me come what may. 

%% file: abstract.tex
\chapter*{Abstract}

 This thesis investigates the metric and tetrad formulations of three gravitational field theories in manifolds with timelike boundaries within the covariant phase space program. With the recently developed relative bicomplex framework, we explore the space of solutions and presymplectic structures associated with each action principle and analyse their equivalence.

 The first action we consider is the Einstein-Hilbert (EH) action with the Gibbons-Hawking-York boundary term. By including the appropriate boundary terms in the variational principles, we show that the metric and tetrad formulations derived from them are equivalent. Furthermore, we show that their solution spaces are the same and that their presymplectic structures and associated charges coincide.

The second action we consider is the Palatini action with the Obukhov boundary term, assuming torsion and non-metricity, and we prove the equivalence between its metric and tetrad formulations. Furthermore, we show that the metric and tetrad-sector of the first-order Palatini formulation are equivalent to the metric and tetrad formulations of the EH action. 

Lastly, we introduce the Hojman-Mukku-Sayed (HMS) action, a generalisation of the Palatini action plus the Holst term in the presence of boundaries with non-metricity and torsion. We prove that the space of solutions of the HMS and Palatini actions coincided and conclude that HMS's metric and tetrad sectors are identical to their corresponding versions of the EH action. Additionally, we prove that the Palatini and HMS Lagrangians are not cohomologically equal despite defining the same space of solutions. Consequently, a careful analysis is required for the presymplectic structures and the charges because they may differ. However, we show that the covariant phase spaces of both theories were equivalent. This sheds light on some open problems regarding the equivalence of their associated charges in different formulations.

%% file: researchoverview.tex
\chapter*{Published and submitted content}
The original contributions of this thesis are based upon the research described in the following publications, listed in chronological order:
\begin{itemize}
\item J. Fernando Barbero G., Juan Margalef-Bentabol, Valle Varo, Eduardo J. S.\\ Villaseñor, \emph{“Covariant phase space for gravity with boundaries: metric vs tetrad formulations”}, Physical Review D, \textbf{104} (2021) 044048. \\This paper is wholly included in the thesis in Chapter \ref{EHaction}.
\item J. Fernando Barbero G., Juan Margalef-Bentabol, Valle Varo, Eduardo J. S.\\ Villaseñor, \emph{“Palatini gravity with nonmetricity, torsion, and boundaries in metric and connection variables”}, Physical Review D, \textbf{104} (2021) 044046. \\This paper is wholly included in the thesis in Chapter \ref{Palatiniaction}.
\item J. Fernando Barbero G, Marc Basquens, Valle Varo, Eduardo J. S.\\ Villaseñor, \emph{“Three roads to the geometric constraint formulation of gravitational theories with boundaries”}, Symmetry \textbf{13} (2021) 1430.  \\ The material from this source included in this thesis is not singled out with typographic means and references.
\item J. Fernando Barbero G, Juan Margalef-Bentabol, Valle Varo, Eduardo J. S. Villaseñor, \emph{“On the on-shell equivalence of general relativity and Holst theories with nonmetricity, torsion, and boundaries”},  Physical Review D, \textbf{105} (2022) 064066. \\This paper is wholly included in the thesis in Chapter \ref{HMS}.
\end{itemize}

%% file: conferences.tex
\chapter*{Other Research Merits}
The doctoral student also attended the following conferences: 
\begin{itemize}
    \item BritGrav21 (12-16 April 2021)
    \item Sixteenth Marcel Grossmann Meeting - MG16 (5-10 July  2021)
\end{itemize}
And gave contributed talks at: 
\begin{itemize}
    \item Junior Seminars Carlos III de Madrid (3 December 2020)
    \item The 7th Conference of the Polish Society on Relativity (20-23 September 2021)
\end{itemize}

%% file: introduction.tex
\chapter{Introduction}

General relativity (GR) theory is the most successful classical field theory for the gravitational field. One of its most outstanding and characteristic features is its manifest background independence, disavowing the absolutes of Newton's space and time. It was first conceived as a theory based on smooth Lorentzian metrics upon a pseudo-Riemannian manifold whose curvature was identified with the gravitational field. The standard method to derive its equations of motion is through a suitable variational principle. The Einstein-Hilbert (EH) action is the variational principle that leads to Einstein's equations for the gravitational field, correctly formulated by Hilbert in \cite{hilbert1915grundlagen}. In this formulation, the metric $g$ is the independent variable and the Lagrangian gives rise to second-order differential equations. 

An alternative geometric formulation of GR based on a dynamical collection of differential forms rather than metrics was developed by Cartan\footnote{According to Akivis \cite{akivis1993elie}, the idea of moving frames can be traced back to the method of moving trihedrons introduced by the Estonian mathematician Martin Bartels ($1769$-$1836$), a teacher of both Gauss and Lobachevsky.} in the early 1920s. These differential forms or \emph{tetrads} (in four dimensions) replaced the metric as the independent variable of the theory. Einstein in 1928 \cite{einstein1928} and Weyl in 1929 \cite{weyl1929} introduced them into GR, obtaining the tetrad Einstein-Hilbert action.

Motivated by geometrical considerations, Einstein envisioned utilising a variational principle where the connection was promoted to the status of independent field variable. For historical reasons, this approach is commonly referred to as Palatini theories (see \cite{ferraris1982variational} for a thorough historical review). The original metric Palatini action is similar to the EH action, but instead considers the Ricci scalar to be a function of the metric and the connection, which is assumed to be torsion-free. Palatini's action was thought not to need any associated surface terms since the Lagrangian no longer included any second derivatives of the metric, since the space of fields considered for the action was Dirichlet \cite{burton1999palatini}. The Palatini action can also be recast as a tetradic first-order theory where its independent fields are the tetrads, denoted by $\e$, and the Lorentz connection, denoted by $\omega$. 

\begin{center}
   \begin{tabular}{| c | c | c |}
\hline
Lagrangian & Metric & Tetrad \\ \hline
Second Order & Einstein-Hilbert ($g$) &  Einstein-Hilbert ($\e$) \\ \hline
First Order & Palatini ($g$, $\nabla$) & Palatini ($\e$, $\omega$) \\ \hline 
\end{tabular} 
\end{center}

The interest in connection-based theories of gravity, such as Palatini or pure-affine theories,  increased due to two related phenomena: the similarities between Yang-Mills gauge theories and the mathematical basis of GR (based on fibre bundles), and the introduction of the Ashtekar program of canonical gravity \cite{ashtekarBook}. The common practice among particle physicists was to regard all gauge theories as fundamentally dependent on connections. 

Around the same period of time, the development of the Arnowitt-Deser-Misner (ADM) formalism showed that GR is a singular Hamiltonian system with two fundamental constraints: the Hamiltonian constraint and the diffeomorphism constraint \cite{adm2}. Those interested in quantum gravity with leanings towards the canonical perspective attempted to quantise this classical geometrodynamical ADM picture by applying the well-known Dirac quantisation procedure to arrive at the Wheeler-DeWitt equation \cite{dewitt1967quantum}. 

Dirac's quantisation of systems with only first class constraints consists of replacing the classical Poisson algebra of the constraints with an algebra of operators acting on a Hilbert space, and determining its kernel to obtain the physical states \cite{DiracBook}. Despite being conceived for mechanical systems and not field theories, Dirac's method is still used as a path to quantisation. In fact, its extension to infinite-dimensional manifolds is almost universally carried out overlooking underlying functional analysis subtleties. 

One of the many problems with quantising the constraints in the Hamiltonian formalism of general relativity in terms of metrics is that they become non-polynomial when promoted to operators. On the other hand, if one recasts the Hamiltonian into the tetrad formalism, arduous second class constraints appear. Second class constraints have non-vanishing Poisson bracket with at least one other constraint, which makes them unsuitable for canonical quantisation in terms of Poisson brackets. Dirac developed a method to generalise Poisson brackets with his so-called Dirac's brackets.\\ In the context of finite-dimensional systems, Sniatycki \cite{sniatycki1974dirac} proved a global theorem that supports the use of Dirac's brackets. However, to our knowledge, there is no infinite-dimensional version of Sniatycki's theorem and in practice, it is difficult to quantise constrained classical field theories without first removing second class constraints.

In 1986, Ashtekar breathed new life into the canonical quantum gravity program by finding a simple Hamiltonian formulation of complex general relativity in terms of an $SO(3)$ Yang-Mills connection and its associated canonical momentum \cite{ashtekar1986new}. Nevertheless, one needed to impose reality conditions. Concurrently, Witten showed that the tetrad Palatini formulation of $2+1$ gravity was equivalent to Chern-Simons theory of the inhomogenous Lie Group $ISO(2,1)$ and thus could be explicitly canonically quantised \cite{wittenquantum}. 

Despite being an incredibly powerful method, the Hamiltonian framework relies on a choice of time that somehow obscures, but still preserves, the covariance of the theory. This spurred some explicitly covariant lines of research in the hopes of tackling the problem of quantum gravity without having to deal with the problem of time \cite{isham1993canonical}. Among these lines of inquiry, geometric quantisation  \cite{woodhouse1997geometric} and the covariant path integral approach \cite{rovelli2015covariant} are particularly relevant for us. 

In 1995, Holst added a term to the Palatini action, which vanished over Palatini's space of solutions. The amplitude of Holst's term is controlled by the \emph{Barbero-Immirzi parameter} $\gamma$, which is of great importance in non-perturbative approaches to quantum gravity such as loop quantum gravity and spinfoam theory \cite{Perez:2012wv}. The Holst term is unique in that, despite not being a total derivative, it does not change the dynamics. Whether the Barbero-Immirzi parameter appears or not in the symplectic structure of the classical field theory has far-reaching consequences in the quantisation of general relativity \cite{dittrich2013role}. In fact,  the canonical formulation derived from the Holst action leads to a family of $SU(2)$ connection formulations related by canonical transformations in the phase space of general relativity, which are labelled by this parameter $\gamma$ \cite{liu2010topological}.

Moving away from the turbulent arenas of quantum gravity that motivate this work, from now on we shall solely focus on the classical treatment of the field theories under consideration. The importance of building a symplectic manifold is that it is an essential ingredient to implement Dirac's quantum conditions within a quantisation program. Our study of the covariant phase space of gravitational field theories with boundaries might be deemed a classical prelude to their quantisation. Depending on the field theory at hand, our procedure will give us a presymplectic rather than symplectic form if the action has gauge symmetries. To obtain a truly symplectic form and prepare the theory for quantisation, we would need to quotient the covariant phase space by the gauge symmetries. Nevertheless, we do not carry out this last step because we are not going to quantise them.

In this thesis, we will apply the covariant phase space program for the three aforementioned gravitational actions in manifolds with boundaries. One of the aims of this program is to characterise the solution spaces for the field equations given by several action principles and build appropriate (pre)symplectic structures on them. The pivotal attribute of this approach is its explicit covariance. The principal novelty of the present work is the utilisation of the recently developed relative bicomplex framework: a method that extends the standard covariant phase space approach to deal with boundaries \cite{CPS}. Relying on this new formalism, we have been able to compute the (pre)symplectic structures and symmetries of gravitational action principles in manifolds with timelike boundaries. This has allowed us to compare different formulations of the same action principle and remove the perceived disparity between the particular use of certain variables, which are the metrics and tetrads. Furthermore, the relative bicomplex allowed us to include the boundary terms and compute the associated (pre)symplectic potentials that ensure the equivalence of the (pre)symplectic forms. 

This thesis is structured as follows. Chapter \ref{historicaloverview} provides a historical overview of the origins and importance of symplectic geometry, first in classical mechanics and later in field theories. Chapter \ref{theCPS} describes the mathematical machinery of the relative bicomplex and summarises its implementation with what we call the Covariant Phase Space (CPS) algorithm. Chapter \ref{EHaction} applies the CPS algorithm to the Einstein-Hilbert action in metric and tetrad variables including the Gibbons-Hawkings-York boundary term. Chapter \ref{Palatiniaction} follows a similar approach for the Palatini action with torsion, nonmetricity and boundaries, both in terms of metric and tetrad variables. Chapter \ref{HMS} presents the Hojman-Mukku-Sayed action, a new generalisation of the Holst action including torsion, nonmetricity and boundaries, both in the metric and tetrad formulations. The main results of the thesis are discussed in Chapter \ref{equivalence}, where the equivalence between the actions and formulations computed in the preceding chapters is shown.

%% file: historicaloverview.tex
\chapter{Historical Overview}\label{historicaloverview}

The concept of symplectic structure first appeared in the works of Joseph-Louis Lagrange and Siméon Denis Poisson on the slow variations of orbital elements of the planets of the solar system carried out from 1808 to 1810. The word \emph{symplectic} is derived from the proto-indo-European root ``plek'' from which  ``com-plex'' originates in Latin and from which $\sigma\upsilon\mu\pi\lambda\epsilon\kappa\tau\iota\kappa \acute{o}\varsigma$ (sym-plek-tikos) does in Greek, which literally translates to English as ``braided'' or ``plaited together''. The German mathematician Hermann Weyl in his book ``The Classical Groups'' decided to replace the Latin root of the \emph{linear $\textbf{complex}$ group} $Sp(n, \mathbb{R})$ with its Greek counterpart, coining the modern name of the \emph{\textbf{symplectic} group}, sparing us much confusion \cite{WeylBook}.  The mathematical point of this linguistic deviation is that the symplectic structure is essentially determined by the subspaces on which it vanishes and that its physical interpretation might be described as the intertwining of the canonical coordinates $p_{i}$ and $q_{i}$ in Hamilton's equations \cite{weinstein1981symplectic}. Because of the physical utility of symplectic geometry, a brief historical detour follows for the interested reader. 

\section{Mechanics}
From 1774 to 1784, several mathematicians and astronomers, notably Laplace and Lagrange, used the laws of classical Newtonian mechanics to calculate the slow variations of the orbital elements of planets in the solar system, without neglecting their masses or the gravitational pull exerted by the other planets \cite{Marle2009}. These orbital elements are six scalar quantities that determine the orbits at any time. Lagrange studied the ``drift'' experienced by the elliptical orbit due to the other planets' gravitational influence and described it with a system of differential equations. By doing so, he not only improved on the results of Laplace, but introduced an important new concept: the \emph{manifold of motions}. The manifold of motions $\mathbf{E}$ consists of the set of all possible Keplerian unperturbed motions of a planet, and he proved that it had locally the structure of a smooth manifold. A member of $\mathbf{E}$ is described by the six numbers describing its orbital motion called \emph{elements}. In his paper of 1808 \cite{Lagrange1808}, Lagrange realised many calculations were simplified when expressed on the set of local coordinates as a composition law of this \emph{manifold of motions}, a law known today as the Lagrange bracket. For any system of elements, $a_{1},..., a_{2n}$, he constructed a bracket $(a_{i}, a_{j})$ with $1 \leqslant i,j \leqslant 2n$, such that it was a function of $\mathbf{E}$. The bracket satisfied the \emph{antisymmetry} condition $(a_{i}, a_{j}) = - (a_{j}, a_{i})$. Then, he defined a real-valued function $\Omega$ on $\mathbf{E}$ to describe the dynamics of the perturbations, called the \emph{disturbing function}, which depended only on the forces exerted by the system:
\begin{equation*}
    \frac{\partial \Omega}{\partial a_{i}} = \sum_{j = 1}^{2n} (a_{i}, a_{j}) \frac{da_{j}}{dt} \\ \hspace{1cm} 1 \leqslant i,j \leqslant 2n.
\end{equation*}
The determinant of the \emph{matrix} whose coefficients are the functions $(a_{i}, a_{j})$ was non-zero, so it was invertible and solving for the time derivatives led to
\begin{equation*}
    \frac{da_{i}}{dt}(t) = \sum_{j=1}^{2n} L_{ij} \frac{\partial \Omega}{\partial a_{j}}\\ \hspace{1cm} 1 \leqslant \textit{i,\;j} \leqslant 2n.
\end{equation*}
Lagrange realised that the antisymmetry of the quantities $L_{ij}$ meant that the total derivative of $\Omega$ with respect to time along a solution of the equation was zero. Thus, the \emph{disturbing function} $\Omega$ was a conserved quantity for the drift motion \cite{Lagrange1809}. He also noticed that the quantities $L_{ij}$ were functions of $a_{1}, ...,a_{2\text{n}}$ but did not give their explicit expression. In 1808, Poisson \cite{Poisson1809} filled this gap by introducing another composition law on the set of differentiable functions defined on $\mathbf{E}$, today called the \emph{Poisson bracket} $\{a_{i}, a_{j}\}$\footnote{The Poisson bracket has another important property, which Lagrange and Poisson ignored: it satisfies the Jacobi identity, discovered by the German mathematician Carl Gustav Jacobi (1804-1851), whose role was crucial in the theory of continuous groups and algebras developed by the Norwegian mathematician Marius Sophus Lie (1842-1899).}. Lagrange understood that the result obtained by Poisson was due to a hitherto unnoticed underlying structure of the equations which govern the slow variation of the orbital parameters and used the Poisson bracket to rewrite the previous equations in the form 
\begin{equation*}
    \frac{da_{i}}{dt}(t) = \sum_{j=1}^{2n} \{ a_{i}, a_{j} \} \frac{\partial \Omega}{\partial a_{j}} \\ \hspace{1cm} 1 \leqslant \textit{i,\;j} \leqslant 2n.
\end{equation*}
The relation between the Lagrange parenthesis and the Poisson bracket would need to wait another twenty years to be unravelled. In 1831, Cauchy, without the explicit use of the word \emph{matrix}, showed that the Lagrange and Poisson brackets of some coordinate functions were inverses of one another. Nevertheless, Lagrange proved his brackets did not depend on time nor the nature of the force exerted on the system, but were a consequence of an intrinsic structure within the manifold of motions. The concept that Lagrange was illuminating was the symplectic structure.

In modern mathematical language these brackets $(a_{i}, a_{j})$ are the components in the local coordinate system formed by the $2n$ parameters  $a_{1}, ...,a_{2\text{n}}$, of the natural symplectic form of the manifold of motions of the mechanical system considered. 

In 1811, in the second edition of his monumental \emph{Mécanique Analytique} \cite{lagrangeBook}, Lagrange gave the expression of these parentheses in terms of what today would be called \emph{canonical coordinates}, $(q_{1}, ..., q_{n}; p_{1}, ..., p_{n})$ for which he found they satisfied certain properties: $(q_{i}, q_{j}) = (p_{i}, p_{j}) = 0$ and $(q_{i}, p_{j})= \delta_{ij}$, and wrote the equations above as
\begin{equation}
    \frac{d q_{i}}{dt} = \frac{\partial \Omega}{\partial p_{i}}, \\\hspace{1cm} \frac{d p_{i}}{dt} = - \frac{\partial \Omega}{\partial q_{i}},
\end{equation}
which are none other than Hamilton's differential equations\footnote{The great Irish mathematician and physicist William Rowan Hamilton (1805-1865) was only six years old when Lagrange published these equations.}. In 1834, Hamilton generalised Lagrange's equations to a general mechanical system. In his two major papers ``On a General Method in Dynamics'' and ``Second Essay on a General Method in Dynamics'' \cite{hamilton1834general, hamilton1835second}, Hamilton presented the formulation of the principle of least action in the time-independent case. In 1842, Jacobi further generalised Hamilton's equations to include time-dependent systems \cite{jacobi1866vorlesungen}. 
The invariant treatment of the Hamiltonian formalism was later studied by Élie Cartan (1869-1951), who realised the importance of having a coordinate-free description of the dynamics of physical systems and revealed the importance of the symplectic form in such an endeavour. His formalism was later a key ingredient of Noether's (1882-1935) theory of infinitesimal symmetries and conservation laws \cite{Chern1952}.

During the 1960s, Jean-Marie Souriau (1922-2012) presented a method for treating dynamical problems with finite degrees of freedom as Hamiltonian systems on symplectic manifolds using the language and techniques of modern symplectic geometry \cite{souriauBook}. He extended Lagrange's results and proved that, under very general assumptions, the set of all possible solutions of any classical mechanical system had a smooth manifold structure and was endowed with a natural symplectic form. He observed that the Lagrangian and Hamiltonian formalisms, in their usual guise, involved a particular reference frame and proposed a new space, which he called the \emph{evolution space} $V$ of the mechanical system \cite{Marle2009}. In his book, Souriau showed that the symmetry properties of $V$ were the same as those of the \emph{manifold of motions}, which we shall rename as $S$. 

The evolution space $(V, \sigma)$ of a classical mechanical system is a $(2N + 1)$-dimensional presymplectic manifold, such that $V= \mathbb{R} \times TQ$ where $Q$ is the configuration manifold and $\sigma$ is the so-called \emph{Lagrange-Souriau presymplectic 2-form}. Therefore, $V$ consists of all the initial conditions that satisfy the  differential equations of motion. Consider the smooth distribution $\mathcal{E}$ on $V$ (i.e. a smooth choice of linear subspaces $\mathcal{E}(t, q,v) \subset \mathbb{R} \times T_{(t,q,v)}(\mathbb{R}\times TQ) \cong T_{t}\mathbb{R}\times T_{(q,v)}TQ$ for each $(t, q, v)\in V$) defined by the equations of motion. Since $\mathcal{E}(t, q, v)$ is $1$-dimensional, it is integrable, and its integral curves are the solutions to the differential equation associated with the mechanical system. This means that $\mathcal{E}$ provides a foliation of $V$ where each leaf is identified with a point of the manifold of motions $x \in S$. 
\begin{EvalBox}{}
\begin{center}
    \begin{tikzpicture}
        %spaces 
        \draw[thick] (0,0) -- (4,0) node[anchor=west]{$\mathbb{R}$};
        \draw[thick]  (4,1) .. controls (5,3) and (4,3) .. (4,3.5);
        \draw[thick]  (0,1) .. controls (1,3) and (0,3) .. (0,3.5);
        \draw[thick]  (0,3.5) -- (4,3.5);
        \draw[thick]  (0,1) -- (4,1);

        \draw[thick] (9,2.25) ellipse (2cm and 1.5cm);
        %curvas en evolution space
        \draw (0.25,1.5) .. controls (2,1.6) and (2,1.25)..  (4.15,1.3);
        \draw (0.45,2.2) .. controls (2,1.7) and (2.5,1.4)..  (4.4,2.2);
        \draw (0.45,2.6) .. controls (2.5,2) and (3.5,2.4)..  (4.45,2.6);
        \draw (0.25,3) .. controls (2,2.8) and (2,2.7)..  (4.3,3);
        %proyecciones
        \draw[thick,dashed] (2,0) -- (2,2.27);
        \filldraw[black] (2,2.28) circle (2pt) node[anchor=south]{$(t, q, v)$};
        \draw[thick] (1.2,2.35) -- (2.9,2.25)node[anchor=north]{$\mathcal{E}$};
        \filldraw[black] (2,0) circle (2pt) node[anchor=north]{$t$};
        %flecha
        \draw [dashed](4.45,2.6) --  (9, 2.6);
        \filldraw[black] (9,2.6) circle (2pt) node[anchor=west]{x};
    \end{tikzpicture}
\end{center}
\end{EvalBox}

Therefore, the manifold of motions is the set of solutions to the differential equations, defined as the evolution space quotient by an isotropic foliation, which is given by a distribution that generates the dynamics.
\begin{align*}
    \text{manifold of motions} = \frac{\text{evolution space}}{\text{isotropic foliation}}
\end{align*}
After doing the quotient, the presymplectic form $\sigma$ projects to a genuine symplectic form on $S$. Souriau stated the closedness of $\sigma$ is a fundamental principle of Mechanics, which he called the \emph{Maxwell principle}. He asserted that physically, the closedness of the symplectic form implies that the laws of physics do not change. In conclusion, the core of Souriau's point of view was identifying the phase space of a dynamical system with its space of classical solutions, and this identification required a conceptual shift in thinking about the system's time evolution. 

\section{Field Theory}
In the early 1950s, the mathematical field of symplectic geometry had proven its physical usefulness as an elegant and powerful tool to generalise both Hamiltonian and Lagrangian formulations of mechanics and provided a route to the ``geometrisation'' of classical physics \cite{ arnol2013mathematical, moser2005notes, abraham2008foundations, gotay1992symplectization}. This success was expected to extend to the study of field theories. It is not completely clear how a formulation of a field theory, via the principle of least action, arises as a limiting case from a formulation of a mechanical system whose number of degrees of freedom is finite, because the generalisation requires knowledge of how to deal with infinite-dimensional manifolds. A rigorous approach was presented by Paul Robert Chernoff (1942-2017) and Jerrold E. Marsden (1942-2010) by modelling symplectic manifolds on Banach or Fréchet spaces, with the caveat of the results being highly dependent on the geometry and hard non-linear analysis\footnote{The formal problem is even more intricate since when choosing an action principle, we need to specify the topology of the base manifold, the domain of the fields and the topology of the space of fields.} \cite{Reyes2004,Chernoff2006}.

Peter J. Olver gave another formal and more algebraic approach \cite{Olver1980}, which generalised the notion of an integral invariant in an infinite-dimensional manifold to better understand the conservational roles of the symmetry groups of the system \cite{OlverBook}. His approach put aside a more geometrical view of the phase space and replaced the symplectic manifold with a \emph{Hamiltonian differential operator}. His viewpoint has been developed and applied with great success to integrable systems, giving rise to the important structure of a \emph{bi-Hamiltonian system}, which encodes the intuitive meaning of integrability for partial differential equations. However, the question of how to construct a phase space for a classical field theory still lingered. 

\subsection{Canonical Approach}

In the late 1950s, R. Arnowitt, S. Deser and C. W. Misner (ADM) proposed a novel Hamiltonian formulation of General Relativity \cite{adm2}. Their seminal work was a forerunner of Wheeler's geometrodynamics and had far-reaching consequences in the quantisation program of the theory \cite{wheeler}. Their formulation, which has been renamed as \emph{canonical} due to its ubiquity, consists of the 3+1 decomposition of spacetime through a foliation by spacelike hypersurfaces, on which the three-dimensional counterparts of the intrinsic curvature and stress-energy tensors are defined. Once the Einstein-Hilbert Lagrangian is recast into the ADM Hamiltonian, the variational principle gives rise to a constrained set of differential equations. In summary, the canonical formalism builds the symplectic and Poisson structures on the phase space of a field theory and requires an explicit choice of time function and the use of the fibre derivative \cite{Sniatycki1984}. Because the field equations of general relativity are hyperbolic, the natural framework to analyse its dynamics is to view them as a Cauchy problem, which must depart from explicit, manifest covariance of the theory since spacetime is being ruptured. In 1987, Crnkovi\v{c} and Witten suggested that although the spacetime symmetry is no longer manifest and not evident from the equations, it must still be present and that a covariant description of the canonical formalism was possible \cite{crnkovic1987covariant}. They found that symmetries could be analysed by Hamiltonian methods, providing crucial insights for the entire framework since there is still an underlying symplectic and Poisson geometry. Nevertheless, the loss of explicit covariance is usually perceived as an imperfection, from a physical point of view, and as a blunder from a mathematical perspective, since one would like to present the dynamics in an intrinsic, global fashion \cite{BojowaldBook}. 

\subsection{Covariant Approach}

Many attempts to develop a fully covariant formalism in classical field theory have been made over the past decades to serve as a starting point for alternative methods of quantisation \cite{Forger2005}.
There are several covariant approaches to the canonical formulation of classical relativistic field theories \cite{Gieres}. The oldest one, based on the covariant phase space, which is defined as the (infinite-dimensional) space of solutions of the equations of motion, will be the basis of the formalism employed in this thesis. Other formalisms, such as the multisymplectic approach \cite{dedonder1935} and the Peierls bracket \cite{peierls1997commutation,dewitt2003global}, will not be discussed here. 

To maintain explicit covariance and avoiding the splitting of spacetime into components, one school of thought developed a geometrical understanding of the structure of the phase space through the development of the \emph{covariant phase space}. The development of the covariant phase space formalism was also prompted by the desire to understand conservation laws in field theories and improve upon the canonical quantisation program of Dirac and Bergmann \cite{Khavkine2014}. The root of the covariant phase space formalism dates back to 1953 when Bergmann and Schiller published a paper in which they studied the implications of general covariance in general relativity or any other second-order covariant theory and wrote the first formulas for the presymplectic current density and its potential \cite{bergmann1953classical}. In 1960, Segal discussed the canonical quantisation of field theories with non-linear hyperbolic equations by identifying their phase space with the space of solutions and endowing it with a Poisson structure in a covariant way \cite{segal1960}. In 1975, Ashtekar and Magnon \cite{ashtekar1982symplectic}, presumably inspired by the ideas of Souriau, generalised his ideas to field theories and continued advertising the identification of the phase space with the space of solutions rather than with that of initial data. In 1986, G. Zuckerman \cite{Zuckerman1987} was the first person to explain, in full generality, how to build phase spaces for Lagrangian theories, with either finite or infinite degrees of freedom, in a covariant way without going through Dirac's theory of constraints. He used formal arguments of the rigorous theory of the variational complex and gave the first step towards finding a description of the dynamics à la Marsden. In 1987, a paper by Crnkovi\v{c} and Witten \cite{crnkovic1987covariant} rediscovered this result and achieved the acceptance of the covariant phase space approach by giving explicit covariant constructions for the symplectic forms of the usual scalar fields, Yang-Mills theories and general relativity. This work has been further elaborated and applied, in particular in the context of gravity, following the works of R. Wald, A. Ashtekar, and their collaborators \cite{lee1990local, wald1990identically, ashtekar1991mechanics}. 
More recent studies of the covariant phase space of field theories without boundaries include \cite{julia2002covariant} and with boundaries \cite{harlow2020covariant}, which are in agreement with our results. In \cite{CPS}, the authors presented the \emph{relative bicomplex framework}, a result of merging an extended version of the \emph{relative framework}, a method in algebraic topology for the incorporation of boundaries, and the \emph{variational bicomplex framework}. The relative bicomplex is able to treat field theories with boundaries in an unambiguous way, completely determining its presymplectic structure, currents and charges by relying on cohomological notions. The following chapter contains all the technical details of this framework and relevant examples in which its utility is made evident. We want to stress that the true potential of this method is its cohomological foundation that provides a unified and consistent treatment of manifolds with boundaries, allowing us to compare Lagrangian pairs in terms of equivalence classes.

%% file: CPS.tex
\chapter{The Covariant Phase Space}\label{theCPS}
The covariant phase space $\mathtt{CPS(\mathbb{S}})$ of a field theory is nothing but the space of solutions to the equations of motion obtained from an action principle $\mathbb{S}$ that naturally equips it with a presymplectic structure. The presence of this presymplectic structure sparked the interest of many physicists and mathematicians \cite{witten1986interacting}. The importance of the presymplectic structure is mainly due to its pivotal role in Dirac's quantisation procedure. Although our study is restricted to classical field theories, it might be considered a preamble to the geometric quantisation of the models studied in this thesis. The reason for considering the presymplectic form of the $\mathtt{CPS(\mathbb{S}})$ rather than the canonical one in the classical phase space (i.e. the cotangent bundle) is due to our desire to understand the Hamiltonian dynamics of Lagrangian field theories without breaking covariance. 

In this chapter, we will present the relative framework \cite{bott1982differential}, the variational bicomplex \cite{anderson1989variational} and the merging of these two into the relative bicomplex \cite{CPS}. The relative bicomplex will allow us to construct the presymplectic potentials, the presymplectic form and the associated conserved currents and charges associated with the $\mathtt{CPS(\mathbb{S}})$ of a field theory defined on a manifold with boundaries. Here we present the CPS algorithm, a systematic procedure to obtain these structures. To illustrate its application, we will study two relevant examples: the scalar field and Yang-Mills theories with boundaries.

The scalar field is a relevant example because it is omnipresent in many effective field theory descriptions, such as the pion or Higgs field. Scalar fields are also employed in the so-called  \emph{quartic interaction} theories, where the potential included in the Lagrangian is proportional to the scalar field exponentiated to the fourth power. Being able to systematically compute the presymplectic form of scalar field theories may help clarify how to quantise them.

In physics, the simplest example of a Yang-Mills theory is Maxwell's theory of electromagnetism. Furthermore, except for gravitation, the most influential theories of modern physics are quantised versions of Yang-Mills theories. These include quantum electrodynamics, the electroweak theory of Salam and Weinberg, the standard model of particle physics, and the GUTs (grand unified theories) proposed in the 1970s as its extensions. Arguably, the most important one is the standard model, which is a gauge theory with a Yang-Mills sector and with a $U(1)\times SU(2)\times SU(3)$ gauge symmetry. A critical point about Yang-Mills theories is that they are renormalisable.

\section{The Relative Framework}
The relative framework was thoroughly developed by Raoul Bott and Loring W. Tu in their book \cite{bott1982differential}, and is based on the idea that Thom classes of an oriented vector bundle can be viewed as relative cohomology classes. This means that for a given pair of smooth manifolds $(M,N)$ related by an inclusion map $\jmath: N \hookrightarrow M$ such that $N$ is of codimension $1$ in $M$. It is possible to define a Grassmann algebra (see \cite{CPS}) such that the $k$-form fields over the manifold pair are elements of
\begin{equation}
    \Omega^{k}(M, N) := \Omega^{k}(M) \oplus \Omega^{k-1}(N).
\end{equation}
This space of $k$-forms is endowed with the following operations. Let  $(\alpha, \beta) \in (M, N)$, $V \in \mathfrak{X}(M)$ such that $\overline{V}:= V |_{\jmath(N)} \in \mathfrak{X}(\jmath(N))$ then, 
\begin{align*}
    &\underline{\mathrm{d}}(\alpha, \beta):= (\mathrm{d}\alpha, \jmath^{*}\alpha - \mathrm{d}\beta),\\
    &\underline{\imath}_{V}(\alpha, \beta):=(\imath_{V}\alpha, - \imath_{\overline{V}}\beta),\\
    & \underline{\mathcal{L}}_{V}(\alpha, \beta):= (\mathcal{L}_{V}\alpha, \mathcal{L}_{\overline{V}}\beta).
\end{align*}
Notice that in particular $\underline{\mathrm{d}}^{2} = 0$, and hence we obtain the relative cohomology  $(\Omega^{k}(M, N), \underline{\mathrm{d}})$. 

Because we are interested in action principles, the integral of top-forms over this relative framework is defined as: 
\begin{equation}\label{lagrangianpairintegral}
    \int_{(M,N)} (\alpha, \beta) := \int_{M} \alpha - \int_{N} \beta\;.
\end{equation}
If $M$ is a manifold with boundary $\partial M$, we will assume that $N \subset \partial M$. Then, the relative boundary is defined as $\underline{\partial}(M, N):= (\partial M  \backslash N, \partial N)$ whose relative inclusion is defined as $\underline{\jmath}: \underline{\partial}(M,N) \hookrightarrow (M,N)$, with the following property  $\underline{\partial}^{2} = 0$.
A theorem we shall repeatedly use is Stokes' theorem. Its relative version is given by 
\begin{equation}
    \int_{(M,N)} \underline{\mathrm{d}}(\alpha, \beta) = \int_{\underline{\partial}(M, N)} \underline{\jmath}^{*}(\alpha, \beta),
\end{equation}
where $\jmath: \partial M \hookrightarrow M$ and $\overline{\jmath}: \partial N \hookrightarrow N$. 
The relative framework allows us to combine forms in manifolds with boundaries as a single entity. This point is crucial because, when dealing with variational problems, the action may need boundary terms to describe a theory with well-posed field equations. Once the pair of forms is defined in this framework, their cohomology is weaved together.

\section{The Variational Bicomplex}

The variational bicomplex naturally generalises the notion of variational calculus using the machinery of jet bundles. The theory of bundles provides a coordinate-free way of defining differential equations and fields in a fibre bundle whose base manifold is spacetime. More precisely, the variational bicomplex is a double de Rham complex of differential forms defined on the infinite jet bundle of a fibre bundle \cite{anderson1989variational}. For a formal and rigorous treatment of the geometry of infinite jet bundles the reader is referred to \cite{saunders1989geometry}.

Let $(E, \pi, M)$ be a vector bundle. A physical field is represented by a section of this bundle, $\phi \in \Gamma(E) \equiv \mathcal{F}$, where $\mathcal{F}$ is the space of fields. Let  $(J^{\infty}(E), \pi^{\infty}_{M}, M)$ be the infinite jet bundle of $M$, such that the fibre at $p \in M$ consists of the equivalence class of local sections sharing all derivatives at $p \in M$. For $r$-independent fields, the local coordinates of an element of the jet bundle are 
\begin{align*}
    j^{\infty}_{p}(\phi) \in J^{\infty}(E) \longrightarrow \{ x^{\alpha}, u^{i}, u^{i}_{\alpha}, u^{i}_{\alpha\beta}, \ldots \},
\end{align*}
where $x^{\alpha}$ are the coordinates of a coordinate patch in a open neighbourhood of $p$, $1 \leq i \leq r$ and $1 \leq \alpha, \beta < \infty$. The coefficients of the coordinates are explicitly, 
\begin{align*}
    & u^{i}_{\alpha}(j^{\infty}_{x}(\phi)) = \frac{\partial \phi^{i}}{\partial x^{\alpha}}(x),
    & u^{i}_{\mu\nu\ldots\alpha}(j^{\infty}_{x}(\phi)) = \frac{\partial^{l}\phi^{i}}{\partial x^{\mu} \partial x^{\nu} \ldots \partial x^{\alpha} }(x),
\end{align*}
where $l$ is a multi-index over $\{ \mu, \nu, \ldots, \alpha \}$. In particular note that, 
\begin{align*}
    &\frac{\partial u^{j}_{\alpha\beta}}{\partial u^{i}_{\mu\nu}}  = \delta_{\alpha}^{(\mu}\delta^{\nu)}_{\beta}\delta^{j}_{i}, &\frac{\partial u^{i}_{\alpha\beta}}{\partial u^{i}_{\alpha\beta}} = \frac{1}{2}.
\end{align*}
Therefore, the jet space of $E$ is the set of symmetrised derivatives of fields such that each component is treated as an independent variable \cite{anderson1989variational}. 

The infinite jet bundle has a natural structure as a differentiable manifold and hence we can define vector fields and $k$-forms upon it. Let $T^{*}_{\phi}\mathcal{F}$ be the cotangent space at $\phi \in \mathcal{F}$. Then an element $\phi^{*} \in T^{*}_{\phi}\mathcal{F}$ is a covector on $\mathcal{F}$ and is given in terms of coordinates as, 
\begin{align*}
    \phi^{*} :=\{\delta u^{i}, \delta u^{i}_{\alpha}, \delta u^{i}_{\alpha\beta}, \cdots \} \;.
\end{align*}
\noindent The variational operator $\delta$, or vertical derivative, is given in coordinates as, 
\begin{align*}
    \delta = \delta u^{i}\frac{\partial}{\partial u^{i}} + \delta u^{i}_{\mu}\frac{\partial}{\partial u^{i}_{\mu}} + \delta u^{i}_{\mu\nu} \frac{\partial}{\partial u^{i}_{\mu\nu}} + \cdots,
\end{align*}
where $\delta^{2} = 0$. This operator defines an exterior derivative on $\mathcal{F}$ and each component of $\phi^{*}$ is a $1$-form in the space of fields. Furthermore, this algebra in the space of fields naturally induces a de Rham complex,
\begin{align*}
    0 \xrightarrow{\delta} \Omega^{0}(\mathcal{F}) \xrightarrow{\delta}\Omega^{1}(\mathcal{F}) \xrightarrow{\delta} \ldots \xrightarrow{\delta}\Omega^{k-1}(\mathcal{F})\xrightarrow{\delta}\Omega^{k}(\mathcal{F}) \xrightarrow{\delta} \cdots.
\end{align*}

\noindent In the base manifold $M$, the exterior derivative or horizontal derivative $\mathrm{d}: \Omega^{k}(M) \xrightarrow{}\Omega^{k+1}(M)$ is given in terms of a basis of the tangent bundle $\{ \frac{\partial}{\partial x^{\mu}}\}$ and the cotangent bundle $\{ \mathrm{d}x^{\mu}\}$ as $\mathrm{d} =\mathrm{d}x^{\mu} \frac{\partial}{\partial x^{\mu}}$. But now, a basis of the tangent bundle is expressed in terms of the components of the infinite jet bundle,
\begin{align*}
    \partial_{\mu} = \frac{\partial}{\partial x^{\mu}} + u^{i}_{\mu} \frac{\partial}{\partial u^{i}} + u^{i}_{\mu\nu}\frac{\partial}{\partial u^{i}_{\nu}} + \cdots.
\end{align*}
The exterior derivative $\mathrm{d}$ also induces a de Rham complex in the base manifold:
\begin{align*}
     0 \xrightarrow{d} \Omega^{0}(M) \xrightarrow{d}\Omega^{1}(M) \xrightarrow{d} \ldots \xrightarrow{d}\Omega^{n-1}(M)\xrightarrow{d}\Omega^{n}(M) \xrightarrow{d} 0.
\end{align*}

\begin{tikzcd}
0 \\
0 \arrow{r}  & \Omega^{(0, k)}(M\times \mathcal{F}) \arrow{r} &\Omega^{(1, k)}(M\times \mathcal{F})\arrow{r}  &\cdots \arrow{r} &\Omega^{(n, k)}(M\times \mathcal{F}) \arrow{r} & 0\\
quad & quad & quad & 1 & quad & quad  \\
0 \arrow{r}  & \Omega^{(0, 1)}(M\times \mathcal{F}) \arrow{r} \arrow{u}  &\Omega^{(1, 1)}(M\times \mathcal{F})\arrow{r}  &\cdots \arrow{r} &\Omega^{(0, 1)}(M\times \mathcal{F}) \arrow{r} & 0\\
0 \arrow{r} & \Omega^{(0, 0)}(M\times \mathcal{F}) \arrow{r} \arrow{u}&\Omega^{(1, 0)}(M\times \mathcal{F})\arrow{r}  &\cdots \arrow{r} &\Omega^{(n, 0)}(M\times \mathcal{F}) \arrow{r} & 0\\
d\arrow{r} & e \arrow{r} & f \arrow{r} &g\\
h\arrow{r} \arrow{u} & i\arrow{u}
\end{tikzcd}

The variational bicomplex consists of considering the product manifold  $M\times \mathcal{F}$ to create a double de Rham complex on $(\Omega^{(r,s)}(M\times \mathcal{F}), \mathbf{d})$ where $\mathbf{d} = \mathrm{d} + (-1)^{r}\delta$ is given by the two differential operators $(\mathrm{d}, \delta)$ such that they commute $[\mathrm{d}, \delta] = 0$ as it is common in the physics literature \cite{wald1990identically} (in contrast to the definition $\{\mathrm{d}, \delta\} = 0$ in more mathematical frameworks). Elements of $\Omega^{(r,s)}(M\times \mathcal{F})$ will be referred to as bi-forms, since they carry two types of graded algebras. 
In this thesis, we will solely work with local sections and hence, although we allow for all derivatives in the jet bundle, each bi-form $\psi \in \Omega^{(r,s)}(M\times \mathcal{F})$ will only depend on a finite (but arbitrary) number of derivatives at $p \in M$, when evaluated with the corresponding vector fields. 

\noindent The operations in the variational bicomplex, for some bi-graded form $\alpha \in \Omega^{(r,s)}(M \times \mathcal{F})$, a vector field $\xi \in \mathfrak{X}(M)$ and a vector field $X \in \mathfrak{X}(\mathcal{F})$ are the following, 
\begin{align*}
    \mathbf{d} \alpha &:= (\mathrm{d} + (-1)^{r}\delta)\alpha = \mathrm{d}\alpha + (-1)^{r}\delta \alpha,  \\
    (\imath_{\xi} + \imath_{\mathrm{X}})\alpha &:= \imath_{\xi}\alpha + \imath_{\mathrm{X}}\alpha, \\
     (\mathcal{L}_{\xi} + \mathcal{L}_{\mathrm{X}})\alpha &:= \mathcal{L}_{\xi}\alpha + \mathcal{L}_{\mathrm{X}} \alpha.
\end{align*}
A vector field $\xi \in \mathfrak{X}(M)$ allows us to define a new associated vector field $\mathrm{X}_{\xi} \in \mathfrak{X}(\mathcal{F})$ as follows. Let $(TM, \tau_{M}, M)$ be the tangent bundle of $M$ and let $(T\mathcal{F}, \tau_{\mathcal{F}}, \mathcal{F})$ be the tangent bundle of $\mathcal{F}$. Then, the Lie derivative of an element of $\phi \in \mathcal{F}$ with respect to an element of $\xi \in \Gamma(\tau_{M})$ is another element of $\mathcal{F}$. This new element $\mathrm{X}_{\xi} \in \Gamma(\tau_{\mathcal{F}})$ defines another vector field which is given in components by, 
\begin{equation}\label{XandXi}
    \mathrm{X}^{I}_{\xi}(\phi) := \mathcal{L}_{\xi}\phi^{I},
\end{equation}
where we made an abuse of notation by identifying via isomorphism that $T_{\phi}\mathcal{F} \cong \mathcal{F}$, where we have assumed that $\mathcal{F}$ is linear. 

\section{The Relative Bicomplex}\label{relativebicomplex}
In \cite{CPS}, the authors merged the relative framework with the variational complex to build a robust and formal approach to deal with field theories with boundaries, coined as the \emph{relative bicomplex}. The relative bicomplex framework attains the formal equivalence between the relative version of a theory with boundaries and a non-relative version of a theory without boundaries. Assuming without loss of generality that $N = \partial M$, the Grassmann algebra of the relative bicomplex is defined combining the algebra of the previous two sections, 
\begin{equation}
    \boxed{\Omega^{(r,s)}\big( (M,\partial M)\times \mathcal{F}\big) := \Omega^{(r,s)}(M\times \mathcal{F}) \oplus \Omega^{(r-1,s)}(\partial M\times \mathcal{F})}
\end{equation}
endowed with an exterior derivative: 
\begin{align*}
    \underline{\mathbf{d}}= \underline{\mathrm{d}} + (-1)^{r}\underline{\delta}
\end{align*}
where $r$ is the number of independent fields of the theory and whose properties are directly given by those of the two graded operators. In the previous section, we gave the properties of $\mathrm{d}$ in the relative framework, namely $\underline{\mathrm{d}}$. The properties and definitions are now extended for the vertical part $\underline{\delta}$, such that for  $(\alpha, \beta),(\gamma, \sigma) \in \Omega^{(r,s)}((M, \partial M)\times \mathcal{F})$ and $X \in \mathfrak{X}(\mathcal{F})$ then,
\begin{align*}
\underline{\delta}(\alpha, \beta) &:= (\delta \alpha, \delta \beta),\\
    \underline{\imath}_{\mathrm{X}}(\alpha, \beta)&:= (\imath_{\mathrm{X}}\alpha, \imath_{\mathrm{X}}\beta), \\
    \underline{\mathcal{L}}_{\mathrm{X}}(\alpha, \beta)&:= (\mathcal{L}_{\mathrm{X}}\alpha, \mathcal{L}_{\mathrm{X}}\beta).
\end{align*}
For a complete list of properties the reader is referred to the original paper \cite{CPS}.

\section{Spacetime}\label{spacetime}
With the help of the relative bicomplex framework, we wish to study the covariant space space of gravitational theories with boundaries. To do so, we must fully specify the base manifold $M$ and the space of fields $\mathcal{F}$. Because the base manifold will be the same for all the theories we consider, let us first define $M$ and delay the definition of $\mathcal{F}$ for each action principle that we will consider. 

Let $M$ be a connected and oriented $4$-dimensional manifold admitting a foliation by Cauchy hypersurfaces. Without loss of generality, $M=I\times \Sigma$ for some interval $I=[t_i,t_f]$, ($t_i<t_f$) and some $(n-1)$-manifold $\Sigma$ with boundary $\partial\Sigma$. Denoting $\Sigma_i=\{t_i\}\times\Sigma$ and $\Sigma_{\!f}=\{t_f\}\times\Sigma$, we split $\partial M$ into three parts
\begin{align*}
    \partial M=\Sigma_{i} \cup \partial_{L} M\cup \Sigma_{f},
\end{align*}
where $\Sigma_{i}$ and $\Sigma_{f}$ are the ``lids'' and $\partial_{L} M:=I\times \partial \Sigma$ is the ``lateral boundary''. The ``corners'' of this manifold are given by $\partial(\partial_{L} M) = \partial\Sigma_i\;\cup\;\partial\Sigma_{\!f}$. The following diagram summarizes relevant notational information about embeddings and the induced geometric objects.
\begin{EvalBox}{}\label{diagram_spacetime}
\begin{center}
    \begin{tikzpicture}
  \matrix (m) [matrix of math nodes,row sep=3em,column sep=4em,minimum width=2em]
  {
       (\Sigma, \gamma, D) & (M, g, \nabla) \\
     (\partial \Sigma, \overline{\gamma}, \overline{D}) & (\partial M, \overline{g}, \overline{\nabla})\\};
  \path[-stealth]
    (m-1-1) edge node [above] {$(\imath, n^{\alpha})$} (m-1-2)
    (m-2-1) edge node [left] {$(\overline{\jmath}, \mu^{a})$} (m-1-1)
    (m-2-2) edge node [right] {$(\jmath, \nu^{\alpha})$} (m-1-2)
    (m-2-1) edge node [below] {$(\overline{\imath}, \overline{m}^{\alpha})$} (m-2-2);
\end{tikzpicture}
\end{center}
\end{EvalBox}
%Because our field theories rest upon an action principle, the initial and final lids will be fixed. Thus, the lateral boundary of $M$ is essentially given solely by $\partial M = \partial_{L}M$. 

The triples shown in the diagram are the manifold, the (non-degenerate) metric or pulled-back metric and its associated Levi-Civita connection. The indices associated with $M$ will be denoted by Greek letters $(\alpha, \beta, \ldots)$, and the indices associated with the $\partial M$ will be denoted by over-lined Greek letters $(\overline{\alpha}, \overline{\beta}, \ldots)$. The indices associated with $\Sigma$ will be denoted by Latin letters $(a,b,\ldots)$ and the ones associated with $\partial \Sigma$ will be denoted by over-lined Latin letters $(\overline{a},\overline{b},\ldots)$. As $(M,g)$ is oriented, we have the metric volume form $\volg$ that assigns the value $1$ to every positive orthonormal basis. The duets that go from one triple to another are the inclusion maps and unit outward pointing vectors to each surface.  We orient $\Sigma$ and $\partial_{L} M$ with $\mathrm{vol}_{\gamma}$ and $\mathrm{vol}_{\overline{g}}$, respectively given by
\begin{align}\label{eq: orientation sigma}
\imath^*(\iota_{U}\volg)=-n_\alpha U^\alpha\mathrm{vol}_{\gamma},\qquad\qquad\qquad \jmath^*(\iota_{U}\volg)=\nu_\alpha U^\alpha\mathrm{vol}_{\overline{g}},
\end{align}
for every vector field $U \in \mathfrak{X}(M)$. These orientations are the ones for which Stokes' theorem holds in its usual form. Finally, $\partial\Sigma$ can be oriented as the boundary of $\Sigma$. Thus $\mathrm{vol}_{\overline{\gamma}}$ is given by,
\begin{align}
    \overline{\jmath}^{*}(\iota_{V}\mathrm{vol}_{\gamma})=\mu_{a} V^{a}\mathrm{vol}_{\overline{\gamma}}, \quad\quad\quad\quad \overline{\imath}^*(\iota_{W}\volgg)=\overline{m}_{\overline{\alpha}} W^{\overline{\alpha}}\mathrm{vol}_{\overline{\gamma}},
\end{align}
where $V \in \mathfrak{X}(\Sigma)$ and $W \in \mathfrak{X}(\partial M)$. Because our field theories rest upon an action principle, the initial and final lids will be fixed.  Also, in the following  we will consider $M$ to be topologically $\mathbb{R}\times \Sigma$ and hence $\partial M = \mathbb{R} \times \partial \Sigma$. This means that the relative boundary of $(M, \partial_{L}M)$ is empty since $\partial M = \partial_{L}M$. Therefore, in this thesis, we are not considering corner terms within the relative bicomplex.
Furthermore, the inclusion of the Cauchy hypersurface and its boundary as a relative pair into the spacetime manifold and its boundary is given by
\begin{equation}\label{underlineimath}
    \underline{\imath}:(\Sigma, \partial \Sigma) \hookrightarrow (M, \partial M).
\end{equation}

\section{CPS Algorithm}\label{cpsalgorithm}
The CPS algorithm is a procedure to unambiguously compute the presymplectic structure of the space of solutions of a field theory with boundaries, making use of the relative bicomplex framework. To this end, the action principle needs to be defined in terms of a Lagrangian pair, but without forgetting that the presymplectic form of the CPS is tied to the choice of action, not to the Lagrangian pair. This means the Lagrangian pairs belong to an equivalence class in the relative bicomplex cohomology. Therefore, two Lagrangian pairs $(L_{1}, \overline{\ell}_{1}),(L_{2}, \overline{\ell}_{2})\in \Omega^{(n, 0)}((M, \partial_{L} M) \times \mathcal{F})$ are in the same equivalence class if and only if there exists $(Y ,\overline{y}) \in \Omega^{(n-1,0)}\big((M, \partial M)\times \mathcal{F} \big)$ such that,
\begin{align*}
    (L_{2}, \overline{\ell}_{2}) = (L_{1}, \overline{\ell}_{1}) + \underline{\mathrm{d}}(Y, \overline{y}) \Longleftrightarrow 
    \begin{cases}
        L_{2} = L_{1} + \mathrm{d}Y, \\
        \overline{\ell}_{2} = \overline{\ell}_{1} + \jmath^{*}Y  - \mathrm{d}\overline{y},
    \end{cases}
\end{align*}
i.e. $ [L_{1}, \overline{\ell}_{1}] = [L_{2}, \overline{\ell}_{2}]$. This cohomological equivalence between Lagrangian pairs will be of importance when comparing presymplectic structures and studying their equivalence. 
With this in mind, a local action is a functional $\mathbb{S}:\mathcal{F} \equiv \Gamma(E) \xrightarrow{} \mathbb{R}$ defined as an integral of a Lagrangian pair $(L, \overline{\ell})$,
\begin{align*}
    \mathbb{S}(\phi) = \int_{(M, \partial_{L} M)} (L, \overline{\ell})(\phi) = \int_{M} L(\phi) - \int_{\partial_{L} M}\overline{\ell}(\phi).
\end{align*}
The variational bicomplex is equipped with a total exterior derivative $\mathbf{d} = \mathrm{d} + (-1)^{r}\delta$, such that $\mathrm{d}$ defines a cochain in $M$ (horizontal part) and $\delta$ defines a cochain in the space of fields $\mathcal{F}$ (vertical part). The variation of the action  $\mathbb{S}(\phi)$
is taken only in terms of the vertical differential. Hence only the vertical derivative corresponds to the usual variational calculus,
%, since $ \underline{\mathrm{d}}(L, \overline{\ell}) = (\mathrm{d}L, \jmath^{*}L - \mathrm{d}\overline{\ell}) = (0,0) \in \Omega^{(n+1, 0)}((M, \partial_{L}M)\times \mathcal{F}$. 
\begin{equation}\label{variationAction}
    \delta \mathbb{S} = \int_{(M, \partial_{L} M)} \underline{\delta}(L, \overline{\ell}) = \int_{(M, \partial_{L} M)} (\delta L, \delta \overline{\ell}) = \int_{M} \delta L - \int_{\partial_{L} M}\delta \overline{\ell}.
\end{equation}
Assuming these variations decompose as \cite{anderson1989variational},
\begin{equation}\label{variationBulk}
    \underline{\delta} (L, \overline{\ell}) = (E_{I}, \overline{b}_{I}) \wedge \delta \phi^{I}  + \underline{\mathrm{d}}(\Theta, \overline{\theta}) \&= (E_{I}\wedge \delta \phi^{I} ,\overline{b}_{I}\wedge \delta \phi^{I}) + (\mathrm{d}\Theta, \jmath^{*}\Theta - \mathrm{d} \overline{\theta}),
\end{equation}
then the equations of motion are given by the so called \textit{Euler forms} $(E_{I}, \overline{b}_{I})$ when imposing the principle of minimal action and $(\Theta, \overline{\theta})$ are the presymplectic potentials. The critical points of the action are given by the fields that satisfy the equations of motion 
\[
(E_{I}, \overline{b}_{I})=0.
\]
The covariant phase space 
\begin{equation}\label{cpspair}
    \mathtt{CPS(\mathbb{S}}) := \big( \mathtt{Sol}(\mathbb{S}), \Omega_{\mathbb{S}}\big)
\end{equation}
is precisely defined as the space of solutions to these equations,
\begin{equation}
     \mathtt{Sol}(\mathbb{S}) = \{ \phi \in \mathcal{F}\; |\; (E_{I}, \overline{b}_{I})(\phi) = 0 \},
\end{equation}
equipped with the presymplectic form,
\begin{equation}\label{presymplecticStructure}
    \Omega_{\mathbb{S}} =  \delta \Big( \int_{\Sigma}  \imath^{*}\Theta - \int_{\partial \Sigma} \overline{\imath}^{*}\overline{\theta} \Big).
\end{equation}
This form is constructed through the variation of the presymplectic potentials and it is independent \emph{on-shell} of the leaf of the foliation and of the Cauchy embeddings \cite{CPS,harlow2020covariant}.

The presymplectic structure can be shown to be the same for any choice of Lagrangian pair within an equivalence class (proof in \cite{CPS}). This means that if we had computed the presymplectic structure given by another Lagrangian pair within the same equivalence class, their presymplectic potentials would be related by,
\begin{equation}\label{presymplecticpotentialequivalence}
    (\Theta_{2}, \overline{\theta}_{2}) = (\Theta_{1}, \overline{\theta}_{1}) + \underline{\mathrm{d}}(Z, \overline{z}) + \underline{\delta}(Y, \overline{y}),
\end{equation}
for some $(Y, \overline{y}) \in \Omega^{(n-1, 0)}((M, \partial M)\times \mathcal{F})$ and $(Z, \overline{z})\in \Omega^{(n-2, 0)}((M, \partial M)\times \mathcal{F})$. The gauge symmetries of the action define gauge orbits on $\mathtt{CPS(\mathbb{S}})$. 

%By declaring the point in these orbits to be equivalent, and taking the appropriate quotient, the resulting space will be equipped with a genuine symplectic form via pull-back.

The study of symmetries is a fundamental element of the CPS formalism. Within the relative bicomplex, several types of symmetries exist:
\begin{enumerate}
    \item A vector field  $\mathrm{X} \in \mathfrak{X}(\mathcal{F})$ is a symmetry of the action if and only if $\mathcal{L}_\mathrm{X}\mathbb{S} = 0$. 
    \item If $\mathrm{X} \in \mathfrak{X}(\mathcal{F})$ is a symmetry of the action then $\underline{\mathcal{L}}_{\mathrm{X}}(L, \overline{\ell})(\phi)$ is $\underline{\mathrm{d}}$-exact over $(M, \partial M)$, but $\underline{\mathcal{L}}_{\mathrm{X}}(L, \overline{\ell})$ might not be exact.
    \item A vector field $\mathrm{X} \in \mathfrak{X}(\mathcal{F})$ is a $\underline{\mathrm{d}}$-symmetry over $((M,\partial M)\times \mathcal{F})$, if \\ $\underline{\mathcal{L}}_{\mathrm{X}}(L, \overline{\ell}) = \underline{\mathrm{d}}(S_{\mathrm{X}}, \overline{s}_{\mathrm{X}})$ for some local   bi-form  $(S_{\mathrm{X}}, \overline{s}_{\mathrm{X}}) \in \Omega^{(n-1,0)}((M, \partial M)\times \mathcal{F})$.
    \item A vector field $\xi \in \mathfrak{X}(M)$ is a symmetry of a Lagrangian pair $(L, \overline{\ell})$ if \\ $\underline{\mathcal{L}}_{\xi}(L, \overline{\ell}) = \underline{\mathcal{L}}_{X_{\xi}}(L, \overline{\ell})$. We then we say that the Lagrangian pair is $\xi$-invariant.
\end{enumerate}

In fact, if $\mathrm{X}$ is a $\underline{\mathrm{d}}$-symmetry of the Lagrangian pair $(L, \overline{\ell})$ then $\mathrm{X}$ is a $\underline{\mathrm{d}}$-symmetry of its entire equivalence class $[L, \overline{\ell}]$. Also, if a Lagrangian pair is $\xi$-invariant, then the associated $\mathrm{X}_{\xi}$ vector field is a symmetry of the action.  Despite $\mathrm{X}_{\xi}$ not being in general a $\underline{\mathrm{d}}$-symmetry, it is still possible to define its associated charges, as we will see later on. 

Noether's theorems tell us that each symmetry has conservation laws associated with it, and  \underline{d}-symmetries are no exception. The generalisation of the Noether's currents \cite{wald1990identically} to the relative bicomplex in which boundaries are considered corresponds to generalised $\mathrm{X}$-currents defined by a $\underline{\mathrm{d}}$-symmetry. These $\mathrm{X}$-currents are bi-forms $(\mathcal{J}_{\mathrm{X}}, \overline{j}_{\mathrm{X}}) \in \Omega^{(n-1, 0)}((M, \partial M)\times \mathcal{F})$ such that, 
\begin{align}
    (\mathcal{J}_{\mathrm{X}}, \overline{j}_{\mathrm{X}}) = (S_{\mathrm{X}}, \overline{s}_{\mathrm{X}}) - \underline{\imath}_{\mathrm{X}}(\Theta, \overline{\theta}),
\end{align}
with associated $\mathrm{X}$-charges given by, 
\begin{align}
    \mathbb{Q}_{\mathrm{X}}:= \int_{\Sigma} \imath^{*}\mathcal{J}_{\mathrm{X}} - \int_{\partial \Sigma} \overline{\imath}^{*} \overline{j}_{\mathrm{X}}.
\end{align}

An important class of these $\mathrm{X}$-currents are those associated with vector fields in the base manifold $\mathrm{X}_{\xi} \in \mathfrak{X}(\mathcal{F})$ (see \ref{XandXi} for the explicit relation between vector fields in the spacetime manifold and on the field space). Therefore, for a given symmetry of the action $\xi \in \mathfrak{X}(M)$, the $\xi$-current of a Lagrangian pair is defined as, 
\begin{equation}
    (\mathcal{J}_{\xi}, \overline{j}_{\xi}) := \underline{\imath}_{\xi}(L, \overline{\ell}) - \underline{\imath}_{\mathcal{\mathrm{X}}_{\xi}}(\Theta, \overline{\theta}),
\end{equation}
and the associated $\xi$-charge is given by, 
\begin{align}\label{chargesformula}
        \mathbb{Q}_{\xi}:= \int_{\Sigma} \imath^{*}\mathcal{J}_{\xi} - \int_{\partial \Sigma} \overline{\imath}^{*} \overline{j}_{\xi}.
\end{align}

The currents and charges given by a Lagrangian pair are associated with any other representative of its equivalence class in the following manner. Let $(L_{1}, \overline{\ell}_{1})$ and $(L_{2}, \overline{\ell}_{2})$ be two Lagrangian pairs in the same equivalence class. Then their $\mathrm{X}$-currents are related by, 
\begin{equation}\label{currentsequiv}
    (\mathcal{J}^{\;2}_{\mathrm{X}}, \overline{j}^{\;2}_{\mathrm{X}}) = (\mathcal{J}^{\;1}_{\mathrm{X}}, \overline{j}^{\;1}_{\mathrm{X}}) + \underline{\mathrm{d}}\big( (A, \overline{a}) - \imath_{\mathrm{X}}(Z, \overline{z})\big),
\end{equation}
for some $(A, \overline{a}) \in \Omega^{(n-2, 0)}((M, \partial_{L}M) \times \mathcal{F})$ and $(Z, \overline{z}) \in \Omega^{(n-2,1)}((M, \partial_{L}M)\times \mathcal{F})$. 

%In fact, although every $\xi$-symmetry is a $\mathrm{X}$-symmetry of a Lagrangian pair, the reverse is not always true. 

It is worth noting that the $\mathrm{X}$-charges, as currently defined in the totality of $\mathcal{F}$, depend on the choice of action and the selected Cauchy embedding. When the equations of motion are computed and the solution space $\mathtt{Sol}(\mathbb{S})$ has been fully characterized, we can define the inclusion $\mathfrak{S}: \mathtt{Sol}(\mathbb{S}) \hookrightarrow \mathcal{F}$ and consider the restriction of a vector field $\mathrm{X}\in \mathfrak{X}(\mathcal{F})$ to the space of solutions $\mathtt{Sol}(\mathbb{S})$. We shall denote this restriction by $\overline{X} \in \mathfrak{X}(\mathtt{Sol}(\mathbb{S}))$ (see proposition III.24 of \cite{CPS} for more details). By pulling-back the $\mathrm{X}$-charges through this inclusion map, the dependence on the Cauchy embedding disappears,
\begin{equation}
    \mathbb{Q}_{\mathrm{\overline{X}}} = \mathfrak{S}^{*}\mathbb{Q}^{\underline{\imath}}_{\mathrm{X}}.
\end{equation}
That is, the $\mathrm{\overline{X}}$-charges solely depend on the choice of action and are independent of the representative within an equivalence class of Lagrangian pairs, as well as the selected Cauchy embedding.

To summarise, the CPS ``algorithm'' is a systematic procedure to fully determine without ambiguity the space of solutions and the presymplectic structure of a field theory defined on a manifold with boundaries. Its main steps are summarised below. 
\begin{EvalBox}{}
\begin{center}
CPS Algorithm
\begin{enumerate}
    \item $\mathbb{S} = \int_{M}L - \int_{\partial M}\overline{\ell}$
    \item $\delta L = E_{I}\wedge \delta \phi^{I} + \mathrm{d}\Theta$ and $\delta \overline{\ell} - \jmath^{*}\Theta = \overline{b}_{I}\wedge \delta \phi^{I} - \mathrm{d}\overline{\theta}$
    \item $\mathtt{Sol}(\mathbb{S}) = \{ \phi \in \mathcal{F} \; | \; (E_{I}, \overline{b}_{I})(\phi) = 0 \}$ 
    \item $\Omega = \delta \Big( \int_{\Sigma} \imath^{*}\Theta - \int_{\partial \Sigma} \overline{\imath}^{*}\overline{\theta} \Big)$ 
    \item $J_{\xi} = \imath_{\xi}L - \imath_{\mathrm{X}_{\xi}}\Theta$ and $\overline{j}_{\xi} = - \imath_{\overline{\xi}}\overline{\ell} - \imath_{\mathrm{X}_{\xi}}\overline{\theta}$
    \item $\mathbb{Q}_{\xi} = \int_{\Sigma} \imath^{*}J_{\xi} - \int_{\partial \Sigma} \overline{\imath}^{*}\overline{j}_{\xi}$
\end{enumerate}
\end{center}
\end{EvalBox}

%%%%%%%%%%%%%%%%%%%%%%%%%
\section{Examples}

To illustrate the steps in the previous algorithm, we will apply them to two background dependent field theories of physical relevance, defined on manifolds with boundaries: the scalar field and pure Yang-Mills. Being background dependent means that there is a fixed structure, which in these examples it will be the metric. These field theories are simple enough to clarify the procedure but not so simple that they lack physical interest. 
\subsection{Scalar Field}
The scalar field is possibly the most studied field theory in physics. Its dynamics are determined by choosing an action principle defined on a specific space of fields, which may include boundary conditions. In this example, we will consider two spaces of fields that we will label as Robin and Dirichlet scalar fields,
\begin{align*}
\mathcal{F}_{\mathcal{R}}:= \Omega^{0}(M), \&\& \mathcal{F}_{\mathcal{D}}:=\left\{\phi \in \Omega^{0}(M) \; |\;  \overline{\phi}:=\jmath^{*}\!\phi=0\right\},
\end{align*}
where $\phi: M \xrightarrow{} \mathbb{R}$ is a real scalar field and the metric $g$ of the base manifold $M$ is fixed.  The action for a massless scalar field in the presence of boundaries is given by,
\begin{align*}
    \mathbb{S}_{\mathtt{SF}}(\phi) = \int_{M}L_{\mathtt{SF}}(\phi) - \int_{\partial_{L}M} \overline{\ell}_{\mathtt{SF}}(\phi).
\end{align*}
The bulk and boundary Lagrangians are, respectively, 
\begin{align*}
    & L_{\mathtt{SF}}(\phi) = \frac{1}{2}g^{\alpha\beta}\nabla_{\alpha}\phi \nabla_{\beta}\phi\;\volg + V(\phi), \\
    &\overline{\ell}_{\mathtt{SF}}(\phi) = \frac{1}{2}f \overline{\phi}^{2}\volgg,
\end{align*}
with $\overline{\phi} :=\jmath^{*}\phi$ and $f: \partial M \xrightarrow{} \mathbb{R}$. If the function $f$ is taken to be zero, then we have Neumann boundary conditions. If, on the other hand, it is non-zero, the integrability of the system is not guaranteed. Taking the variations of the Lagrangians with respect to $\phi$ results in, 
\begin{align*}
    & \delta  L_{\mathtt{SF}}(\phi)  = E_{\mathtt{SF}}(\phi) \wedge \delta \phi + \mathrm{d} \Theta_{\mathtt{SF}}(\phi),\\
    & \delta \overline{\ell}_{\mathtt{SF}}(\phi) - \jmath^{*} \Theta_{\mathtt{SF}}(\phi)= b_{\mathtt{SF}}(\phi) \wedge \;\delta \phi.
\end{align*}
The presymplectic potentials are given by,
\begin{align}
   (\nabla^{\alpha}\phi)\volg &=: \nabla_{\alpha} B^{\alpha}\volg = \mathcal{L}_{B}\volg = \mathrm{d}(\imath_{B}\volg),\\
     \overline{\theta}_{\mathtt{SF}}(\phi) &= 0,
\end{align}
and the Euler forms are,
\begin{align*}
    & E_{\mathtt{SF}}(\phi) =   - \big(\nabla^{\beta}\nabla_{\beta}\phi - V'(\phi)\big) \volg, \\
   & b_{\mathtt{SF}}(\phi) = \big(f \; \overline{\phi} - \jmath^{*}(\mathcal{L}_{\nu}\phi)\big)\volg,
\end{align*}
such that $E_{\mathtt{SF}}(\phi) = 0$ and $b_{\mathtt{SF}}(\phi)=0$ are the equations of motion. The solution spaces for the two types of boundary conditions are
\begin{align*}
&   \mathtt{Sol}_{\mathcal{R}}( \mathbb{S}_{\mathtt{SF}}):= \{ \phi \in  \mathcal{F}_{\mathcal{N}} | \; \mathrm{E}_{\mathtt{SF}}(\phi) = 0, \overline{b}_{\mathtt{SF}}(\phi) = 0\},
\\
& \mathtt{Sol}_{\mathcal{D}}( \mathbb{S}_{\mathtt{SF}}):= \{ \phi \in \mathcal{F}_{\mathcal{D}} | \; \mathrm{E}_{\mathtt{SF}}(\phi) = 0\}.
\end{align*}
The presymplectic form is given by the CPS algorithm (\ref{cpsalgorithm}), 
\begin{align*}
    \Omega_{\mathtt{SF}}(\phi) = \delta \Big ( \int_{\Sigma} \imath^{*}\Theta_{\mathtt{SF}}(\phi)- \int_{\partial \Sigma} \overline{\imath}^{*}\overline{\theta}_{\mathtt{SF}}(\phi) \Big),
\end{align*}
where by computing the variations of the presymplectic potentials we arrive at, 
\begin{align*}
    & \delta \Theta_{\mathtt{SF}}(\phi) = \delta (\imath_{B}\volg) = \delta(B^{\alpha} n_{\alpha} \wedge \mathrm{vol}_{\gamma})= \delta \phi\wedge n^{\alpha}\nabla_{\alpha}\delta \phi \;\mathrm{vol}_{\gamma}=  \delta \phi \wedge \mathcal{L}_{n}\delta \phi \;\mathrm{vol}_{\gamma},
\end{align*}
and since the presymplectic potential at the boundary is zero, the presymplectic form is given by,
\begin{align*}
    \Omega_{\mathtt{SF}}(\phi) =   \int_{\Sigma} \imath^{*}\big(\delta \phi \wedge \mathcal{L}_{n}\delta \phi \;\mathrm{vol}_{\gamma}\big).
\end{align*}
For a vector field $\xi \in \mathfrak{X}(M)$ and its associated vector field $\mathrm{X}_{\xi} \in \mathfrak{X}(\mathcal{F})$, the $\xi$-currents at the bulk are given by,
\begin{align*}
    \mathcal{J}_{\mathtt{SF}}(\phi) &= \iota_{\xi}L_{\mathtt{SF}}(\phi) - \iota_{\mathrm{X}_{\xi}}\Theta_{\mathtt{SF}}(\phi) \\
    &= \iota_{\xi}\big(\frac{1}{2}g^{\alpha\beta}\nabla_{\alpha}\phi \nabla_{\beta}\phi\;\volg + V(\phi)\volg\big) - \iota_{\mathrm{X}_{\xi}}(\imath_{B}\volg)\\
    &= \big(\frac{1}{2}g^{\alpha\beta}\nabla_{\alpha}\phi \nabla_{\beta}\phi+ V(\phi)\big)\iota_{\xi}\volg - \iota_{\mathrm{X}_{\xi}}(\imath_{B}\volg)\\
    &= \big(\frac{1}{2}g^{\alpha\beta}\nabla_{\alpha}\phi \nabla_{\beta}\phi+ V(\phi)\big)\xi^{\alpha} n_{\alpha} \mathrm{vol}_{\gamma} - \iota_{\mathrm{X}_{\xi}}(\delta \phi \nabla^{\alpha}\phi  \wedge n_{\alpha}\mathrm{vol}_{\gamma})\\
    &= \big(\frac{1}{2}g^{\alpha\beta}\nabla_{\alpha}\phi \nabla_{\beta}\phi+ V(\phi)\big)\xi^{\alpha} n_{\alpha} \mathrm{vol}_{\gamma} - \xi^{\sigma}\nabla_{\sigma}\phi \nabla^{\alpha}\phi  \; \wedge n_{\alpha}\mathrm{vol}_{\gamma},
\end{align*}
where $ \iota_{\mathrm{X}_{\xi}}(\delta \phi) = \mathcal{L}_{\mathrm{X}_{\xi}}\phi = \mathcal{L}_{\xi}\phi = \xi^{\alpha}\nabla_{\alpha}\phi$ and $\iota_{\mathrm{X}_{\xi}}$ only affects $\delta \phi$ since the other terms are $0$-forms in the space of fields. \noindent The current at the boundary is computed similarly,
\begin{align*}
    \overline{j}_{\mathtt{SF}}(\phi) &= -\iota_{\xi}\overline{\ell}_{\mathtt{SF}}(\phi) - \iota_{\mathrm{X}_{\xi}}\overline{\theta}_{\mathtt{SF}}(\phi)\\
    &= -\iota_{\xi}\big( \frac{1}{2}f \overline{\phi}^{2}\volgg \big) = - \frac{1}{2}f \overline{\phi}^{2} \xi^{\overline{\alpha}}\overline{m}_{\overline{\alpha}} \mathrm{vol}_{\overline{\gamma}}.
\end{align*}
Hence the charge is given by,
\begin{align*}
    \mathbb{Q}_{\mathtt{SF}}(\phi) = \int_{\Sigma} \imath^{*}\mathcal{J}_{\mathtt{SF}}(\phi) - \int_{\partial\Sigma} \overline{\imath}^{*}\overline{j}_{\mathtt{SF}}(\phi),
\end{align*}
explicitly,
\begin{align*}
    \mathbb{Q}_{\mathtt{SF}}(\phi) =  \int_{\Sigma}  n_{\alpha}\xi_{\beta} \Big(\nabla^{\alpha}\phi \nabla^{\beta}\phi - g^{\alpha\beta}\big(\frac{1}{2}\nabla^{\sigma}\phi\nabla_{\sigma}\phi + V(\phi)\big)\Big) \mathrm{vol}_{\gamma} + \frac{1}{2} \int_{\partial \Sigma}\overline{\imath}^{*}\big(f \overline{\phi}^{2} \xi^{\overline{\alpha}}\overline{m}_{\overline{\alpha}} \mathrm{vol}_{\overline{\gamma}}\big).
\end{align*}
This is the charge associated with the Robin space of fields. The charge associated with Dirichlet is obtained equivalently by setting $\overline{\phi} = 0$ in the previous expression. 

\subsection{Yang-Mills}
In this section, we shall to consider a pure Yang-Mills theory based upon the symmetry group $G=SU(2)$ with background flat Minkowski metric. For a simpler example, one may consider $G=U(1)$ and arrive at Maxwell's equations. For this model, let us define the Neumann and Dirichlet field spaces,
\begin{align*}
\mathcal{F}_{\mathcal{N}}:=\Omega^{1}(M), && \mathcal{F}_{\mathcal{D}}:=\left\{  \A \in \Omega^{1}(M)\; |\; \overline{\A}:=\jmath^{*}\!\A = 0\right\}.
\end{align*}

Let $A$ be the connection $1$-form and let $\F$ be its curvature or field strength. The Yang-Mills action is defined as,
\begin{align*}
    \mathbb{S}_{\mathtt{YM}}(\A) = \int_{M}L_{\mathtt{YM}}(\A) - \int_{\partial M}\overline{\ell}_{\mathtt{YM}}(\A),
\end{align*}
where the bulk and boundary Lagrangians are,
\begin{align*}
    &L_{\mathtt{YM}}(\A) = - \frac{1}{2}\mathrm{tr}(\F\wedge \star\; \F), \\
    &\overline{\ell}_{\mathtt{YM}}(\A)  = 0.
\end{align*}
Computing their variations results in
\begin{align*}
    & \delta L_{\mathtt{YM}}(\A) = E_{\mathtt{YM}}(\A)\wedge \delta \A + \mathrm{d}\Theta_{\mathtt{YM}}(\A), && \delta \overline{\ell}_{\mathtt{YM}}(\A) - \jmath^{*}\Theta_{\mathtt{YM}}(\A) =  b_{\mathtt{YM}}(\A)\wedge \delta \A + \mathrm{d} \overline{\theta}_{\mathtt{YM}}(\A),\\
    & E_{\mathtt{YM}}(\A) = - \mathrm{D}\star \F,  && b_{\mathtt{YM}}(\A) = \jmath^{*}(\star \F), \\
    & \Theta_{\mathtt{YM}}(\A) = - \mathrm{tr}(\delta \A \wedge \star \F), && \overline{\theta}_{\mathtt{YM}}(\A) = 0.
\end{align*}
The presymplectic form given by the CPS algorithm is
\begin{align*}
    \Omega_{\mathtt{YM}}(\A) = \delta \Big( \int_{M} \imath^{*}\Theta_{\mathtt{YM}}(\A) - \int_{\partial M} \overline{\imath}^{*}\overline{\theta}_{\mathtt{YM}}(\A) \Big).
\end{align*}
Explicitly computing the variations of the presymplectic potentials gives,
\begin{equation}
     \Omega_{\mathtt{YM}}(\A)= \int_{M} \imath^{*} \mathrm{tr}(\delta \A \wedge \star D \delta \A),
\end{equation}
where we see that the presymplectic form is degenerate. The $\xi$-currents are given by
\begin{align*}
    \mathcal{J}_{\mathtt{YM}}(\A) &= \iota_{\xi}L_{\mathtt{YM}}(\A) - \iota_{\mathrm{X}_{\xi}} \Theta_{\mathtt{YM}}(\A) \\
    &= -\frac{1}{2}\iota_{\xi}\mathrm{tr}(\F \wedge \star \F)  + \iota_{\mathrm{X}_{\xi}}\mathrm{tr}(\delta \A \wedge \star \; \F) \\
    &= -\frac{1}{2}\iota_{\xi}\mathrm{tr}(\F \wedge \star \F)  + \mathrm{tr}(\mathcal{L}_{\xi}A \wedge \star \; \F) \\
    &= -\frac{1}{2}\iota_{\xi}\mathrm{tr}(\F \wedge \star \F) + \mathrm{tr}(\iota_{\xi}\F\wedge \star  \F) + \mathrm{tr}(D\iota_{\xi}\A \wedge \star \F),
\end{align*}
since $\imath_{\mathrm{X}_{\xi}}\star \F = 0$, and where we have used that $\imath_{\mathrm{X}_{\xi}} \delta \A = \mathcal{L}_{X_\xi}\A=\mathcal{L}_{\xi}\A$, since $\A$ is a $0$-form in the space of fields and also $\mathcal{L}_{\xi}A = \imath_{\xi}\F + D\imath_{\xi}\A$. The ${X_\xi}$-presymplectic current at the boundary is trivially zero since $\overline{\ell}_{\mathtt{YM}} = 0$ and $\overline{\theta}_{\mathtt{YM}} = 0$. The associated charges are given only by the current at the bulk,
\begin{align*}
    \mathbb{Q}_{\mathtt{YM}}(\A) &= \int_{\Sigma}\imath^{*} \mathcal{J}_{\mathtt{YM}}(\A).
\end{align*}
Integrating by parts, using $\alpha \wedge \star \beta = \langle \alpha, \beta \rangle\volg$  and $\imath^{*}(\iota_{\xi}\volg) = - n_{\alpha}\xi^{\alpha}\volg$, the charge is explicitly computed to be, 
\begin{align*}
     &\mathbb{Q}_{\mathtt{YM}}(\A) = \int_{\Sigma}\imath^{*} \bigg(-\frac{1}{2}\iota_{\xi}\mathrm{tr}(\F \wedge \star \F) + \mathrm{tr}(\iota_{\xi}\F\wedge \star  \F) + \mathrm{tr}(D\iota_{\xi}\A \wedge \star \F)\bigg) \\
     &= \int_{\Sigma}\imath^{*}\mathrm{tr} \bigg( -\frac{1}{2}\langle \F,\F \rangle \iota_{\xi}\volg  + \langle \iota_{\xi}\F, \F \rangle \volg + \mathrm{D}(\iota_{\xi}\A\wedge \star \F) - \iota_{\xi}\A \wedge \mathrm{D}\star \F \bigg) \\
    &= \int_{\Sigma}\imath^{*}\mathrm{tr} \bigg( -\frac{1}{2}\langle \F,\F \rangle \imath_{\xi}\volg  + \langle \imath_{\xi}\F, \F \rangle \volg  + \imath_{\xi}\A \wedge E_{\mathtt{YM}}(\A) \bigg) + \int_{\partial \Sigma} \overline{\jmath}^{*}\overline{\imath}^{*}\Big(\mathrm{tr}(\iota_{\xi}\A\wedge \star \F)\Big)\\
    &= \int_{\Sigma} \frac{1}{2}\mathrm{tr}\langle \F,\F \rangle n_{\alpha}\xi^{\alpha}\mathrm{vol}_{\gamma} + \mathrm{tr}(\langle n \wedge \iota_{\xi}\F, \F \rangle)\iota_{n}\volg  \!+ \!\int_{\Sigma} \overline{\imath}^{*} \mathrm{tr}(\imath_{\xi}\A \wedge E_{\mathtt{YM}}(\A))\! +\! \int_{\partial \Sigma} \overline{\jmath}^{*}\overline{\imath}^{*}\mathrm{tr}(\iota_{\xi} \A \wedge \star \F) \\
    &= \int_{\Sigma} n_{\alpha} \xi_{\alpha}( \tensor{\F}{^\beta_\tau}\tensor{\F}{^\alpha^\tau} - \frac{1}{4}g^{\alpha\beta}\tensor{\F}{^\tau^\delta}\tensor{\F}{_\tau_\delta}\big) \mathrm{vol}_{\gamma} + \int_{\Sigma} \imath^{*} \mathrm{tr}(\iota_{\xi}\A \wedge E_{\mathtt{YM}}(\A)) + \int_{\partial \Sigma} \overline{\jmath}^{*}\overline{\imath}^{*}\mathrm{tr}(\iota_{\xi} \A \wedge \star \F),
\end{align*}
where the last term contains the Euler form at the boundary, $b_{\mathtt{YM}}(\A) = \jmath^{*}(\star \F)$, and thus the last two terms vanish identically over solutions, as expected.

%% file: EH.tex
\chapter{The Einstein-Hilbert Action}\label{EHaction}

The Einstein field equations are the stationarity conditions for the Einstein-Hilbert (EH) action. The spacetime manifold upon which GR was initially formulated was boundary-less, but the variation of EH action included a surface term. This surface term did not allow the Hamiltonian to be derived directly from a covariant action. Nevertheless, physicists ignored the surface term for most derivations of the gravitational Hamiltonian, but this meant the variational problem was not well-defined. Gibbons, Hawking and York realised that introducing a counter term that cancelled this boundary term made the variational principle well-defined. They also proved that for asymptotically flat spacetimes, when going to the Hamiltonian formalism, the energy of this new action principle agreed with the usual ADM definition. Therefore, the boundary term we consider for the metric EH action is this Gibbons-Hawking-York Lagrangian at the boundary, written equivalently in the tetrad formalism by a change of variables. Boundaries are therefore crucial when choosing variational principles and studying their equivalence. 

The supposed inequivalence of tetrad and metric formulations of GR may be attributed, as we shall discuss later on, to an erroneous choice of bulk and boundary Lagrangians that do not describe the same dynamics. In this chapter, we will use the CPS algorithm described in the previous chapter (\ref{cpsalgorithm}) to study the EH action in both the metric and tetrad formalisms. We will discuss the solution spaces they define1 and build the symplectic structures. With these results, we will show the equivalence between both formulations in chapter $6$. 

%%%% METRIC VARIABLES %%%%%%%%
\section{Metric Variables}
Let $M$ be a differentiable manifold with boundary $\partial M$ as described in section \ref{spacetime}. 
Let $\text{Met}(M)$ be the space of metrics on $M$. Let us consider the Neumann and Dirichlet field spaces on $M$ defined as,
\begin{align*}
\mathcal{F}_{\mathcal{N}}:=\mathrm{Met}(M), && \mathcal{F}_{\mathcal{D}}:=\left\{ g\in\mathrm{Met}(M)\ |\ \overline{g}:=\jmath^*\!g\text{ fixed}\right\},
\end{align*}
that contain metrics satisfying Neumann and Dirichlet boundary conditions and where $\overline{g}$ is the induced metric at the lateral boundary given by the inclusion $\jmath: \partial_{L} M \hookrightarrow M$. According to (\ref{lagrangianpairintegral}) we will consider an action of the form,
\begin{equation}
\mathbb{S}^{\mathtt{(m)}}_{\mathtt{EH}}(g) = \int_{M} L_{\mathtt{EH}}(g) - \int_{\partial_{L}M} \bar{\ell}_{\mathtt{GHY}}(g),
\end{equation}
where $g$ will belong to either $\mathcal{F}_{N}$ or $\mathcal{F}_{D}$, and where we take the following Lagrangian pair,
\begin{align}
&L_{\mathtt{EH}}(g) = (\R - 2\Lambda) \volg, \\
&\overline{\ell}_{\mathtt{GHY}}(g) = - 2 \K \volgg,
\end{align}
in which $\R$ is the Ricci scalar of the metric $g$, $\Lambda$ is the cosmological constant, $\volg$ is the volume form of $M$, $\volgg$ is the volume form of its lateral boundary,  $\K$ is the trace of the extrinsic curvature $\K_{\overline{\alpha} \overline{\beta}}$ of the lateral boundary, defined as,
\begin{equation*}
\K_{\overline{\alpha} \overline{\beta}} :=  \jmath^{\alpha}_{\overline{\alpha}}\jmath^{\beta}_{\overline{\beta}}\nabla_{\alpha} \nu_{\beta},
\end{equation*}
and $\nu_{\alpha}\nu^{\alpha} = 1$ is the unit outer-pointing normal vector to $\partial_{L}M$. 
The equations of motion are computed by taking the variations of the action with respect to the dynamical field. 
The variation of the Lagrangian at the bulk has the form shown in (\ref{variationBulk}),
\begin{align*}
\delta L_{\text{EH}}(g) &= \mathrm{E}_{\mathtt{EH}}^{\alpha\beta}(g) \wedge \delta g_{\alpha\beta} + \de \Theta^{\mathtt{(m)}}_{\mathtt{EH}}.
\end{align*} 
The Euler forms  $\mathrm{E}_{\mathtt{EH}}^{\alpha\beta}(g)$ and the presymplectic potentials $\Theta^{\mathtt{(m)}}_{\mathtt{EH}}$ at the bulk are, 
\begin{align*}
\mathrm{E}_{\mathtt{EH}}^{\alpha\beta}(g)&:= \mathcal{E}_{\mathtt{EH}}^{\alpha\beta}(g) \volg= - \big( \R^{\alpha\beta}  - \frac{1}{2}\R g^{\alpha\beta} + \Lambda g^{\alpha\beta} \big) \volg,\\
\Theta^{\mathtt{(m)}}_{\mathtt{EH}} &:= \imath_{W}\volg, \\
W^{\alpha} &:= \nabla^{\beta}(g^{\alpha \sigma} \delta g_{\sigma \beta}) - \nabla^{\alpha}\delta g = (g^{\lambda \beta}g^{\alpha \kappa} - g^{\lambda \alpha}g^{\kappa \beta}) \nabla_{\lambda} \delta g_{\kappa \beta}. \end{align*}
At the boundary, the variation is
\begin{align*}
\delta \overline{\ell}_{\mathtt{GHY}}(g)  - \jmath^{*}\Theta^{\mathtt{(m)}}_{\mathtt{EH}} = \overline{b}_{\mathtt{EH}}^{\overline{\alpha} \overline{\beta}}(g) \wedge \delta \overline{g}_{\overline{\alpha}\overline{\beta}} - \de \overline{\theta}^{\mathtt{(m)}}_{\mathtt{EH}},
\end{align*}
and the Euler form $\overline{b}_{\mathtt{EH}}^{\overline{\alpha}\overline{\beta}}(g)$ and the presymplectic potentials $\overline{\theta}^{\mathtt{(m)}}_{\mathtt{EH}}$ at the boundary are
\begin{align*}
\overline{b}_{\mathtt{EH}}^{\overline{\alpha}\overline{\beta}}(g) &:= \overline{\mathfrak{b}}_{\mathtt{EH}}^{\overline{\alpha}\overline{\beta}} \volgg = (\K^{\overline{\alpha}\overline{\beta}}  - \overline{g}^{\overline{\alpha}\overline{\beta}}\K)\volgg,\\
\overline{\theta}^{\mathtt{(m)}}_{\mathtt{EH}} &= \imath_{\overline{V}}\volgg, \\
\overline{V}^{\overline{\alpha}} &:= - \jmath^{\beta}_{\overline{\beta}}(\overline{g}^{\overline{\alpha}\overline{\beta}}\nu^{\lambda}\delta g_{\lambda\beta}).
\end{align*}
The variation of the Lagrangian pair allows us to compute the variation of the action,
\begin{align*}
&\delta \mathbb{S}^{\mathtt{(m)}}_{\mathtt{EH}} =  \int_{M} \delta L_{\mathtt{EH}}(g) - \int_{\partial_{L}M} \delta \ell_{\mathtt{GHY}}(g) \\
&= \int_{M} \bigg(\mathrm{E}_{\mathtt{EH}}^{\alpha\beta}(g) \wedge \delta g_{\alpha\beta} + \de \Theta^{\mathtt{(m)}}_{\mathtt{EH}}\bigg)-  \int_{\partial_{L} M}\bigg( \overline{b}_{\mathtt{EH}}^{\overline{\alpha}\overline{\beta}}(g) \wedge \delta \overline{g}_{\overline{\alpha}\overline{\beta}} - \de \overline{\theta}^{\mathtt{(m)}}_{\mathtt{EH}} +  \jmath^{*}\Theta^{\mathtt{(m)}}_{\mathtt{EH}} \bigg) \\
&=  \int_{M} \mathrm{E}_{\mathtt{EH}}^{\alpha\beta}(g) \wedge \delta g_{\alpha\beta} - \int_{\partial M} \overline{b}_{\mathtt{EH}}^{\overline{\alpha}\overline{\beta}}(g)\wedge \delta \overline{g}_{\overline{\alpha}\overline{\beta}}  + \int_{M} \de \Theta^{\mathtt{(m)}}_{\mathtt{EH}}  - \int_{\partial M}\jmath^{*}\Theta^{\mathtt{(m)}}_{\mathtt{EH}} + \int_{\partial_{L} M} \de \overline{\theta}^{\mathtt{(m)}}_{\mathtt{EH}}. 
\end{align*} 
By Stokes' theorem, the third and fourth terms in the last line of the previous expression cancel each other, and the fifth term  is also zero since $\partial(\partial_{L} M) = \varnothing$. Therefore the principle of least action implies that $\big(\mathrm{E}_{\mathtt{EH}}^{\alpha\beta}(g),\overline{{b}}_{\mathtt{EH}}^{\overline{\alpha}\overline{\beta}}(g)\big)= 0$.  
Once the equations of motion have been computed, it is possible to study the solutions associated with the selected space of dynamical fields, $\mathcal{F}_{N}$ and $\mathcal{F}_{D}$,  
\begin{align*}
& \mathtt{Sol}_{\mathcal{N}}( \mathbb{S}^{\mathtt{(m)}}_{\mathtt{EH}}):= \{ g \in  \mathcal{F}_{\mathcal{N}}  | \; \mathrm{E}_{\mathtt{EH}}^{\alpha\beta}(g) = 0, \overline{b}_{\mathtt{EH}}^{\alpha \beta}(g) = 0\},
\\
& \mathtt{Sol}_{\mathcal{D}}( \mathbb{S}^{\mathtt{(m)}}_{\mathtt{EH}}):= \{ g \in \mathcal{F}_{\mathcal{D}} | \; \mathrm{E}_{\mathtt{EH}}^{\alpha\beta}(g) = 0\}. 
\end{align*}
Because we are interested in fully constructing the covariant phase space of these spaces of solutions, the next step is to compute the presymplectic form associated with the action $\mathbb{S}_{\mathtt{EH}}$ on them.  The CPS algorithm gives us a method to compute this structure without any ambiguities by varying the presymplectic potentials and integrating over a Cauchy hypersurface (\ref{presymplecticStructure}),
\begin{equation}
\Omega^{\mathtt{(m)}}_{\mathtt{EH}} = \delta \Big ( \int_{\Sigma} \imath^{*}\Theta^{\mathtt{(m)}}_{\mathtt{EH}}- \int_{\partial \Sigma} \overline{\imath}^{*}\overline{\theta}^{\mathtt{(m)}}_{\mathtt{EH}} \Big).
\end{equation}
From \cite{CPS} we know this structure to be independent of the chosen Lagrangian representatives and, \emph{on-shell}, of the embeddings  $\imath: \Sigma \hookrightarrow M$ and $\overline{\imath}: \partial\Sigma \hookrightarrow \partial M$. Each term is explicitly given by,
\begin{align*}
    \delta \Theta^{\mathtt{(m)}}_{\mathtt{EH}} &= \delta(\iota_{W}\volg) = \iota_{\delta W}\volg - \iota_{W}\delta \volg = \iota_{\{\delta W - \frac{1}{2}W\delta g\}}\volg, \\
    \delta\overline{\theta}^{\mathtt{(m)}}_{\mathtt{EH}} &= \delta ( \iota_{\overline{V}}\volgg) = \iota_{\delta \overline{V}}\volgg - \iota_{\overline{V}}\delta \volgg = \iota_{\{\delta \overline{V} - \frac{1}{2}\overline{V}\delta g\}}\volgg.
\end{align*}
The vector field contracted with the volume in the first expression is explicitly
 \begin{align}\label{eq:variationW}
     \delta W^{\alpha}- \frac{1}{2}W^{\alpha}\delta g = -\frac{1}{2}\delta^{\alpha \eta \sigma}_{\lambda \beta \zeta}g^{\beta\rho}g^{\zeta\phi}\delta g_{\eta\rho}\wedge\nabla^{\lambda}\delta g_{\sigma \phi} + \frac{1}{2}g^{\beta\mu}g^{\sigma\lambda}\nabla_{\lambda}(\delta g_{\alpha \beta}\wedge \delta g_{\mu\sigma}),
 \end{align}
 where we have used the generalised Kronecker delta,  
 \begin{align*}
    \delta^{\lambda\eta\sigma}_{\alpha \beta \xi} := \delta^{\lambda}_{\alpha}\delta^{\eta}_{\beta}\delta^{\sigma}_{\xi} + \delta^{\eta}_{\alpha}\delta^{\sigma}_{\beta}\delta^{\lambda}_{\xi} + \delta^{\sigma}_{\alpha}\delta^{\lambda}_{\beta}\delta^{\eta}_{\xi} - \delta^{\eta}_{\alpha}\delta^{\lambda}_{\beta}\delta^{\sigma}_{\xi} - \delta^{\lambda}_{\alpha}\delta^{\sigma}_{\beta}\delta^{\eta}_{\xi} - \delta^{\sigma}_{\alpha}\delta^{\eta}_{\beta}\delta^{\lambda}_{\xi},
\end{align*}
and the vector field contracted with the second term is given by 
\begin{align}\label{eqn:variationV}
    \delta \overline{V}^{\overline{\alpha}}  \!- \frac{1}{2}\overline{V}^{\overline{\alpha}} \delta \overline{g}& = \! - \delta \jmath^{\beta}_{\overline{\beta}}(\overline{g}^{\overline{\alpha}\overline{\beta}}\nu^{\gamma}\delta g_{\beta\gamma}) \!+ \frac{1}{2}\jmath^{\beta}_{\overline{\beta}}(\overline{g}^{\overline{\alpha}\overline{\beta}}\nu^{\gamma}\delta g_{\beta\gamma}) \wedge \delta \overline{g} \! = \frac{1}{2}\jmath^{\beta}_{\overline{\beta}}(\overline{g}^{\overline{\alpha}\overline{\beta}}\nu^{\sigma}\delta g_{\sigma\beta}) \wedge \delta \overline{g}.
\end{align}
Making use of (\ref{eq:variationW}) and (\ref{eqn:variationV}), the variations of the presymplectic potentials are explicitly written as,
 \begin{align}
    \delta \Theta^{\mathtt{(m)}}_{\mathtt{EH}} &=  \frac{1}{2}n^{\alpha}\{ \delta^{\lambda \eta \sigma}_{\alpha \beta \zeta}g^{\beta\rho}g^{\zeta\phi}\delta g_{\eta\rho}\wedge\nabla_{\lambda}\delta g_{\sigma \phi} - g^{\beta\mu}g^{\sigma\lambda}\nabla_{\lambda}(\delta g_{\alpha \beta}\wedge \delta g_{\mu \sigma})\}\text{vol}_{\gamma}, \\
    \delta\overline{\theta}^{\mathtt{(m)}}_{\mathtt{EH}} &=  \frac{1}{2}g^{\sigma\lambda}\overline{m}^{\beta}\nu^{\gamma} \delta g_{\gamma \beta}\wedge \delta g_{\sigma \lambda} \text{vol}_{\gamma},
\end{align}
where the identities $\volg = - n \wedge \text{vol}_{\gamma}$ and $\volgg = \overline{m}_{\alpha} \wedge \text{vol}_{\gamma}$ have been used to simplify the expression (\ref{diagram_spacetime}). The presymplectic form is then computed to be,
\begin{align*}
\Omega^{\mathtt{(m)}}_{\mathtt{EH}} &= \frac{1}{2}\int_{\Sigma} (n^{\alpha} \delta^{\lambda \eta \sigma}_{\alpha \beta \zeta}g^{\beta\rho}g^{\zeta\phi}\delta g_{\eta\rho}\wedge\nabla_{\lambda}\delta g_{\sigma \phi} - n^{\alpha} g^{\beta\mu}g^{\sigma\lambda}\nabla_{\lambda}(\delta g_{\alpha \beta}\wedge \delta g_{\mu \sigma}))\text{vol}_{\gamma} \\
& \quad- \frac{1}{2}\int_{\partial \Sigma}  \overline{m}_{\overline{\alpha}}\overline{g}^{\overline{\alpha}\overline{\beta}}\jmath^{\beta}_{\overline{\beta}}(\iota_{\nu} \delta g)_{\beta}\wedge \delta g \; \mathrm{vol}_{\overline{\gamma}}.
\end{align*}
The second term at the bulk of the symplectic form can be written as a surface term since,
\begin{align}\label{something1}
    n^{\alpha} g^{\beta\mu}g^{\sigma\lambda}\nabla_{\lambda}(\delta g_{\alpha \beta}\wedge \delta g_{\mu \sigma}) = n^{\alpha}\nabla^{\sigma}(g^{\beta\mu}\delta g_{\alpha \beta}\wedge \delta g_{\mu \sigma}) =: n^{\alpha}\nabla^{\sigma}(t_{\alpha\sigma}),
\end{align}
where we have defined the antisymmetric tensor $t_{\alpha\sigma}$. The last object in (\ref{something1}) can be written in the form,
\begin{align*}
    n^{\alpha}\nabla^{\sigma}(t_{\alpha\sigma}) &= n^{\alpha}(\epsilon n^{\lambda}n^{\sigma} + \imath^{\lambda}_{a}\imath^{\sigma}_{b}\gamma^{ab}) \nabla_{\lambda}(t_{\alpha\sigma})
    = \imath^{\lambda}_{a} \imath^{\sigma}_{b}\gamma^{ab} [ \nabla_{\lambda}(n^{\alpha} t_{\alpha\sigma}) - t_{\alpha\sigma}\nabla_{\lambda}n^{\alpha}] \\
    &= \imath^{\lambda}_{a} \imath^{\sigma}_{b}\gamma^{ab}  \nabla_{\lambda}(n^{\alpha} t_{\alpha\sigma})- \imath^{\lambda}_{a} \imath^{\sigma}_{b}\gamma^{ab}t_{\alpha\sigma}\nabla_{\lambda}n_{\beta}g^{\beta \alpha} \\
    &= \imath^{\lambda}_{a} \imath^{\sigma}_{b}\gamma^{ab}  \nabla_{\lambda}(n^{\alpha} t_{\alpha\sigma}) -  \imath^{\lambda}_{a} \imath^{\sigma}_{b}\gamma^{ab}t_{\alpha\sigma}(\epsilon n^{\alpha}n^{\beta} + \imath^{\alpha}_{c}\imath^{\beta}_{b}\gamma^{cd})\nabla_{\lambda}n_{\beta} \\
    &= \imath^{\lambda}_{a} \imath^{\sigma}_{b}\gamma^{ab}  \nabla_{\lambda}(n^{\alpha} t_{\alpha\sigma}) - \imath^{\sigma}_{b}\imath^{\alpha}_{c}t_{\alpha \sigma}\gamma^{cd} \K_{ab} \\
    &= \gamma^{ab} \mathrm{D}_{a}(\imath^{\sigma}_{b}n^{\alpha}t_{\alpha\sigma}) + \epsilon \K_{ab}(n^{\alpha}n^{\sigma}t_{\alpha\sigma}) - \tensor{T}{_b^d}\K_{ab} \\
    &= D^{b}(\imath^{\sigma}_{b}n^{\alpha}t_{\alpha\sigma}):= D^{b}(\imath^{\sigma}_{b}n^{\alpha} g^{\beta\mu}\delta g_{\alpha\beta}\wedge \delta g_{\mu\sigma}),
\end{align*}
where we have used the definition of the extrinsic curvature, the identity $g^{\lambda\sigma} =  n^{\lambda}n^{\sigma} + \imath^{\lambda}_{a}\imath^{\sigma}_{b}\gamma^{ab}$ and the fact that $n^{\beta}\nabla_{\lambda}n_{\beta} = 0 $. If we use this along with Stokes' theorem, we can write the symplectic structure as,
\begin{align*}
\Omega^{\mathtt{(m)}}_{\mathtt{EH}} &= \frac{1}{2}\int_{\Sigma} (n^{\alpha} \delta^{\lambda \eta \sigma}_{\alpha \beta \zeta}g^{\beta\rho}g^{\zeta\phi}\delta g_{\eta\rho}\wedge\nabla_{\lambda}\delta g_{\sigma \phi})\text{vol}_{\gamma} \\
&\quad-\frac{1}{2}\int_{\partial \Sigma} (\mu^{\sigma}n^{\alpha}g^{\beta\mu} - \overline{m}^{\beta}\nu^{\alpha}g^{\mu\sigma}) \delta g_{\alpha\beta} \wedge \delta g_{\mu\sigma} \mathrm{vol}_{\overline{\gamma}}.
\end{align*}
The Noether current associated with a vector field  $\xi$ on $M$ is defined by
\begin{align}\label{currentEHmetric}
    \mathcal{J}^{\mathtt{(m)}}_{\mathtt{EH}}:&= \iota_{\xi}L_{\mathtt{EH}}(g) - \iota_{X_{\xi}}\Theta^{\mathtt{(m)}}_{\mathtt{EH}} \\
    &= \iota_{\xi}(\R-2\Lambda)\volg- \iota_{X_{\xi}}(\iota_{W}\volg) \\
    &= \iota_{\{ \xi (\R - 2\Lambda) - \iota_{X_{\xi}}W \}} \volg = \star_{g}\{ \xi(\R - 2\Lambda) - \iota_{X_{\xi}}W\},
\end{align}
where $\star_{g}$ is the Hodge-star operator, 
\begin{align*}
\xi^{\alpha}(\R- 2\Lambda) &= g^{\alpha\beta}\xi_{\beta}(\R - 2 \Lambda) = 2(\mathcal{E}^{\alpha\beta}_{\mathtt{EH}}(g) + \R^{\alpha\beta})\xi_{\beta} = 2\mathcal{E}^{\alpha\beta}_{\mathtt{EH}}(g)\xi_{\beta} + 2\R^{\alpha\beta}\xi_{\beta},\\
\iota_{X_{\xi}}W^{\alpha} &= (g^{\lambda \beta}g^{\alpha \kappa} - g^{\lambda \alpha}g^{\kappa \beta})\nabla_{\lambda}(\iota_{X_{\xi}}\delta g_{\kappa \beta}) \\
    &= (g^{\lambda \beta}g^{\alpha \kappa} - g^{\lambda \alpha}g^{\kappa \beta})(g_{\kappa \sigma}\nabla_{\lambda}\nabla_{\beta}\xi^{\sigma} + g_{\beta \sigma}\nabla_{\lambda}\nabla_{\kappa}\xi^{\sigma}) \\
    &= 2[\nabla^{\alpha}, \nabla^{\beta}]\xi_{\beta} + \nabla_{\beta}(\nabla^{\beta}\xi^{\alpha} - \nabla^{\alpha}\xi^{\beta}) \\
    &= 2\tensor{\R}{_\beta^\sigma^\alpha^\beta}\xi_{\sigma} - \nabla_{\beta}(\de\xi)^{\alpha\beta} = 2\tensor{\R}{^\sigma^\alpha}\xi_{\sigma} + \nabla_{\beta}(\de\xi)^{\beta \alpha}.
\end{align*}
Substituting these expressions back in (\ref{currentEHmetric}) gives,  
\begin{align*}
    \mathcal{J}^{\mathtt{(m)}}_{\mathtt{EH}} &= \star_{g}\{ \xi_{\alpha}(\R - 2\Lambda) - \iota_{X_{\xi}}W_{\alpha}\} \\ &= \star_{g}\{ 2(\mathcal{E}_{\mathtt{EH}}(g))_{\alpha\beta} \xi^{\beta} + 2\R_{\alpha\beta}\xi^{\beta} - (2\tensor{\R}{_\sigma_\alpha}\xi^{\sigma} + \nabla^{\beta}(\de\xi)_{\beta \alpha})\}\\
    &= 2\star_{g}(\mathcal{E}_{\mathtt{EH}}(g))_{\alpha\beta}\xi^{\beta} - \star_{g}\nabla^{\beta}(\de\xi)_{\beta\alpha} = 2 \star_{g}\{\imath_{\xi} \mathcal{E}_{\mathtt{EH}}(g)\} + \de\mathcal{Q}^{\mathtt{(m)}}_{\mathtt{EH}},
\end{align*}
where $\mathcal{Q}^{\mathtt{(m)}}_{\mathtt{EH}}:= \star_{g}\de \xi$ is the $\xi$-Noether potential of the current at the bulk. The associated current at the boundary is given by
\begin{align*}
    \overline{j}^{\mathtt{(m)}}_{\mathtt{EH}} &= -\iota_{\overline{\xi}}\overline{\ell}^{\mathtt{(m)}}_{\mathtt{EH}} - \iota_{X_{\overline{\xi}}}\overline{\theta}_{\mathtt{EH}}(g) \\
    &=  \iota_{\overline{\xi}}(2\overline{\K}\volgg) - \iota_{X_{\xi}}(\iota_{\overline{V}}\volgg) = \iota_{\{ 2\xi\overline{K} - \iota_{X_{\xi}}\overline{V}\}}\volgg.
\end{align*}
The vector contracted with the volume form may be alternatively written as,
\begin{align*}
    2\overline{\K}\;\overline{\xi}^{\overline{\alpha}} - \iota_{X_{\xi}}\overline{V}^{\overline{\alpha}} &= 2\overline{\xi}^{\overline{\alpha}}\overline{\K} + \overline{g}^{\overline{\alpha}\overline{\beta}}\jmath^{\beta}_{\overline{\beta}}(\nu^{\sigma}(\nabla_{\sigma}\xi_{\beta} - \nabla_{\beta}\xi_{\sigma}))\\ &\overset{\vec{\xi}\perp\vec{\nu}}{=} 2\overline{\xi}^{\overline{\alpha}}\overline{\K} + \overline{g}^{\overline{\alpha}\overline{\beta}} \jmath^{\beta}_{\overline{\beta}}(\nu^{\sigma}(\de\xi)_{\sigma \beta} - 2 \xi_{\sigma}\nabla_{\beta}\nu^{\sigma}) \\
    &= 2\overline{\xi}^{\overline{\alpha}}\overline{\K} - \overline{g}^{\overline{\alpha}\overline{\beta}} \jmath^{\beta}_{\overline{\beta}}(2 \xi^{\sigma}\K_{\beta\sigma}) + \overline{g}^{\overline{\alpha}\overline{\beta}}\jmath^{\beta}_{\overline{\beta}}(\nu^{\sigma}(\de \xi)_{\sigma\beta}) \\
    &= -2(\iota_{\overline{\xi}}\overline{\mathfrak{b}}_{\mathtt{EH}})^{\overline{\alpha}} + \jmath^{\beta}_{\overline{\beta}}(\iota_{\nu}\de\xi)^{\overline{\alpha}}.
\end{align*}
Therefore, by plugging this expression back into the current at the boundary we get,
\begin{align}
    \overline{j}^{\mathtt{(m)}}_{\mathtt{EH}}
     =   \star_{\overline{g}}\{-2(\iota_{\overline{\xi}}\overline{\mathfrak{b}}_{\mathtt{EH}}(g)) + \jmath^{*}(\iota_{\nu}\de\xi)\} = -2\star_{\overline{g}}\{\iota_{\overline{\xi}}\overline{\mathfrak{b}}_{\mathtt{EH}}(g)\} + \jmath^{*}\star_{\overline{g}}(\iota_{\nu}\de\xi).
\end{align}
The Noether charges associated with a $\xi$-vector field for the Einstein-Hilbert action at the bulk and at the boundary are thus given by,
\begin{align}
\mathcal{J}^{\mathtt{(m)}}_{\mathtt{EH}} &= \de \mathcal{Q}^{\mathtt{(m)}}_{\mathtt{EH}} + 2 \star_{g}\{\imath_{\xi} \mathcal{E}_{\mathtt{EH}}(g)\},\\
\overline{j}^{\mathtt{(m)}}_{\mathtt{EH}} &= \jmath^{*}\mathcal{Q}^{\mathtt{(m)}}_{\mathtt{EH}} - 2 \star_{\overline{g}}\{\imath_{\overline{\xi}} \overline{\mathfrak{b}}_{\mathtt{EH}}(g)\}.
\end{align}
The $\xi$-charge associated with the Einstein-Hilbert action is,
\begin{align*}
\mathbb{Q}^{\mathtt{(m)}}_{\mathtt{EH}} &= \int_{\Sigma}\imath^{*}\mathcal{J}^{\mathtt{(m)}}_{\mathtt{EH}} - \int_{\partial \Sigma} \overline{\imath}^{*}   \overline{j}^{\mathtt{(m)}}_{\mathtt{EH}} \\
&=  \int_{\Sigma} \imath^{*} (\de \mathcal{Q}^{\mathtt{(m)}}_{\mathtt{EH}} + 2 \star_{g}\{\imath_{\xi} \mathcal{E}_{\mathtt{EH}}(g)\}) -  \int_{\partial \Sigma} \overline{\imath}^{*}(\jmath^{*}\mathcal{Q}^{\mathtt{(m)}}_{\mathtt{EH}} - 2 \star_{\overline{g}}\{\imath_{\overline{\xi}} \overline{\mathfrak{b}}_{\mathtt{EH}}(g)\})\\
&= \int_{\Sigma} 2\star_{g}\{ \iota_{\xi}\mathcal{E}_{\mathtt{EH}}(g)\} + \int_{\partial\Sigma} 2\star_{\overline{g}}\{\iota_{\overline{\xi}}\mathfrak{\overline{b}}_{\mathtt{EH}}(g)\},
\end{align*}
where the first and third terms cancel each other by Stokes' theorem so that the charges only depend on the density of the Euler forms $\mathcal{E}_{\mathtt{EH}}(g)$ and $\mathfrak{\overline{b}}$.
If we evaluate the last expression in $\mathtt{Sol}_{\mathcal{N}}( \mathbb{S}^{\mathtt{(m)}}_{\mathtt{EH}})$, they are found to be zero, as expected, since there are no Noether charges associated with a diffeomorphism invariant field theory. On the other hand, if we evaluate it over $\mathtt{Sol}_{\mathcal{D}}( \mathbb{S}^{\mathtt{(m)}}_{\mathtt{EH}})$, the Noether charges are solely dependent on the boundary equations,
\begin{align*}
&(\mathbb{Q}^{\mathtt{(m)}}_{\mathtt{EH}})^{\mathcal{D}} = 0, & (\mathbb{Q}^{\mathtt{(m)}}_{\mathtt{EH}})^{\mathcal{N}}  = 2\int_{\partial\Sigma} \star_{\overline{g}}\{\iota_{\overline{\xi}}\;\mathfrak{\overline{b}}_{\mathtt{EH}}(g)\},
\end{align*}
since fixing the metric at the boundary certainly breaks diffeomorphism invariance.

\section{Tetrad Variables}
The Einstein-Hilbert action in terms of non-degenerate tetrad variables in the second order formalism takes $\e^{I}$ as the dynamical variable (for $I = 1, \ldots, 4)$. The spin-connection associated with the tetrad (see Appendix \ref{appendix} for more details) is not an independent variable and thus its variations are expressed in terms of the tetrads. 
The field spaces that we will use to work with tetrad variables are of the Neumann and Dirichlet types
\begin{align}
&\mathcal{F}_{N}  = \{ \e^{I}_{\alpha} \in \Omega^{1}(M) | \;\;  g = \e^{\alpha}_{I}\e^{I}_{\alpha} \in \text{Met}(M) \},\\
&\mathcal{F}_{D}  = \{ \e^{I}_{\alpha} \in \Omega^{1}(M) |\;\;  \e^{I}_{\alpha} \in \mathcal{F}_{N} \;\; \text{and} \;\; \overline{\e}:=\jmath^{*}\e \; \; \; \text{is fixed} \}.
\end{align} 
The Einstein-Hilbert action in terms of tetrads is
\begin{equation}
\mathbb{S}^{\mathtt{(t)}}_{\mathtt{EH}}(\e) = \int_{M} L_{\mathtt{EH}}(\e) - \int_{\partial_{L}M} \overline{\ell}_{\mathtt{GHY}}(\e), 
\end{equation}
where the Lagrangians are derived from the metric EH action by a change of variables, 
\begin{align*}
&L_{\mathtt{EH}}(\e) := L_{\mathtt{EH}}(g = \mathrm{Tr}(\e \otimes \e)), \\
&\overline{\ell}_{\mathtt{GHY}}(\e) := \overline{\ell}_{\mathtt{GHY}}(g = \mathrm{Tr}(\e \otimes \e)),
\end{align*}
and $g_{\alpha\beta} = \eta_{IJ}\e^{I}_{\alpha}\e^{J}_{\beta}$ and $E^{\alpha}_{I}\e{I}_{\beta} = g^{\alpha}_{\beta}$ with $E^{\alpha}_{I}\e^{J}_{\alpha} = \eta^{J}_{I}$ (see appendix \ref{appendix} for further details). Therefore, the Lagrangians in tetrad variables are
\begin{align}
&L_{\mathtt{EH}}(\e) = \frac{1}{2} \epsilon_{IJKL} \big( \F^{IJ} - \frac{\Lambda}{3} \e^{I} \wedge \e^{J}\big) \wedge \e^{K} \wedge \e^{L}, \\
&\overline{\ell}_{\mathtt{GHY}}(\e) = -\epsilon_{IJKL} \N^{I}(\de \N^{J} + \tensor{\overline{\omega}}{^J_K}\wedge \N^{K}) \wedge \overline{\e}^{K} \wedge \overline{\e}^{L},
\end{align}
where $\Lambda$ is the cosmological constant, $\N^{I} = \iota_{\nu}\e^{I}$ is the internal normal scalar, $F_{IJ} = \mathrm{d} \omega_{IJ}+ \omega_{IK} \wedge \tensor{\omega}{^K_J}$ is the curvature, $\overline{\omega}:= \jmath^{*}\omega$ is the pullback of the connection to the boundary,  and $\epsilon_{IJKL}$ is the totally antisymmetric tensor. The equations of motion are obtained by varying the action with respect to the tetrads.  The variation of the bulk Lagrangian is given by
%VARIATIONS  
\begin{align*}
\delta L_{\mathtt{EH}}(\e)  &= \mathrm{E}_{\mathtt{EH}}(\e)_{I}\wedge \delta \e^{I} + \de \Theta^{\mathtt{(t)}}_{\mathtt{EH}},
\end{align*}
where
\begin{align*}
&\mathrm{E}_{\mathtt{EH}}(\e)_{I}:= \frac{1}{2}\epsilon_{IJKL}\e^{J} \wedge F^{KL},\\
&\Theta^{\mathtt{(t)}}_{\mathtt{EH}}(\e) := \frac{1}{2}\epsilon_{IJKL}\e^{I} \wedge\e^{J} \wedge \delta \omega^{KL}.
\end{align*}
The variation of the boundary Lagrangian is given by, 
\begin{align*}
 &\delta \overline{\ell}_{\mathtt{GHY}}(\e) - \jmath^{*}\Theta^{\mathtt{(t)}}_{\mathtt{EH}} =  \overline{b}_{\mathtt{EH}}(\e)_{I}  \wedge \delta \overline{\e}^{I} - \de \overline{\theta}^{\mathtt{(t)}}_{\mathtt{EH}},
\end{align*}
with
\begin{align*}
 \overline{b}_{\mathtt{EH}}(\e)_{I}  &:= \epsilon_{IJKL} \big(\N^{J} \wedge \de \N^{K} - \frac{1}{2} \overline{\omega}^{JK}\big) \wedge \overline{\e}^{L} - 2 \N_{I}\epsilon_{JKLR}\gamma^{JK}N^{L}\wedge \overline{\e}^{R},\\
\overline{\theta}^{\mathtt{(t)}}_{\mathtt{EH}} &:= \epsilon_{IJKL}\overline{\e}^{I} \wedge \overline{\e}^{J}\wedge \N^{K} \wedge \delta \N^{L}.
\end{align*}
Therefore, the space of solutions of the action $\mathbb{S}^{\mathtt{(t)}}_{\mathtt{EH}}$ in the Neumann and Dirichlet field spaces is
\begin{align*}
& \mathtt{Sol}_{\mathcal{N}}( \mathbb{S}^{\mathtt{(t)}}_{\mathtt{EH}}):= \{ \e \in  \mathcal{F}_{\mathcal{N}} | \; \mathrm{E}_{\mathtt{EH}}(\e)_{I} = 0, \overline{b}_{\mathtt{EH}}(\e)_{I} = 0\},
\\
& \mathtt{Sol}_{\mathcal{D}}( \mathbb{S}^{\mathtt{(t)}}_{\mathtt{EH}}):= \{ \e \in \mathcal{F}_{\mathcal{D}} | \; \mathrm{E}_{\mathtt{EH}}(\e)_{I} = 0\}.
\end{align*}

\noindent The symplectic form is computed as usual via the symplectic potentials, 
\begin{equation*}
\Omega^{\mathtt{(t)}}_{\mathtt{EH}} = \delta \Big ( \int_{\Sigma} \imath^{*}\Theta^{\mathtt{(t)}}_{\mathtt{EH}}- \int_{\partial \Sigma} \overline{\imath}^{*}\overline{\theta}^{\mathtt{(t)}}_{\mathtt{EH}} \Big),
\end{equation*}
where 
\begin{align*}
    \delta \Theta^{\mathtt{(t)}}_{\mathtt{EH}} &:= \frac{1}{2}\epsilon_{\mathtt{IJKL}}\delta (\e^{K} \wedge\e^{\mathtt{L}})\wedge \delta \omega^{IJ},\\
    \delta \overline{\theta}^{\mathtt{(t)}}_{\mathtt{EH}}&:= \epsilon_{IJKL} \big( \delta (\overline{\e}^{K}\wedge \overline{\e}^{L})\wedge N^{I} + \overline{\e}^{K}\wedge\overline{\e}^{L} \wedge \delta N^{J} \big) \wedge \delta N^{J}.
\end{align*}
%SYMPLECTIC
From these we can explicitly get, 
\begin{align*}
\Omega^{\mathtt{(t)}}_{\mathtt{EH}} &= \frac{1}{2}\int_{\Sigma} \epsilon_{IJKL} \delta (\e^{K} \wedge \e^{L}) \wedge  \delta \omega^{IJ} \\
& \quad -  \int_{\partial \Sigma} \epsilon_{IJKL} \big( \delta (\overline{\e}^{K}\wedge \overline{\e}^{L})\wedge N^{I} + \overline{\e}^{K}\wedge\overline{\e}^{L} \wedge \delta N^{I} \big) \wedge \delta N^{J}.
\end{align*}

%CURRENT
\noindent The $\xi$-currents associated with this symplectic structure are given by
\begin{align*}
\mathcal{J}^{\mathtt{(t)}}_{\mathtt{EH}}&:= \imath_{\xi}L_{\mathtt{EH}}(\e)- \imath_{\mathrm{X}_{\xi}}\Theta^{\mathtt{(t)}}_{\mathtt{EH}} \\
&= \imath_{\xi} \Big(  \frac{1}{2} \epsilon_{IJKL} \big( \F^{IJ} - \frac{\Lambda}{3} \e^{I} \wedge \e^{J}\big) \wedge \e^{K} \wedge \e^{L} \Big) -  \imath_{\mathrm{X}_{\xi}}\Big( \frac{1}{2}\epsilon_{IJKL}\e^{K} \wedge\e^{L} \wedge\delta \omega^{IJ} \Big)\\
&= \frac{1}{2}\epsilon_{IJKL}(\imath_{\xi}\e^{I} \wedge \e^{J} \wedge F^{KL}) - \de \Big(\frac{1}{2}\epsilon_{IJKL}\e^{K}\wedge \e^{L} \wedge \imath_{\xi}\omega^{IJ} \Big) \\
&= \imath_{\xi}e^{I} \wedge \mathrm{E}_{\mathtt{EH}}(\e)_{I} + \de \mathcal{Q}^{\mathtt{(t)}}_{\mathtt{EH}}.
\end{align*}
Furthermore, we find that $\mathcal{J}^{\mathtt{(t)}}_{\mathtt{EH}}$ is exact over the space of solutions with the $\xi$-potential being $\mathcal{Q}^{\mathtt{(t)}}_{\mathtt{EH}} = \frac{1}{2}\epsilon_{IJKL}\e^{I}\wedge \e^{J} \wedge \imath_{\xi}\omega^{KL}$. In a similar way, we compute the current for the boundary,
\begin{align*}
    \overline{\mathfrak{j}}^{\mathtt{(t)}}_{\mathtt{EH}}  &:= - \imath_{\xi}\overline{\ell}_{\mathtt{EH}}(\e) - \imath_{\mathrm{X}_{\xi}}\overline{\theta}^{\mathtt{(t)}}_{\mathtt{EH}}
    \\&= -\epsilon_{IJKL}\big( \N^{I}\wedge(\mathcal{L}_{\xi} - \mathcal{L}_{X_{\xi}})\N^{J} \wedge \overline{\e}^{K}  - (4\N^{I} \wedge \de \N^{J} - 2\overline{\omega}^{IJ}) \wedge \imath_{\xi}\overline{\e}^{K} \big) \wedge \overline{\e}^{L} \\
&= \imath_{\overline{\xi}}\overline{\e}^{I}\wedge \overline{b}_{\mathtt{EH}}(\e)_{I} + \jmath^{*}\mathcal{Q}^{\mathtt{(t)}}_{\mathtt{EH}}.
\end{align*}
Thus, the presymplectic $\xi$-currents are given by
\begin{align}
    &\mathcal{J}^{\mathtt{(t)}}_{\mathtt{EH}} = \imath_{\xi}e^{I} \wedge \mathrm{E}_{\mathtt{EH}}(\e)_{I} + \de \mathcal{Q}^{\mathtt{(t)}}_{\mathtt{EH}}, &&\  \overline{\mathfrak{j}}^{\mathtt{(t)}}_{\mathtt{EH}} = \imath_{\overline{\xi}}\overline{\e}^{I}\wedge \overline{b}_{\mathtt{EH}}(\e)_{I} + \jmath^{*}\mathcal{Q}^{\mathtt{(t)}}_{\mathtt{EH}}.
\end{align}
The associated charges are
\begin{align*}
    \mathbb{Q}^{\mathtt{(t)}}_{\mathtt{EH}} &= \int_{\Sigma}\imath^{*}\mathcal{J}^{\mathtt{(t)}}_{\mathtt{EH}} - \int_{\partial \Sigma}  \overline{\imath}^{*}\overline{\mathfrak{j}}^{\mathtt{(t)}}_{\mathtt{EH}}
    \\ &= \int_{\Sigma}\imath^{*}\big( \imath_{\xi}\e^{I} \wedge \mathrm{E}_{\mathtt{EH}}(\e)_{I}  + \de \mathcal{Q}^{\mathtt{(t)}}_{\mathtt{EH}} \big) - \int_{\partial \Sigma} \overline{\imath}^{*} \big(\imath_{\overline{\xi}}\overline{\e}^{I}\wedge \overline{b}_{\mathtt{EH}}(\e)_{I} + \jmath^{*}\mathcal{Q}^{\mathtt{(t)}}_{\mathtt{EH}} \big) \\
    &=  \int_{\Sigma}\imath^{*} \imath_{\xi}\big(\e^{I} \wedge \mathrm{E}_{\mathtt{EH}}(\e)_{I}\big)  - \int_{\partial \Sigma} \overline{\imath}^{*} \big( \imath_{\overline{\xi}}\overline{\e}^{I}\wedge \overline{b}_{\mathtt{EH}}(\e)_{I} \big).
\end{align*}

\noindent In $\mathtt{Sol}_{\mathcal{N}}( \mathbb{S}^{\mathtt{(t)}}_{\mathtt{EH}})$, the charges vanish as expected. In the Dirichlet solution space, $\mathtt{Sol}_{\mathcal{D}}( \mathbb{S}^{\mathtt{(t)}}_{\mathtt{EH}})$, $\overline{\e}^{I}$ is fixed at the boundary and the charges only vanish \emph{on-shell}.

%% file: PALATINI.tex
\chapter{The Palatini Action}\label{Palatiniaction}
An additional set of gravitational theories known as metric-affine or Palatini theories assume the metric and the connection to be equally independent and fundamental. These theories make the Lagrangian first order in derivatives, so the Palatini formalism is sometimes called a first-order formalism. Palatini theories may also be described in tetrad variables, where the independent variables are the tetrads and the spin-connection. The addition of the appropriate surface terms to the Palatini action was done by Obukhov in \cite{obukhov1987palatini}. The traditional Palatini formalism assumes the connection to be torsion-free and metric-compatible. In this chapter, we will study Palatini theories in full generality, taking into account torsion, non-metricity and boundaries. We will study their spaces of solutions in terms of metric and tetrad variables and improve previous results regarding the variational treatment of the problem by relying on the covariant phase space.  Furthermore, the equivalence between the Einstein-Hilbert and Palatini actions and between their metric and tetrad formulations in the presence of boundaries has remained an open problem until now \cite{jacobson2015black, oliveri2020boundary, de2018gauge}. 

\section{Metric Variables}
Let $M$ be a four dimensional manifold, $X, Y, Z \in \mathfrak{X}(M)$ vector fields, and let $\widetilde{\nabla}$ be a completely general affine connection on the manifold. Let its associated Riemann and torsion tensors be defined as 
\begin{align*}
    &\widetilde{\R}(X, Y)Z:= \widetilde{\nabla}_{X}\widetilde{\nabla}_{Y}Z - \widetilde{\nabla}_{Y}\widetilde{\nabla}_{X}Z - \widetilde{\nabla}_{[X,Y]}Z \quad &&\equiv \quad (\widetilde{\R}(X, Y)Z)^\alpha =: \tensor{\widetilde{\R}}{^\alpha_\beta_\gamma_\delta}Z^{\beta}X^{\gamma}Y^{\delta},\\
    &\widetilde{T}(X, Y):= \widetilde{\nabla}_{X}Y - \widetilde{\nabla}_{Y}X - [X,Y] \quad && \equiv \quad (\widetilde{T}(X,Y))^{\alpha} =: \tensor{\widetilde{T}}{^\alpha_\beta_\gamma}X^{\beta}Y^{\gamma}.
\end{align*}
The components of the Riemann tensor satisfy the following identity
\begin{align*}
    \tensor{\widetilde{\R}}{^\alpha_\beta_\gamma_\delta}Z^{\beta} = \widetilde{\nabla}_{\gamma}\widetilde{\nabla}_{\delta}Z^{\alpha} - \widetilde{\nabla}_{\delta}\widetilde{\nabla}_{\gamma}Z^{\alpha} + \tensor{\widetilde{T}}{^\lambda_\gamma_\delta}\widetilde{\nabla}_{\lambda}Z^{\alpha}
\end{align*}
Let us consider a metric $g$ on $M$. The connection $\widetilde{\nabla}$ in its full generality has torsion and is not metric-compatible. The non-metricity tensor is defined as
\begin{align*}
    \widetilde{M}(X,Y)Z:&= Z(g(X,Y)) - g(\widetilde{\nabla}_{Z}X, Y) - g(X, \widetilde{\nabla}_{Z}Y) 
    \\
    &= Z^{\alpha}\widetilde{\nabla}_{\alpha}(g_{\beta\gamma}X^{\beta}Y^{\gamma}) - g_{\beta\gamma}Z^{\alpha}(\widetilde{\nabla}_{\alpha}X^{\beta})Y^{\gamma} - g_{\beta\gamma}X^{\beta}Z^{\alpha}(\widetilde{\nabla}_{\alpha}Y^{\gamma}) \\
    &= (\widetilde{\nabla}_{\alpha}g_{\beta\gamma})Z^{\alpha}X^{\beta}Y^{\gamma} = \widetilde{M}_{\alpha\beta\gamma}Z^{\alpha}X^{\beta}Y^{\gamma},
\end{align*}
and hence non-metricity implies that $\widetilde{M}_{\alpha\beta\gamma} \neq 0$. Let $\widetilde{\nabla}$ and $\nabla$ be two connections on $M$, whose difference is a tensor field on $M$
\begin{align*}
    Q(X,Y):= \widetilde{\nabla}_{X}Y - \nabla_{Y}X \quad \equiv \quad Q(X,Y)^{\alpha} =: \tensor{Q}{^\alpha_\beta_\gamma}X^{\beta}Y^{\gamma}.
\end{align*}
The tensor $Q$ allows us to relate the non-metricity, the torsion, the Riemann curvature tensor, the Ricci tensor and the Ricci scalar of both connections according to
\begin{align*}
    &\widetilde{M}_{\alpha\beta\gamma} = M_{\alpha\beta\gamma} - g_{\lambda\gamma}\tensor{Q}{^\lambda_\alpha_\beta} - g_{\beta\lambda}\tensor{Q}{^\lambda_\alpha_\gamma}, \\
   &\tensor{\widetilde{T}}{^\alpha_\beta_\gamma} = \tensor{T}{^\alpha_\beta_\gamma} + \tensor{Q}{^\alpha_\beta_\gamma} - \tensor{Q}{^\alpha_\gamma_\beta}, \\
   &\tensor{\widetilde{\R}}{^\alpha_\beta_\gamma_\delta} = \tensor{\R}{^\alpha_\beta_\gamma_\delta} + \nabla_{\gamma}\tensor{Q}{^\alpha_\delta_\beta} - \nabla_{\delta}\tensor{Q}{^\alpha_\gamma_\beta} + \tensor{Q}{^\alpha_\gamma_\lambda}\tensor{Q}{^\lambda_\delta_\beta} - \tensor{Q}{^\alpha_\delta_\lambda}\tensor{Q}{^\lambda_\gamma_\beta} + \tensor{T}{^\lambda_\gamma_\delta}\tensor{Q}{^\alpha_\lambda_\beta},\\
   &\tensor{\widetilde{\R}}{_\beta_\delta} := \tensor{\widetilde{\R}}{^\alpha_\beta_\alpha_\delta} = \tensor{\R}{_\beta_\delta} + \nabla_{\alpha}\tensor{Q}{^\alpha_\delta_\beta} - \nabla_{\delta}\tensor{Q}{^\alpha_\alpha_\beta} + \tensor{Q}{^\alpha_\alpha_\lambda}\tensor{Q}{^\lambda_\delta_\beta} - \tensor{Q}{^\alpha_\delta_\lambda}\tensor{Q}{^\lambda_\alpha_\beta} + \tensor{T}{^\lambda_\alpha_\delta}\tensor{Q}{^\alpha_\lambda_\beta}, \\
   &\widetilde{\R} = g^{\beta\delta}\widetilde{\R}_{\beta\delta} = \R + g^{\beta\delta}\nabla_{\alpha}\tensor{Q}{^\alpha_\delta_\beta} - g^{\beta\delta}\nabla_{\delta}\tensor{Q}{^\alpha_\alpha_\beta} + \tensor{Q}{^\alpha_\alpha_\lambda}\tensor{Q}{^\lambda^\beta_\beta} - \tensor{Q}{^\alpha^\beta_\lambda}\tensor{Q}{^\lambda_\alpha_\beta} + \tensor{T}{^\lambda_\alpha^\beta}\tensor{Q}{^\alpha_\lambda_\beta}.
\end{align*}
Alternatively, we may choose to write $Q$ in terms of the components of the torsion and non-metricity,
\begin{align*}
    Q_{\alpha\beta\gamma} &= \frac{1}{2}\big( \widetilde{T}_{\alpha\beta\gamma} - T_{\alpha\beta\gamma} - \widetilde{T}_{\beta\gamma\alpha} + T_{\beta\gamma\alpha} + \widetilde{T}_{\gamma\alpha\beta} - T_{\gamma\alpha\beta}\big) \\
    &\quad + \frac{1}{2}\big(M_{\beta\gamma\alpha} - \widetilde{M}_{\beta\gamma\alpha} + M_{\gamma\alpha\beta} - \widetilde{M}_{\gamma\alpha\beta}   - M_{\alpha\beta\gamma} + \widetilde{M}_{\alpha\beta\gamma}\big).
\end{align*}

The extrinsic curvature of the lateral boundary $\partial_{L} M$ is generalised using the metric $g$ and the connection $\widetilde{\nabla}$. As stated previously, the inclusion map of the boundary will be denoted by $\jmath: \partial_{L}M \hookrightarrow M$, the induced metric at the boundary will be denoted by $\overline{g}:=\jmath^{*}g$ and the unit outer normal vector to the lateral boundary of $M$ will be the vector $\nu^{\alpha}$. The generalisation of the extrinsic curvature is then given by \cite{obukhov1987palatini},
\begin{align*}
    \widetilde{\K}_{\overline{\alpha}\overline{\beta}}:=\frac{1}{2}\jmath^{\alpha}_{\overline{\alpha}}\jmath^{\beta}_{\overline{\beta}}(\widetilde{\nabla}_{\alpha}\nu_{\beta}+g_{\alpha\gamma}\widetilde{\nabla}_{\beta}\nu^{\gamma}).
\end{align*}
Being a generalisation of the associated extrinsic curvature of the Levi-Civita connection, $\widetilde{\K}_{\alpha\beta}$ is not necessarily symmetric.  In terms of the Levi-Civita connection $\nabla$ and the $Q$ tensor, the extrinsic curvature may be expressed as
\begin{align}\label{extrinsicGeneralised}
    \widetilde{\K}_{\alpha\beta} = \K_{\alpha\beta} + \frac{1}{2}\jmath^{\alpha}_{\overline{\alpha}}\jmath^{\beta}_{\overline{\beta}}(\nabla_{\alpha}\nu_{\beta} - \tensor{Q}{^\gamma_\alpha_\beta} + g_{\alpha\gamma}\nabla_{\beta}\nu^{\gamma} + g_{\alpha\gamma}\tensor{Q}{^\gamma_\beta_\lambda}\nu^{\lambda}).
\end{align}
With all of these ingredients unravelled, we are now in a position to discuss the Palatini action in metric variables.

Traditionally, the Palatini action is written in terms of a metric and a connection. But because of the simplicity of working with tensors, we choose to work with the variables $(g,Q)$. This is possible because there is a bijection between the connection $\widetilde{\nabla}$ and the $Q$-difference tensor once we have fixed a connection. From now on, we will fix the connection $\nabla$ to be the Levi-Civita connection of the metric g, while keeping $\widetilde{\nabla}$ completely general. The space of fields for the Palatini action is thus given by, 
\begin{equation}\label{palatiniSpaceofFields}
    \mathcal{F}_{\mathtt{PT}} := \{ g \;\;|\;\;\jmath^{*}g \;\; \text{is timelike}\} \times \mathfrak{T}^{2}_{1}(M),
\end{equation}
where $\mathfrak{T}^{2}_{1}(M)$ is the space of tensor fields of type $(2,1)$. The metric Palatini action is
\begin{align}\label{Palatiniactionmetric}
    \mathbb{S}^{\mathtt{(m)}}_{\mathtt{PT}}(g,Q) = \int_{M}L_{\mathtt{PT}}(g,Q) - \int_{\partial M} \overline{\ell}_{\mathtt{PT}}(g, Q),
\end{align}
where the Lagrangians are given by
\begin{align}
    &L_{\mathtt{PT}}(g,Q) = (\widetilde{\R} - 2\Lambda)\volg, \\    
    &\overline{\ell}_{\mathtt{PT}}(g, Q) = - 2 \widetilde{\K}\volgg,
\end{align}
in which $\widetilde{\K}$ is the trace of the extrinsic curvature from (\ref{extrinsicGeneralised}). 
A closer analysis of the terms that appear in these Lagrangians reveals it is possible to rewrite them as those of the Einstein-Hilbert action  plus a  Q-dependent coupling term,
\begin{align*}
    L_{\mathtt{PT}}(g,Q) &= (\mathring{\R} + \nabla_{\alpha}\tensor{Q}{^\alpha^\beta_\beta} - \nabla_{\delta}\tensor{Q}{^\alpha_\alpha^\delta} + \tensor{Q}{^\alpha_\alpha_\lambda}\tensor{Q}{^\lambda^\beta_\beta} - \tensor{Q}{^\alpha^\beta_\lambda}\tensor{Q}{^\lambda_\alpha_\beta} - 2\Lambda)\volg \\
    &= (\R - 2\Lambda)\volg +  (\tensor{Q}{^\alpha_\alpha_\lambda}\tensor{Q}{^\lambda^\beta_\beta} - \tensor{Q}{^\alpha^\beta_\lambda}\tensor{Q}{^\lambda_\alpha_\beta}\big)\volg + \mathrm{d}(\iota_{Z}\volg) \\
    &= L_{\mathtt{EH}}(g) + L_{\mathtt{CP}}(g,Q) ,
\end{align*}
where $Z^{\alpha} := \tensor{Q}{^\alpha^\beta_\beta}-\tensor{Q}{^\beta_\beta^\alpha}$. For the Lagrangian at the boundary, a similar situation occurs,
\begin{align*}
    \overline{\ell}_{\mathtt{PT}}(g, G) &= - 2 \big( \K + \frac{1}{2}(\tensor{Q}{^\alpha_\alpha_\beta} - \tensor{Q}{_\beta_\alpha^\alpha})\nu^\alpha \big) \volgg \\
    &= -2\K\volgg - (\tensor{Q}{^\alpha_\alpha_\beta} - \tensor{Q}{_\beta_\alpha^\alpha})\nu^{\alpha}\volgg \\
    &= \overline{\ell}_{\mathtt{GHY}}(g) + \jmath^{*}(\iota_{Z}\volg).
\end{align*}
We now compute the variations of the Lagrangian Palatini pair explicitly. First, in the bulk we have,
\begin{align*}
    &\delta L_{\mathtt{PT}}(g,Q) = \delta L_{\mathtt{EH}}(g) + \delta L_{\mathtt{CP}}(g,Q),
\end{align*}
where, 
\begin{align*}
    \delta L_{\mathtt{EH}}(g) &= - \big( \R^{\alpha\beta}  - \frac{1}{2}\R g^{\alpha\beta} + \Lambda g^{\alpha\beta} \big) \volg  + \mathrm{d}(\iota_{W}\volg), \\
    \delta L_{\mathtt{CP}}(g,Q) &=  (\tensor{Q}{^\lambda^\alpha_\sigma}\tensor{Q}{^\sigma_\lambda^\beta} - \tensor{Q}{^\gamma_\gamma_\sigma}\tensor{Q}{^\sigma^\alpha^\beta} + \frac{1}{2}(\tensor{Q}{^\lambda_\lambda_\sigma}\tensor{Q}{^\sigma^\gamma_\gamma} - \tensor{Q}{^\lambda^\gamma_\sigma}\tensor{Q}{^\sigma_\lambda_\gamma})g^{\alpha\beta})\volg \wedge \delta g_{\alpha\beta}, \\
     & \quad +  \big( \delta^{\alpha}_{\gamma}\tensor{Q}{^\sigma^\beta_\beta} + g^{\alpha\sigma}\tensor{Q}{^\lambda_\lambda_\gamma} - \tensor{Q}{^\sigma_\gamma^\alpha}- \tensor{Q}{^\alpha^\sigma_\gamma}\big)\volg \wedge \delta \tensor{Q}{^\gamma_\alpha_\sigma}+ \mathrm{d}(\delta\iota_{Z}\volg), \\
      W^{\alpha} &=  (g^{\alpha\mu}g^{\beta\lambda} - g^{\alpha\lambda}g^{\beta\mu})\nabla_{\lambda}\delta g_{\beta\mu}.
\end{align*}
Therefore, the total variation of the Palatini Lagrangian at the bulk is given by,
\begin{align*}
        \delta  L_{\mathtt{PT}}(g,Q) &= (E^{\alpha\beta}_{\mathtt{EH}}(g) + E^{\alpha\beta}_{\mathtt{CP}}(g,Q))\wedge \delta g_{\alpha\beta} +  \tensor{(E_{\mathtt{CP}})}{_\gamma^\alpha^\sigma}(g,Q) \wedge \delta \tensor{Q}{^\gamma_\alpha_\sigma} + \mathrm{d}\big(\Theta_{\mathtt{PT}}(g,Q)\big),
\end{align*}
where,
\begin{align*}
    (E_{\mathtt{CP}})^{\alpha\beta}(g,Q) &:= \;\mathcal{E}^{\alpha\beta}_{\mathtt{CP}}(g,Q) \volg \\
    &=  \big(\tensor{Q}{^\lambda^\alpha_\sigma}\tensor{Q}{^\sigma_\lambda^\beta} - \tensor{Q}{^\gamma_\gamma_\sigma}\tensor{Q}{^\sigma^\alpha^\beta} + \frac{1}{2}(\tensor{Q}{^\lambda_\lambda_\sigma}\tensor{Q}{^\sigma^\gamma_\gamma} - \tensor{Q}{^\lambda^\gamma_\sigma}\tensor{Q}{^\sigma_\lambda_\gamma})g^{\alpha\beta}\big)\volg,\\
    \tensor{(E_{\mathtt{CP}})}{_\gamma^\alpha^\sigma}(g,Q)&:= \tensor{(\mathcal{E}_{\mathtt{CP}})}{_\gamma^\alpha^\sigma}(g,Q)\volg \\
    &= \big( \delta^{\alpha}_{\gamma}\tensor{Q}{^\sigma^\beta_\beta} + g^{\alpha\sigma}\tensor{Q}{^\lambda_\lambda_\gamma} - \tensor{Q}{^\sigma_\gamma^\alpha}- \tensor{Q}{^\alpha^\sigma_\gamma}\big) \volg,\\
    \Theta^{\mathtt{(m)}}_{\mathtt{PT}}&:= \Theta^{\mathtt{(m)}}_{\mathtt{EH}} + \Theta^{\mathtt{(m)}}_{\mathtt{CP}} = \iota_{W}\volg + \delta(\iota_{Z}\volg).
\end{align*}
The variation of the Lagrangian at the boundary is
\begin{align*}
    \delta \overline{\ell}_{\mathtt{PT}}(g, Q) - \jmath^{*}\Theta^{\mathtt{(m)}}_{\mathtt{PT}} &= \delta \overline{\ell}_{\mathtt{GHY}}(g) + \delta \jmath^{*}(\iota_{Z}\volg) - \jmath^{*}\Theta^{\mathtt{(m)}}_{\mathtt{EH}} - \jmath^{*}\Theta^{\mathtt{(m)}}_{\mathtt{PT}} \\
    &= \delta \overline{\ell}_{\mathtt{GHY}}(g) - \jmath^{*}\Theta^{\mathtt{(m)}}_{\mathtt{EH}} = \overline{b}_{\mathtt{EH}}^{\overline{\alpha}\overline{\beta}}(g) \wedge \delta \overline{g}_{\overline{\alpha}\overline{\beta}} - \de \overline{\theta}^{\mathtt{(m)}}_{\mathtt{EH}},
\end{align*}
with $\jmath^{*}\Theta^{\mathtt{(m)}}_{\mathtt{PT}}$ = $\delta \jmath^{*}(\iota_{Z}\volg)$ and $
\overline{\theta}^{\mathtt{(m)}}_{\mathtt{CP}}\&=0$, so we have at the boundary the same variations as in the Einstein-Hilbert action. We continue by solving the equations of motion. Let us consider the equation associated with the variations of $Q$,
\begin{align*}\label{CP_METRIC_Q}
    \tensor{(E_{\mathtt{CP}})}{_\gamma^\alpha^\sigma}(g,Q):=  \big( \delta^{\alpha}_{\gamma}\tensor{Q}{^\sigma^\beta_\beta} + g^{\alpha\sigma}\tensor{Q}{^\lambda_\lambda_\gamma} - \tensor{Q}{^\sigma_\gamma^\alpha}- \tensor{Q}{^\alpha^\sigma_\gamma}\big) \volg = 0.
\end{align*}
Standard algebraic manipulations lead us to the general solution for $Q$, which in turn also solves the second term of the other equation of motion obtained by performing the variations of $g$, 
\begin{equation*}
    Q^{\alpha\beta\gamma}_{0} = g^{\alpha \gamma} U^{\beta} 
     \quad \Longrightarrow \quad \tensor{(E_{\mathtt{CP}})}{_\gamma^\alpha^\sigma}(g,Q) = 0 \quad\text{and}\quad (E_{\mathtt{CP}})^{\alpha\beta}(g) = 0.
\end{equation*}
Now, let us consider the full equation of motion associated with the variations of $g$
\begin{align*}
    E^{\alpha\beta}_{\mathtt{EH}}(g) + E^{\alpha\beta}_{\mathtt{CP}}(g,Q) = 0.
\end{align*}
The solution for $Q$ solves $E^{\alpha\beta}_{\mathtt{CP}}(g,Q)=0$, so then we only need to impose $g$ to be part of the solution space of the Einstein-Hilbert action so that the first term $E^{\alpha\beta}_{\mathtt{EH}}(g)$ is also zero. In short, this implies that the solution space of Palatini is
\begin{align*}
    \mathtt{Sol}_\mathcal{N}(\mathbb{S}^{\mathtt{(m)}}_{\mathtt{PT}}) = \{ (g_{\alpha\beta}, \delta^\alpha_\gamma U_\beta) \mid  g \in \mathtt{Sol}(\mathbb{S}^{\mathtt{(m)}}_{\mathtt{EH}}),  \; U_{\beta} \in \Omega^{1}(M) \}.
\end{align*}
The study of the space of solutions associated with the boundary is trivial since we have obtained the same equations of motion as those from the EH action, whose solutions are already well-known. \\
The following identities hold on the space of solutions obtained above 
\begin{align*}
    &\tensor{\widetilde{\R}}{^\alpha_\beta_\mu_\nu} = \tensor{\R}{^\alpha_\beta_\mu_\nu} + g^{\alpha}_{\beta}(\de U)_{\mu\nu}, && \widetilde{\K}_{\alpha\beta} = \K_{\alpha\beta}, \\
    &\tensor{\widetilde{\R}}{_\beta_\nu} = \tensor{\R}{_\beta_\nu} + (\de U)_{\beta\nu}, && \widetilde{\K} = \K, \\
    & \widetilde{\R} = \R,\\
    & \widetilde{M}_{\alpha \beta \gamma} = - 2 g_{\beta\gamma}U_{\alpha},
    && \tensor{\widetilde{T}}{^\gamma_\alpha_\beta} = \delta^{\gamma}_{\beta}U_{\alpha} - \delta^{\gamma}_{\alpha}U_{\beta}.
\end{align*}
The last two equations imply that, over solutions, $\widetilde{M}_{\alpha \beta \gamma} = 0 $ if and only if $\tensor{\widetilde{T}}{^\gamma_\alpha_\beta} = 0$. This means that if we choose to work with a connection that has no torsion, then we are necessarily imposing the metricity condition and vice versa. Finally, the presymplectic form is given by
\begin{align*}
\Omega^{\mathtt{(m)}}_{\mathtt{PT}} &= \delta \Big( \int_{\Sigma} \imath^{*}\Theta^{\mathtt{(m)}}_{\mathtt{PT}}- \int_{\partial \Sigma} \overline{\imath}^{*}\overline{\theta}^{\mathtt{(m)}}_{\mathtt{PT}} \Big).
\end{align*}

\section{Tetrad Variables}
The 1-form connection defined by the Levi-Civita connection $\nabla$ corresponding to the metric $g_{\alpha\beta}=\eta_{IJ}\e^{I}_{\alpha}\e^{J}_{\beta}$ is given by
\begin{align*}
    \tensor{\omega}{_\mu^K_I}:= e^{K}_{\alpha}\nabla_{\mu}E^{\alpha}_{I}.
\end{align*}
A generic connection $1$-form with torsion and non-metricity will be defined in terms of its covariant derivative,
\begin{align*}
    \tensor{\widetilde{\omega}}{_\mu^K_I}:= e^{K}_{\alpha}\widetilde{\nabla}_{\mu}E^{\alpha}_{I}.
\end{align*}
Given two spin-connections $\omega$ and $\widetilde{\omega}$, their covariant derivatives are related by a $(2,1)$-tensor $Q$, 
\begin{align*}
    \tensor{Q}{^\beta_\mu_\alpha} = \tensor{(\widetilde{\nabla} - \nabla)}{^\beta_\mu_\alpha}= E^{\beta}_{K}e^{J}_{\alpha}(\tensor{\widetilde{\omega}}{_\mu^K_J} - \tensor{\omega}{_\mu^K_J}) =: \tensor{\varphi(e, \widetilde{\omega})}{^\beta_\mu_\alpha},
\end{align*}
which allows us to define the map,
\begin{equation}\label{mapPalatini}
    \Psi (e, \widetilde{\omega}) := (\Phi(e), \varphi(e, \widetilde{\omega})),
\end{equation}
where $\Psi: (e, \widetilde{\omega}) \mapsto (g, Q)$ and $\Phi: e \mapsto g$.
The Lagrangian pair is given by this change of variables, 
\begin{align*}
    L_{\mathtt{PT}}(\e, \widetilde{\omega}) &:= L_{\mathtt{PT}}\big(g = \mathrm{Tr}(\e \otimes \e), Q = \varphi(e, \widetilde{\omega})\big) \\
\overline{\ell}_{\mathtt{PT}}(\e, \hat{\omega}) &:= \overline{\ell}_{\mathtt{PT}}\big(g = \mathrm{Tr}(\e \otimes \e), Q = \varphi(e, \widetilde{\omega})\big),
\end{align*}
where $g_{\alpha\beta} = \eta_{IJ}\e^{I}_{\alpha}\e^{J}_{\beta}$, $E^{\alpha}_{I}\e^{I}_{\beta} = g^{\alpha}_{\beta}$ and $E^{\alpha}_{I}\e^{J}_{\alpha} = \eta^{J}_{I}$ (see Appendix \ref{appendix} for more details on this). 
For convenience, it is easier to work out the solution spaces if we decompose the spin-connection into its antisymmetric and symmetric parts,
\begin{align*}
    \widetilde{\omega}_{IJ} = \widehat{\omega}_{IJ} + S_{IJ}.
\end{align*}
This splitting  allows us to rewrite the Lagrangian in terms of $(e, \widehat{\omega}, S)$,
\begin{equation}
\mathbb{S}^{\mathtt{(t)}}_\mathtt{PT} = \int_{M} L_\mathtt{PT}(e, \widehat{\omega}, S) - \int_{\partial M} \ell_\mathtt{PT}(e, \widehat{\omega}).
\end{equation}
The Lagrangians are then given by
\begin{align*}
    &L_\mathtt{PT}(e, \widehat{\omega}, S) := \frac{1}{2}\epsilon_{IJKL}\Big( \widehat{F}^{IJ} - \frac{\Lambda}{6}\e^{I}\wedge\e^{J} + \tensor{S}{^I_M}\wedge \tensor{S}{^M^J}\Big)\wedge \e^{K}\wedge \e^{L},   \\
    &\ell_\mathtt{PT}(e, \widehat{\omega}, S):= -\frac{1}{2}\epsilon_{IJKL}\big(2 \N^{I}\de \N^{J} - (\jmath^{*}\widehat{\omega})^{IJ}\big)\wedge \overline{\e}^{K} \wedge \overline{\e}^{J},
\end{align*}
with $N^{I}:= \iota_{\nu}e^{I}$, and where the curvature of the spin-connection $\widehat{\omega}$ is given as usual by $\widehat{F}_{IJ} = \mathrm{d}\widehat{\omega}_{IJ} + \tensor{\widehat{\omega}}{_I_K} \wedge \tensor{\widehat{\omega}}{^K_J}$. 
The variations of this Lagrangian pair are, 
\begin{align*}
    & \delta L_\mathtt{PT}(e, \widehat{\omega}, S) \! =   \mathrm{E}^{(1)}_{\mathtt{PT}}(e, \widehat{\omega}, S)_{I} \wedge \delta \e^{I} \!+ \mathrm{E}^{(2)}_{\mathtt{PT}}(e, \widehat{\omega}, S)_{IJ}\wedge \delta \widehat{\omega}^{IJ}\! + \mathrm{E}^{(3)}_{\mathtt{PT}}(e, \widehat{\omega}, S)_{IJ} \wedge \delta S^{IJ} \!+ \de \Theta^{\mathtt{(t)}}_{\mathtt{PT}},\\
    &\delta \ell_\mathtt{PT}(e, \widehat{\omega}, S) - \jmath^{*}\Theta^{\mathtt{(t)}}_{\mathtt{PT}}= \overline{b}_{\mathtt{PT}}(\e)_{I} \wedge \delta \overline{\e}^{I} - \de \Theta^{\mathtt{(t)}}_{\mathtt{PT}},
\end{align*}
whose Euler forms and the presymplectic potential at the bulk are,
\begin{align*}
    \mathrm{E}^{(1)}_{\mathtt{PT}}(e, \widehat{\omega}, S)_{I}&:= \epsilon_{IJKL}\big( \widehat{F}^{IJ} + \tensor{S}{^I_M} \wedge \tensor{S}{^M^J} - \frac{\Lambda}{3}\e^{I}\wedge \e^{J}\big) \wedge \e^{K}, \\
    \mathrm{E}^{(2)}_{\mathtt{PT}}(e, \widehat{\omega}, S)_{IJ}&:= \frac{1}{2}\widehat{D}(\epsilon_{IJKL}\e^{I}\wedge \e^{J}) = \epsilon_{IJKL}(\mathrm{d}\e^{I}\wedge \e^{J}  + \tensor{\widehat{\omega}}{^I_K}\wedge\e^{K}\wedge \e^{J}), \\
    \mathrm{E}^{(3)}_{\mathtt{PT}}(e, \widehat{\omega}, S)_{IJ} &:= \frac{1}{2}(\epsilon_{IJKL}\delta^{R}_{M} + \epsilon_{IJKL}\delta^{R}_{J})\tensor{S}{_R^I}\wedge \e^{K} \wedge \e^{L},\\
    \Theta^{\mathtt{(t)}}_{\mathtt{PT}}&:=\frac{1}{2}\epsilon_{IJKL}\e^{I} \wedge \e^{J} \wedge \delta \widehat{\omega}^{KL}.
\end{align*}
At the boundary, the Euler forms and the presymplectic potential are given by 
\begin{align*}
    \overline{b}_{\mathtt{PT}}(e, \widehat{\omega})_{I}&:= \epsilon_{IJKL}(2\N^{K}\de N^{L} - (\jmath^{*}\widehat{\omega})^{KL}) \wedge \overline{\e}^{J} + 2\epsilon_{MJKL}\N_{I}\N^{L} (\iota_{\overline{E}^{J}} \de \overline{e}^{K})\wedge \overline{e}^{M}, \\
    \overline{\theta}^{\mathtt{(t)}}_{\mathtt{PT}}&:= \epsilon_{IJKL}\overline{\e}^{I}\wedge \overline{\e}^{J}\wedge \N^{K} \wedge \delta \N^{L}.
\end{align*}
Let us now solve the equations of motion. We begin by expanding $S_{IJ} = S_{IJK}e^{K}$ and using the unique decomposition $S_{IJK}= \cancel{S}_{IJK} + \eta_{IJ}U_{K}$ with $\cancel{S}_{IJK} = 0$. Plugging this into $\mathrm{E}^{(3)}_{\mathtt{PT}}(e, \widehat{\omega}, S)_{IJ} = 0$ we get, 
\begin{equation*}
    \cancel{S}_{RLI} + \cancel{S}_{ILR} - \tensor{\cancel{S}}{_R_J^J}\eta_{LI} - \tensor{\cancel{S}}{_I_J^J}\eta_{LR} = 0,
\end{equation*}
and taking its trace we can see that $\tensor{\cancel{S}}{_I_J^J} = 0$ which, in turn, implies $\cancel{S}_{IJK} = 0$. The general solution to $\mathrm{E}^{(3)}_{\mathtt{PT}}(e, \widehat{\omega}, S)_{IJ} = 0$ is then $S_{IJK} = \eta_{IJ}U_{K}$, with arbitrary $U_{K}$. Similarly for $\mathrm{E}^{(2)}_{\mathtt{PT}}(e, \widehat{\omega}, S)_{IJ} = 0$ we obtain the solution $\widehat{\omega}_{IJ} = \omega_{IJ}$. Thus,
\begin{align*}
    \mathtt{Sol}_{\mathcal{N}}(\mathbb{S}^{\mathtt{(t)}}_{\mathtt{PT}} )= \{ (e^{I}_{\alpha}, \omega^{IJ}_{\mu} + \eta^{IJ}U_{\mu}) \mid  e^{I}_{\alpha} \in \mathtt{Sol}(\mathbb{S}^{\mathtt{(t)}}_{\mathtt{EH}}), \;  U_{\mu} \;  \text{arbitrary}\}.
\end{align*}

We see again that the tetrad sector of Palatini is equivalent to tetrad-EH and that the boundary plays no role in the connection sector. As in the metric case, the Dirichlet or Neumann boundary conditions for tetrads are incorporated in $\mathtt{Sol}(\mathbb{S}_{\mathtt{EH}}(\e))$, which were discussed in the previous section. The presymplectic form is given by,
\begin{align*}
    \Omega^{\mathtt{(t)}}_{\mathtt{PT}} = \delta \bigg( \int_{\Sigma}\imath^{*}\Theta^{\mathtt{(t)}}_{\mathtt{PT}} - \int_{\partial \Sigma} \overline{\imath}^{*} \overline{\theta}^{\mathtt{(t)}}_{\mathtt{PT}}\bigg).
\end{align*}
\noindent Explicitly, 
\begin{align*}
    \Omega^{\mathtt{(t)}}_{\mathtt{PT}} &= \int_{\Sigma}\epsilon_{IJKL}\delta \e^{K}\wedge \e^{L} \wedge \delta \widehat{\omega}^{IJ} - \int_{\partial \Sigma}\epsilon_{IJKL}(2\delta \overline{\e}^{K}\wedge \overline{\e}^{L}\wedge \N^{I} + \overline{\e}^{K}\wedge\overline{\e}^{L}\wedge\delta \N^{I})\wedge \delta \N^{J}.
\end{align*}
As we can see, the vectors of the form $(0, \mathbb{W}) \in \mathrm{T}\mathtt{Sol}(\mathbb{S}^{\mathtt{(t)}}_{\mathtt{PT}})$ correspond to degenerate directions of $\Omega^{\mathtt{(t)}}_{\mathtt{PT}}$. 

%% file: HMS.tex
\chapter{The Hojman-Mukku-Sayed Action}\label{HMS}
In this chapter, we present the Hojman-Mukku-Sayed (HMS) action \cite{hojman1980parity}, a generalisation of the Holst action \cite{holst1995} in which non-metricity and torsion are considered on manifolds with timelike boundaries. It is named after the work of Hojman, Mukku and Sayed, who considered the Palatini variational principle in metric variables along with a parity-violating term. In the tetrad formalism, HMS reduces to the Holst action when imposing metricity. We generalise the Holst action to include metricity and torsion, and derive a new boundary Lagrangian in the metric formalism necessary to guarantee the equivalence between the metric and tetrad formulations of the HMS action. This new boundary Lagrangian reduces to the one given by Bodendorfer \emph{et al.} in \cite{Bodendorfer:2013hla} when imposing metricity. 

In the metric formalism, the introduction of torsion and non-metricity on gravitational theories started with the works of Hehl \cite{hehl1974spin} and Trautman \cite{trautman1972einstein}. A variational treatment allowing torsion and non-metricity was soon after carried out by Sandberg \cite{sandberg1975torsion}. He showed the invariance of the action under a projective transformation of the connection and how it was not determined by Einstein's field equations. In a different spirit, Floreanini and Percacci \cite{floreanini1990palatini} constructed an action by adding the appropriate terms so that the conditions of metricity and torsion were obtained as dynamical equations. A historical overview of metric Palatini theories or metric affine theories can be found in \cite{ferraris1982variational, burton1999palatini}. 

In the tetrad formalism, the counterpart to the projective character of the connection in Palatini theories was shown to be a gauge symmetry by Julia and Silva \cite{Julia:1998ys}. The Holst action with torsion, but assuming metricity, was studied by Dadhich and Pons \cite{dadhich2012equivalence}. In the presence of inner boundaries, the Holst action has been studied for isolated horizons by \cite{Ashtekarisolated, AshtekarCorichi, Chatterjee:2008if, Engle:2009vc}. In \cite{Bodendorfer:2013hla}, the authors proposed the introduction of a boundary Lagrangian to the Holst action allowing for torsion but assuming metricity. This boundary term is the generalisation of other boundary terms considered in \cite{ashtekar1991lectures, ashtekar2008asymptotics, Wieland:2020gno, Pranzetti2016, CorichiActionsTopologicalTerms}. 

With the introduction of the HMS action, we aim at completing the study of the Holst action with torsion, non-metricity and boundaries for both the metric and tetrad formulations and give response to the following three unanswered questions: What is the covariant phase space associated with the HMS action in the metric and tetrad formalisms? Do the metric and tetrad formalisms provide equivalent formulations? Are the HMS action and presymplectic potentials cohomologically equivalent to those for Palatini theories?

This chapter is thus devoted to the study of the covariant phase space of the HMS action in both the tetrad and metric formalisms using the relative bicomplex (\ref{relativebicomplex}) and the CPS algorithm (\ref{cpsalgorithm}). The results obtained in this chapter will allow us to show the equivalence of the metric and tetrad formalisms and then proceed to study their cohomological equivalence with Palatini theories in Chapter \ref{equivalence}. 

\section{Metric Variables}
Let $M$ be a differentiable manifold with boundary $\partial M$ as explained in section \ref{spacetime}. The space of fields for the HMS action is the same as the one for the Palatini action, 
\begin{equation}\label{HMSSpaceofFields}
    \mathcal{F}_{\mathtt{HMS}} := \mathcal{F}_{\mathtt{PT}} = \{ g \;\;|\;\;\jmath^{*}g \;\; \text{is timelike}\} \times \mathfrak{T}^{2}_{1}.
\end{equation}
A Lagrangian pair define the action,
\begin{equation}\label{HMSaction}
    \mathbf{S}^{\mathtt{(m)}}_{\mathtt{HMS}}(g,Q) = \int_{\mathcal{M}} L_{\mathtt{HMS}}(g,Q) - \int_{\partial_{L}\mathcal{M}} \overline{\ell}_{\mathtt{HMS}}(g,Q),
\end{equation}
where the bulk and boundary Lagrangians are given by
\begin{align}
    L_{\mathtt{HMS}}(g,Q) &:= \big(\widetilde{\R} - 2\Lambda - \frac{1}{2\gamma} \volg^{\alpha\beta\mu\nu} \widetilde{\R}_{\alpha\beta\mu\nu}\big)\volg,\\
    \overline{\ell}_{\mathtt{HMS}}(g,Q) &:= -2\widetilde{\K}\volgg + \frac{1}{\gamma}\jmath^{*}(\iota_{q}\volg),
\end{align}
where $\gamma \in \mathbb{R}$ is a constant, $q^{\mu}:= (\volg)^{\mu\alpha\beta\nu}Q_{\alpha\beta\nu}$ and the HMS-Lagrangians are just the Palatini Lagrangians with a parity-violating term, in analogy with the Holst term for tetrad variables that we shall consider later. 

It is convenient to split the HMS Lagrangian pair into the following parts before proceeding with its variations,
\begin{align*}
     L_{\mathtt{HMS}}(g,Q)  &= L_{\mathtt{EH}}(g) + L_{\mathtt{CP}}(g,Q) -\frac{1}{\gamma} L_{\mathtt{CH}}(g,Q) + \mathrm{d}(\iota_{Z}\volg + \frac{1}{\gamma}\iota_{q}\volg),\\
     \overline{\ell}_{\mathtt{HMS}}(g,Q) &=\overline{\ell}_{\mathtt{PT}}(g, Q) + \frac{1}{\gamma}\jmath^{*}\iota_{q}\volg =  \overline{\ell}_{\mathtt{GHY}}(g) + \jmath^{*}(\iota_{Z + q/\gamma}\volg),
\end{align*}
where,
\begin{align*}
    L_{\mathtt{EH}}(g) &:= (\R - 2\Lambda)\volg, \\
    L_{\mathtt{CP}}(g,Q) &:= (\tensor{Q}{^\alpha_\alpha_\lambda}\tensor{Q}{^\lambda^\beta_\beta} - \tensor{Q}{^\alpha^\beta_\lambda}\tensor{Q}{^\lambda_\alpha_\beta}\big)\volg, \\
     L_{\mathtt{CH}}(g,Q) &:= (\volg)^{\alpha\beta\mu\nu}Q_{\alpha\mu\sigma}\tensor{Q}{^\sigma_\nu_\beta}\volg, \\
     \overline{\ell}_{\mathtt{GHY}}(g) &:= - 2\K\volgg, \\
    Z^{\alpha} &:= \tensor{Q}{^\alpha^\beta_\beta} - \tensor{Q}{_\beta^\beta^\alpha},\\
    W^{\alpha} &:= (g^{\lambda \beta}g^{\alpha \kappa} - g^{\lambda \alpha}g^{\kappa \beta}) \nabla_{\lambda} \delta g_{\kappa \beta}.
\end{align*}
With these terms at hand, in order to get the variations of the Holst action we only need to compute the variation of the $L_{\mathtt{CH}}(g,Q)$ and $\Theta_{\mathtt{HMS}}(g,Q)$ terms, since the variations of the other two components were computed in Chapter \ref{Palatiniaction}. The variations at the bulk are given by
\begin{align}
    &\delta L_{\mathtt{HMS}}(g,Q)  = \delta L_{\mathtt{EH}}(g) + \delta L_{\mathtt{CP}}(g,Q) -\frac{1}{\gamma}\delta L_{\mathtt{CH}}(g,Q) + \mathrm{d}\Theta^{\mathtt{(m)}}_{\mathtt{HMS}},
\end{align}
\noindent explicitly 
\begin{align*}
    \delta L_{\mathtt{EH}}(g) &= (E_{\mathtt{EH}})^{\alpha\beta}(g)\wedge \delta g_{\alpha\beta} + \mathrm{d}\Theta_{\mathtt{EH}}(g), \\
    \delta L_{\mathtt{CP}}(g) &= (E_{\mathtt{CP}})^{\alpha\beta}(g,Q)\wedge \delta g_{\alpha\beta} + \tensor{(E_{\mathtt{CP}})}{_\alpha^\beta^\sigma}(g,Q) \wedge \delta \tensor{Q}{^\alpha_\beta_\sigma},\\
    \delta L_{\mathtt{CH}}(g,Q) &= (E_{\mathtt{CH}})^{\alpha\beta}(g,Q)\wedge\delta g_{\alpha\beta} + \tensor{(E_{\mathtt{CH}})}{_\alpha^\beta^\sigma}(g,Q)\wedge \delta \tensor{Q}{^\alpha_\beta_\sigma},\\
    \Theta^{\mathtt{(m)}}_{\mathtt{HMS}}&:=\Theta^{\mathtt{(m)}}_{\mathtt{PT}}+ \frac{1}{\gamma}\delta \iota_{\vec{q}}\volg =\Theta^{\mathtt{(m)}}_{\mathtt{EH}}+ \delta (\iota_{Z}\volg +  \iota_{q/\gamma}\volg),
\end{align*}
whose Euler forms are given by 
\begin{align*}
    &(E_{\mathtt{CH}})^{\alpha\beta}(g,Q) = \frac{1}{2}\volg^{\mu\nu\eta\xi}\big(\delta^{\beta}_{\xi}(\tensor{Q}{_\mu^\alpha^\sigma}\tensor{Q}{_\sigma_\eta_\nu} - \tensor{Q}{_\mu_\nu^\sigma}\tensor{Q}{_\sigma_\eta^\alpha}) + \delta^{\alpha}_{\eta}Q_{\mu\xi\sigma}\tensor{Q}{^\sigma^\beta_\nu} + g^{\alpha\beta}Q_{\mu\eta\sigma}\tensor{Q}{^\sigma_\xi_\nu},\\
    & \quad  + \delta^{\alpha}_{\xi}(\tensor{Q}{_\mu^\beta^\sigma}\tensor{Q}{_\sigma_\eta_\nu} - \tensor{Q}{_\mu_\nu^\sigma}\tensor{Q}{_\sigma_\eta^\beta}) + \delta^{\beta}_{\eta}Q_{\mu\xi\sigma}\tensor{Q}{^\sigma^\alpha_\nu} + g^{\alpha\beta}Q_{\mu\eta\sigma}\tensor{Q}{^\sigma_\xi_\nu}\big)\volg, \\
    &\tensor{(E_{\mathtt{CH}})}{_\alpha^\beta^\sigma}(g,Q) = \volg^{\xi\mu\beta\nu}(g_{\xi\alpha}\tensor{Q}{^\sigma_\nu_\mu} - \delta^{\sigma}_{\xi}Q_{\nu\mu\alpha})\volg, \\
    &(E_{\mathtt{EH}})^{\alpha\beta}(g) = - (\R^{\alpha\beta} - \frac{1}{2}\R g^{\alpha\beta} + \Lambda g^{\alpha\beta})\volg, \\
    &(E_{\mathtt{CP}})^{\alpha\beta} = \frac{1}{2}\big( \tensor{Q}{^\gamma^\alpha_\sigma}\tensor{Q}{^\sigma_\gamma^\beta} \!+ \tensor{Q}{^\gamma^\beta_\sigma}\tensor{Q}{^\sigma_\gamma^\alpha}\! - \tensor{Q}{^\mu_\mu_\sigma}(\tensor{Q}{^\sigma^\alpha^\beta}\!+ \tensor{Q}{^\sigma^\beta^\alpha})\!+ \!g^{\alpha\beta}(\tensor{Q}{^\mu_\mu_\sigma}\tensor{Q}{^\sigma^\gamma_\gamma} \!- \!\tensor{Q}{^\gamma^\tau_\sigma}\tensor{Q}{^\sigma_\gamma_\tau})\big)\volg,\\
    & \tensor{(E_{\mathtt{CP}})}{_\alpha^\beta^\sigma}(g,Q) = \big( \delta^{\beta}_{\alpha}\tensor{Q}{^\sigma^\gamma_\gamma} + g^{\beta\sigma}\tensor{Q}{^\mu_\mu_\alpha} - \tensor{Q}{^\sigma_\alpha^\beta} - \tensor{Q}{^\beta^\sigma_\alpha}\big)\volg.
\end{align*}
There is no contribution to the presymplectic potential from the coupling term because it has no derivatives. The full expression for the variation of the bulk HMS Lagrangian is,
\begin{align}
    \delta L_{\mathtt{HMS}}(g,Q)  &=  (E^{(1)}_{\mathtt{HMS}})^{\alpha\beta}\wedge \delta g_{\alpha\beta} +  E^{(2)}_{\mathtt{HMS}}\wedge \delta \tensor{Q}{^\alpha_\beta_\sigma} + \mathrm{d}\Theta^{\mathtt{(m)}}_{\mathtt{HMS}},
\end{align}
where, 
\begin{align}
    &(E^{(1)}_{\mathtt{HMS}})^{\alpha\beta} := \big(E_{\mathtt{EH}}(g) + E_{\mathtt{CP}}(g) - \frac{1}{\gamma}E_{\mathtt{CH}}(g,Q)\big)^{\alpha\beta},\\
    &\tensor{(E^{(2)}_{\mathtt{HMS}})}{_\alpha^\beta^\sigma}(g,Q) := \tensor{\big( E_{\mathtt{CP}}(g,Q)- \frac{1}{\gamma}E_{\mathtt{CH}}(g,Q) \big)}{_\alpha^\beta^\sigma}.
\end{align}
At the boundary, the variations are
\begin{align*}
        \delta \overline{\ell}_{\mathtt{HMS}}(g,Q) - \jmath^{*}\Theta^{\mathtt{(m)}}_{\mathtt{HMY}} &= \delta \overline{\ell}_{\mathtt{GHY}}(g) + \delta\jmath^{*}(\iota_{Z + q/\gamma}\volg)-\jmath^{*}\Theta^{\mathtt{(m)}}_{\mathtt{EH}} - \jmath^{*}\delta(\iota_{Z + q/\gamma}\volg)\\
        & = \delta\overline{\ell}_{\mathtt{GHY}}(g)-\jmath^{*}\Theta^{\mathtt{(m)}}_{\mathtt{EH}}.
\end{align*}
To study the space of solutions of the HMS-action we need the irreducible decomposition of the torsion and non-metricity tensors given by,
\begin{equation*}
    \tensor{\widetilde{T}}{^\alpha_\beta_\sigma} = ^{(1)}\tensor{\widetilde{T}}{^\alpha_\beta_\sigma} + ^{(2)}\tensor{\widetilde{T}}{^\alpha_\beta_\sigma} + ^{(3)}\tensor{\widetilde{T}}{^\alpha_\beta_\sigma}
    \begin{cases}
    & ^{(1)}\tensor{\widetilde{T}}{^\alpha_\beta_\sigma}:= \frac{1}{2}(\tensor{\widetilde{T}}{^\alpha_\beta_\alpha}\delta^{\alpha}_{\sigma} - \tensor{\widetilde{T}}{^\alpha_\sigma_\alpha}\delta^{\alpha}_{\beta}), \\
    & ^{(2)}\tensor{\widetilde{T}}{^\alpha_\beta_\sigma}:= g^{\alpha\mu}\widetilde{T}_{[\mu\beta\sigma]}, \\
    & ^{(3)}\tensor{\widetilde{T}}{^\alpha_\beta_\sigma} := \tensor{\widetilde{T}}{^\alpha_\beta_\sigma} - ^{(1)}\tensor{\widetilde{T}}{^\alpha_\beta_\sigma} -  ^{(2)}\tensor{\widetilde{T}}{^\alpha_\beta_\sigma},
    \end{cases}
\end{equation*}
where all of its components are antisymmetric in the last two indices, the components $^{(1)}\tensor{\widetilde{T}}{^\alpha_\beta_\sigma}$ and $^{(2)}\tensor{\widetilde{T}}{^\alpha_\beta_\sigma}$ are traceless, and $^{(3)}\tensor{\widetilde{T}}{^\alpha_\beta_\sigma}$ satisfies the cyclic identity, 
\begin{equation}
    ^{(3)}\widetilde{T}_{[\alpha\beta\sigma]}=0.
\end{equation}
For convenience, we further define
\begin{equation}
    \widetilde{d}_{\mu} := 4^{(2)}\widetilde{T}^{\alpha\beta\sigma}(\volg)_{\alpha\beta\sigma\mu}.
\end{equation}
The general irreducible decomposition of a $3$-tensor antisymmetric in its last two indices is given by two $1$-forms and another $3$-tensor. Therefore, the torsion is completely characterized by these variables,
\begin{align}
    \tensor{\widetilde{T}}{^\alpha_\beta_\sigma} \quad \longleftrightarrow\quad \{\tensor{\widetilde{T}}{^\sigma_\beta_\sigma},\Tilde{d}_{\mu}, ^{(3)}\tensor{\widetilde{T}}{^\alpha_\beta_\sigma}\}.
\end{align}
The general irreducible decomposition of the non-metricity tensor is given by
\begin{align*}
     \widetilde{M}_{\alpha\beta\sigma} = ^{(1)}\!\!\widetilde{M}_{\alpha\beta\sigma} + ^{(2)}\!\!\widetilde{M}_{\alpha\beta\sigma} + 
    ^{(3)}\!\!\widetilde{M}_{\alpha\beta\sigma} + 
    ^{(4)}\!\!\widetilde{M}_{\alpha\beta\sigma},
\end{align*}
where
\begin{align*}
    ^{(1)}\widetilde{M}_{\alpha\beta\sigma} &:=\frac{1}{4}g_{\beta\sigma}\tensor{\widetilde{M}}{_\alpha_\lambda^\lambda}, \\
    ^{(2)}\widetilde{M}_{\alpha\beta\sigma} &:= \frac{1}{36}(g_{\beta\sigma}\delta^{\mu}_{\alpha} - 2g_{\alpha\beta}\delta^{\mu}_{\sigma} - 2g_{\alpha\sigma}\delta^{\mu}_{\beta})(\tensor{\widehat{M}}{_\alpha_\lambda^\lambda} - 4 \tensor{\widetilde{M}}{^\lambda_\beta_\lambda}),\\
     ^{(3)}\widetilde{M}_{\alpha\beta\sigma} &:= \widetilde{M}_{(\alpha\beta\sigma)} - \frac{1}{18}(g_{\alpha\beta}\delta^{\mu}_{\sigma} + g_{\sigma\alpha}\delta^{\mu}_{\beta} + g_{\beta\sigma}\delta^{\mu}_{\alpha})(\tensor{\widetilde{M}}{_\mu_\lambda^\lambda} + 2 \tensor{\widetilde{M}}{^\lambda_\mu_\lambda}),\\
     ^{(4)}\widetilde{M}_{\alpha\beta\sigma}&:=  \widetilde{M}_{\alpha\beta\sigma}- ^{(1)}\!\!\widetilde{M}_{\alpha\beta\sigma} - ^{(2)}\!\!\widetilde{M}_{\alpha\beta\sigma} - ^{(3)}\!\!\widetilde{M}_{\alpha\beta\sigma}.
\end{align*}
The irreducible decomposition of a $3$-tensor symmetric in its last two indices is given by two $1$-forms $\tensor{\widetilde{M}}{_\alpha_\beta^\beta}$, $\tensor{\widetilde{M}}{^\alpha_\beta_\alpha}$, a completely symmetric and traceless $3$-tensor $^{(3)}\tensor{\widetilde{M}}{_\alpha_\beta_\sigma}$ and a $3$-tensor $^{(4)}\tensor{\widetilde{M}}{_\alpha_\beta_\sigma}$ symmetric in its last two indices, which is traceless and satisfies the cyclic identity. This means the non-metricity tensor irreducible components are,
\begin{align}
    \tensor{\widetilde{M}}{_\alpha_\beta_\sigma} \quad \longleftrightarrow \quad \{ \tensor{\widetilde{M}}{_\alpha_\beta^\beta}, \tensor{\widetilde{M}}{^\alpha_\beta_\alpha}, ^{(3)}\tensor{\widetilde{M}}{_\alpha_\beta_\sigma}, ^{(4)}\tensor{\widetilde{M}}{_\alpha_\beta_\sigma}\}.
\end{align}
In order to solve the equations, we will switch variables. We showed before that $Q$ could be expressed in terms of the torsion and the metricity tensors of two connections. In this case, one of the connections is that of Levi-Civita, hence it is torsionless and has vanishing non-metricity. The expression for $Q$ is therefore,
\begin{equation}
   Q_{\alpha\beta\gamma} =\frac{1}{2}(\widetilde{T}_{\alpha\beta\gamma} - \widetilde{T}_{\beta\gamma\alpha} + \widetilde{T}_{\gamma\alpha\beta} + \widetilde{M}_{\alpha\beta\gamma} - \widetilde{M}_{\beta\gamma\alpha} - \widetilde{M}_{\gamma\alpha\beta}) .
\end{equation}
We proceed by using the irreducible decompositions given in the Appendix (\ref{appendix_irreducible}) to solve the equations of motion. Let us start with the equation of motion associated with the variations of $Q$ in the HMS action, 
\begin{equation}
    \tensor{(E^{(2)}_{\mathtt{HMS}})}{_\alpha^\beta^\sigma}(g,Q)= 0  \quad \longrightarrow \quad \tensor{(E^{(2)}_{\mathtt{HMS}})}{_\alpha^\beta^\sigma}(g,\widetilde{T}, \widetilde{M}) = 0.
\end{equation}
Explicitly, the change from $Q$ to the irreducible components of $\widetilde{T}$ and $\widetilde{M}$ is the rather long expression, 
\begin{align*}
    \tensor{Q}{^\alpha_\beta_\sigma} &= \frac{1}{3}(\tensor{\widetilde{T}}{^\lambda^\alpha_\lambda}g_{\beta\sigma} - \tensor{\widetilde{T}}{^\lambda_\sigma_\lambda}\delta^{\alpha}_{\beta})  + \frac{1}{2}(^{(2)}\tensor{\widetilde{T}}{^\alpha_\beta_\sigma} - 2^{(3)}\tensor{\widetilde{T}}{_\beta_\sigma^\alpha} - ^{(3)}\!\tensor{\widetilde{M}}{^\alpha_\beta_\sigma} + 2^{(4)}\tensor{\widetilde{M}}{^\alpha_\beta_\sigma}) \\
    & \quad \frac{1}{36}\big( (2\tensor{\widetilde{M}}{^\lambda_\sigma_\lambda} - 5 \tensor{\widetilde{M}}{_\sigma_\lambda^\lambda})\delta^{\alpha}_{\beta} + (2\tensor{\widetilde{M}}{^\lambda_\beta_\lambda} - 5 \tensor{\widetilde{M}}{_\beta_\lambda^\lambda})\delta^{\alpha}_{\sigma} + (7\tensor{\widetilde{M}}{^\alpha_\lambda^\lambda} - 10 \tensor{\widetilde{M}}{^\lambda^\alpha_\lambda})g_{\beta\sigma}\big).
\end{align*}
Therefore, the equation of motion in terms of $(g,Q)$ goes from,
\begin{equation*}
    \tensor{(\mathcal{E}^{(2)}_{\mathtt{HMS}})}{_\alpha^\beta^\sigma}(g,Q) = (\delta^{\beta}_{\alpha}\tensor{Q}{^\sigma^\lambda_\lambda} + g^{\beta\sigma}\tensor{Q}{^\lambda_\lambda_\alpha} - \tensor{Q}{^\sigma_\alpha^\beta} - \tensor{Q}{^\beta^\sigma_\alpha} - \frac{1}{\gamma}(\volg)^{\xi\mu\beta\nu}(g_{\xi\alpha}\tensor{Q}{^\sigma_\nu_\mu} - \delta^{\sigma}_{\xi}Q_{\nu\mu\alpha})\big)\volg,
\end{equation*}
to the following expression in terms of $(g,\widetilde{T}, \widetilde{M})$,
\begin{align*}
    \tensor{(E^{(2)}_{\mathtt{HMS}})}{^\alpha^\beta^\sigma}(g,\widetilde{T}, \widetilde{M}) &= (\frac{2}{3}\tensor{\widetilde{T}}{^\lambda^\sigma_\lambda} + \frac{4}{9}\tensor{\widetilde{M}}{^\sigma_\lambda^\lambda} - \frac{7}{9}\tensor{\widetilde{M}}{^\lambda^\sigma_\lambda})g^{\alpha\beta} - (\frac{2}{3}\tensor{\widetilde{T}}{^\lambda^\alpha_\lambda} + \frac{2}{3}\tensor{\widetilde{M}}{^\alpha_\lambda^\lambda} + \frac{1}{9}\tensor{\widetilde{M}}{^\lambda^\alpha_\lambda})g^{\sigma\beta}\\
    & \quad - \frac{1}{18}(\tensor{\widetilde{M}}{^\beta_\lambda^\lambda} - 4\tensor{\widetilde{M}}{^\lambda^\beta_\lambda})g^{\sigma\alpha} - ^{(2)}\!\widetilde{T}^{\alpha\beta\sigma}  - ^{(3)}\!\widetilde{T}^{\beta\sigma\alpha} + ^{(3)}\!\widetilde{M}^{\alpha\beta\sigma} + ^{(4)}\!\widetilde{M}^{\alpha\beta\sigma}\\
    & \quad - \frac{1}{2\gamma} \bigg(\frac{2}{3}(2\tensor{\widetilde{T}}{^\lambda_\mu_\lambda} + \tensor{\widetilde{M}}{_\mu_\lambda^\lambda} - \tensor{\widetilde{M}}{^\lambda_\mu_\lambda})\delta^{\sigma}_{\nu}\delta^{\alpha}_{\rho} + ^{(2)}\!\tensor{\widetilde{T}}{^\sigma_\mu_\nu} + {}^{(3)}\tensor{\widetilde{T}}{^\sigma_\mu_\nu})\delta^{\alpha}_{\rho} \\
    & \quad + (^{(2)}\tensor{\widetilde{T}}{^\alpha_\mu_\nu} + ^{(3)}\!\tensor{\widetilde{T}}{^\alpha_\mu_\nu} - 2{}^{(4)} \tensor{\widetilde{M}}{_\nu_\mu^\alpha})\delta^{\sigma}_{\rho} \bigg )(\volg)^{\rho\beta\mu\nu} = 0.
\end{align*}
Because of the irreducible decompositions, we know we have to solve for the seven variables,
\begin{equation}\label{variables}
    \{\tensor{\widetilde{T}}{^\sigma_\beta_\sigma},\Tilde{d}_{\mu}, ^{(3)}\tensor{\widetilde{T}}{^\alpha_\beta_\sigma},\tensor{\widetilde{M}}{_\alpha_\beta^\beta}, \tensor{\widetilde{M}}{^\alpha_\beta_\alpha}, ^{(3)}\tensor{\widetilde{M}}{_\alpha_\beta_\sigma}, ^{(4)}\tensor{\widetilde{M}}{_\alpha_\beta_\sigma}\}.
\end{equation}
Let us first look at the traces of this equation and its contraction with the volume form,
\begin{align*}
     \tensor{(E^{(2)}_{\mathtt{HMS}})}{_\alpha^\alpha^\sigma}(g,\widetilde{T}, \widetilde{M}) = 0 \quad &\longrightarrow \quad 2 \tensor{\widetilde{T}}{^\lambda^\sigma_\lambda} + \frac{3}{2}\tensor{\widetilde{M}}{^\sigma_\beta^\beta} - 3 \tensor{\widetilde{M}}{^\alpha^\sigma_\alpha} - \frac{1}{8\gamma}\Tilde{d}^{\sigma} = 0,\\
     \tensor{(E^{(2)}_{\mathtt{HMS}})}{^\sigma^\beta_\beta}(g,\widetilde{T}, \widetilde{M}) = 0 \quad &\longrightarrow \quad 2 \tensor{\widetilde{T}}{^\lambda^\sigma_\lambda} + \frac{1}{2}\tensor{\widetilde{M}}{^\sigma_\beta^\beta} + \tensor{\widetilde{M}}{^\alpha^\sigma_\alpha} - \frac{1}{8\gamma}\Tilde{d}^{\sigma} = 0,\\
    (\volg)_{\alpha\beta\sigma\mu}\tensor{(E^{(2)}_{\mathtt{HMS}})}{^\alpha^\beta^\sigma}(g,\widetilde{T}, \widetilde{M}) = 0 \quad &\longrightarrow \quad 4 \tensor{\widetilde{T}}{^\lambda_\mu_\lambda} + 2 \tensor{\widetilde{M}}{_\mu_\beta^\beta} - 2 \tensor{\widetilde{M}}{^\alpha_\mu_\alpha} + \frac{\gamma}{4}\Tilde{d}_{\mu}  = 0.
\end{align*}
The solution to this system of linear equations is,
\begin{align*}
    \tensor{\widetilde{M}}{^\sigma_\beta^\beta} = - 8 U^{\sigma}, \quad\quad \tensor{\widetilde{M}}{^\alpha^\sigma_\alpha} = - 2U^{\sigma}, \quad\quad \tensor{\widetilde{T}}{^\lambda^\sigma_\lambda} = 3 U^{\sigma}, \quad\quad \Tilde{d}^{\sigma} = 0,
\end{align*}
for an arbitrary vector field $U^{\sigma}$. 
The other three equations are solved by substituting the four solutions into  $\tensor{(E^{(2)}_{\mathtt{HMS}})}{^\alpha^\beta^\sigma}(g,\widetilde{T}, \widetilde{M}) = 0$ which gives
\begin{align*}\label{firstsolve}
    \tensor{(E^{(2)}_{\mathtt{HMS}})}{^\alpha^\beta^\sigma}(g,\widetilde{T}, \widetilde{M}) &= {}^{(3)}\!\tensor{\widetilde{M}}{^\alpha^\beta^\sigma}+ ^{(4)}\!\tensor{\widetilde{M}}{^\alpha^\beta^\sigma}  - ^{(3)}\tensor{\widetilde{T}}{^\beta^\sigma^\alpha} \\
    & \quad - \frac{1}{2\gamma}\Big((\volg)^{\alpha\beta\mu\nu}{}^{(3)}\tensor{\widetilde{T}}{^\sigma_\mu_\nu} - (\volg)^{\sigma\beta\mu\nu}(^{(3)}\tensor{\widetilde{T}}{^\alpha_\mu_\nu} - 2{}^{(4)}\tensor{\widetilde{M}}{_\nu_\mu^\alpha})\Big).
\end{align*}
By completely symmetrising this expression, we are only left with the first term, thus,
\begin{equation*}
    {}^{(3)}\widetilde{M}^{(\alpha\beta\sigma)}=0.
\end{equation*}
Five out of the seven solutions have now been found. We plug this new solution again into the field equations, symmetrise the $(\alpha,\sigma)$ indices and use the cyclicity of ${}^{(4)}\tensor{\widetilde{M}}{}$ to get,
\begin{align*}
    {}^{(4)}\tensor{\widetilde{M}}{^\alpha^\beta^\sigma} + \frac{1}{\gamma}(\volg^{\alpha\beta\mu\nu}\delta^{\sigma}_{\kappa} + \volg^{\alpha\sigma\mu\nu}\delta^{\beta}_{\kappa})\;{}^{(4)}\tensor{\widetilde{M}}{_\mu_\nu^\kappa} = 0.
\end{align*}
By applying recursively this equation and using the properties of ${}^{(4)}\tensor{\widetilde{M}}{^\alpha^\beta^\sigma}$ (\ref{appendix}) we arrive at,
\begin{align}
    {}^{(4)}\tensor{\widetilde{M}}{^\alpha^\beta^\sigma} = -\frac{9}{\gamma^{2}}{}^{(4)}\tensor{\widetilde{M}}{^\alpha^\beta^\sigma} \quad \longrightarrow \quad {}^{(4)}\tensor{\widetilde{M}}{^\alpha^\beta^\sigma} = 0.
\end{align}
It only remains to solve for ${}^{(3)}\tensor{\widetilde{T}}{^\alpha^\beta^\sigma}$. We plug the solutions already found into the general equations, which results in,
\begin{align}
    {}^{(3)}\tensor{\widetilde{T}}{^\alpha^\beta^\sigma}  = \frac{1}{2\gamma}(\volg^{\alpha\beta\mu\nu}\delta^{\sigma}_{\kappa} - \volg^{\alpha\sigma\mu\nu}\delta^{\beta}_{\kappa}){}^{(3)}\tensor{\widetilde{T}}{_\mu_\nu^\kappa}.
\end{align}
Solving recursively leads to
\begin{align}
    {}^{(3)}\tensor{\widetilde{T}}{^\alpha^\beta^\sigma} =   - \frac{1}{\gamma^{2}}{}^{(3)}\tensor{\widetilde{T}}{_\alpha_\beta^\sigma} \quad \longrightarrow \quad {}^{(3)}\tensor{\widetilde{T}}{^\alpha^\beta^\sigma} = 0.
\end{align}
By solving the equation of motion $\tensor{(E^{(2)}_{\mathtt{HMS}})}{^\alpha^\beta^\sigma}(g,\widetilde{T}, \widetilde{M})=0$ we have computed the value of the independent variables (\ref{variables}),
\begin{equation*}
    \tensor{\widetilde{T}}{^\sigma_\beta_\sigma}\! = 3U^{\sigma}, \;\; \tensor{\widetilde{M}}{^\sigma_\beta^\beta} \!= - 8U^{\sigma}, \;\; \tensor{\widetilde{M}}{^\alpha^\sigma_\alpha} \!= -2 U^{\sigma}, \; \Tilde{d}_{\mu} \!=  ^{(3)}\!\!\tensor{\widetilde{T}}{^\alpha_\beta_\sigma} = ^{(3)}\!\!\tensor{\widetilde{M}}{_\alpha_\beta_\sigma} = ^{(4)}\!\!\tensor{\widetilde{M}}{_\alpha_\beta_\sigma} = 0,
\end{equation*}
where $U^\sigma$ is an arbitrary vector field in $M$.
These necessary conditions are also sufficient to completely solve the first equation of motion $\tensor{(E^{(1)}_{\mathtt{HMS}})}{^\alpha^\beta}(g,\widetilde{T}, \widetilde{M})=0$. These solutions imply for the original variables $(g,Q)$ that,
\begin{equation}
\tensor{Q}{^\alpha_\beta_\sigma}=\delta^\alpha_\sigma U_\beta.
\end{equation}
This result implies that, when substituted into the equation of motion for the variation of $g$, only the term from the Einstein-Hilbert part remains,
\begin{align}
    (E^{(1)}_{\mathtt{HMS}})^{\alpha\beta} = \big(E_{\mathtt{EH}}(g) + E_{\mathtt{CP}}(g,Q) - \frac{1}{\gamma}E_{\mathtt{CH}}(g,Q)\big)^{\alpha\beta} \overset{\mathtt{Sol}}{=} E^{\alpha\beta}_{\mathtt{EH}}(g) = 0.
\end{align}
Thus, $(g, Q)$ is a solution for the HMS action if and only if $\tensor{Q}{^\alpha_\beta_\sigma}=\delta^\alpha_\sigma U_\beta$ and $g$ satisfies the Einstein-Hilbert equations i.e.,
\begin{equation}
    \mathtt{Sol}(\mathbb{S}^{\mathtt{(m)}}_{\mathtt{HMS}}) = \{ (g_{\alpha\beta}, \delta^{\alpha}_{\gamma}U_{\beta}  | \; g \in \mathtt{Sol}(\mathbb{S}^{\mathtt{(m)}}_{\mathtt{EH}}) , \; U_{\beta} \ \in \Omega^{1}(M) \}.
\end{equation}
The presymplectic structure is given as usual by,
\begin{align*}
\Omega^{\mathtt{(m)}}_{\mathtt{HMS}}&= \delta \Big ( \int_{\Sigma} \imath^{*}\Theta^{\mathtt{(m)}}_{\mathtt{HMS}}- \int_{\partial \Sigma} \overline{\imath}^{*}\overline{\theta}^{\mathtt{(m)}}_{\mathtt{HMS}}\Big).
\end{align*}
Explicitly, the variation of the presymplectic  potentials is
\begin{align*}
    & \delta \Theta^{\mathtt{(m)}}_{\mathrm{HMS}} = \delta \Theta^{\mathtt{(m)}}_{\mathrm{EH}}+ \delta^{2} (\iota_{Z}\volg +  \iota_{q/\gamma}\volg) = \delta \Theta^{\mathtt{(m)}}_{\mathrm{EH}}, \\
    & \delta \overline{\theta}^{\mathtt{(m)}}_{\mathtt{HMS}}= \delta \overline{\theta}^{\mathtt{(m)}}_{\mathtt{EH}},
\end{align*}
where $Z^{\alpha} = \tensor{Q}{^\alpha^\beta_\beta} - \tensor{Q}{^\beta^\alpha_\beta}$ and  $\delta^{2} = 0$. Therefore the presymplectic form derived from the HMS action on the solution space is simply,
\begin{align*}
\Omega^{\mathtt{(m)}}_{\mathtt{HMS}}&= \Big ( \int_{\Sigma} \imath^{*}  \delta\Theta^{\mathtt{(m)}}_{\mathtt{EH}}- \int_{\partial \Sigma} \overline{\imath}^{*} \delta\overline{\theta}^{\mathtt{(m)}}_{\mathtt{EH}} \Big) =  \Omega^{\mathtt{(m)}}_{\mathtt{EH}}.
\end{align*}
The $\xi$-currents given by the CPS algorithm (\ref{cpsalgorithm}) are, 
\begin{align*}
    \mathcal{J}^{\mathtt{(m)}}_{\mathtt{HMS}} &= \iota_{\xi}L_{\mathtt{HMS}} - \iota_{X_{\xi}}\Theta^{\mathtt{(m)}}_{\mathtt{HMS}} \\
    &=\iota_{\xi}( L_{\mathtt{EH}}(g) + L_{\mathtt{CP}}(g,Q) -\frac{1}{\gamma} L_{\mathtt{CH}}(g,Q) ) - \iota_{\mathrm{X}_{\xi}}(\Theta^{\mathtt{(m)}}_{\mathtt{EH}}+ \delta (\iota_{Z + q/\gamma}\volg )) \\
    &= \iota_{\xi}\Big( L_{\mathtt{EH}}(g) + (\tensor{Q}{^\sigma_\sigma_\lambda}\tensor{Q}{^\lambda^\beta_\beta} - \tensor{Q}{^\sigma^\beta_\lambda}\tensor{Q}{^\lambda_\sigma_\beta})\volg + \mathrm{d}(\iota_{Z + q/\gamma}\volg) -\frac{1}{\gamma} (\volg)^{\alpha\beta\mu\nu}\tensor{Q}{_\alpha_\mu_\sigma}\tensor{Q}{^\sigma_\nu_\beta}\volg\Big) \\
    & \quad -\iota_{\mathrm{X}_{\xi}}\Big(\Theta^{\mathtt{(m)}}_{\mathtt{EH}}+ \delta (\iota_{Z + q/\gamma}\volg )\Big).
\end{align*}
Because they do not depend on any background objects, the fourth and sixth terms of the last equality are actually equal because on them $\mathcal{L}_{\xi} = \mathcal{L}_{\mathrm{X}_{\xi}}$, so using Cartan's formula we see that,
\begin{equation*}
\iota_{\xi}\mathrm{d}(\iota_{Z + q/\gamma}\volg) = \mathcal{L}_{\xi} \iota_{Z + q/\gamma}\volg - \mathrm{d}\iota_{\xi}\iota_{Z + q/\gamma}\volg  = \iota_{X_{\xi}}\delta(\iota_{Z + q/\gamma}\volg)- \mathrm{d}\iota_{\xi}\iota_{Z + q/\gamma}\volg,
\end{equation*}
where the last term goes to the boundary by Stokes' theorem.
Therefore, the current at the bulk is
\begin{align*}
    \mathcal{J}^{\mathtt{(m)}}_{\mathtt{HMS}}  &= \mathcal{J}^{\mathtt{(m)}}_{\mathtt{EH}} + \iota_{\xi}\big((\tensor{Q}{^\sigma_\sigma_\lambda}\tensor{Q}{^\lambda^\beta_\beta} - \tensor{Q}{^\sigma^\beta_\lambda}\tensor{Q}{^\lambda_\sigma_\beta})(\volg) -\frac{1}{\gamma} (\volg)^{\alpha\beta\mu\nu}\tensor{Q}{_\alpha_\mu_\sigma}\tensor{Q}{^\sigma_\nu_\beta}\volg \big).
\end{align*}
The current at the boundary is, 
\begin{align*}
    \overline{\mathfrak{j}}^{\mathtt{(m)}}_{\mathtt{HMS}} &= - \iota_{\xi}\overline{\ell}_{\mathtt{HMS}}(g,Q) - \iota_{\mathrm{X}_{\xi}}\overline{\theta}_{\mathtt{HMS}}(g,Q)\\
    &=\iota_{\xi}(\overline{\ell}_{\mathtt{EH}}(g) + \jmath^{*}\iota_{Z + q/\gamma}\volg) - \iota_{\mathrm{X}_{\xi}}(\overline{\theta}_{\mathtt{EH}}(g))\\
    &= \overline{\mathfrak{j}}^{\mathtt{(m)}}_{\mathtt{EH}} - \iota_{\xi}(\jmath^{*}\iota_{Z + q/\gamma}\volg).
\end{align*}
We can now compute the charges,
\begin{align*}
    \mathbb{Q}^{\mathtt{(m)}}_{\mathtt{HMS}} &= \int_{\Sigma}\imath^{*}\mathcal{J}^{\mathtt{(m)}}_{\mathtt{HMS}} - \int_{\partial \Sigma}\overline{\imath}^{*}\overline{\mathfrak{j}}^{\mathtt{(m)}}_{\mathtt{HMS}}\\
    &= \int_{\Sigma}\imath^{*}\mathcal{J}^{\mathtt{(m)}}_{\mathtt{EH}} + \iota_{\xi}\big(\tensor{Q}{^\sigma_\sigma_\lambda}\tensor{Q}{^\lambda^\beta_\beta} - \tensor{Q}{^\sigma^\beta_\lambda}\tensor{Q}{^\lambda_\sigma_\beta})(\volg)_{\alpha\tau\kappa\rho} -\frac{1}{\gamma} (\volg)^{\alpha\beta\mu\nu}\tensor{Q}{_\alpha_\mu_\sigma}\tensor{Q}{^\sigma_\nu_\beta}\volg \big) \\
    & \quad - \int_{\partial \Sigma}\overline{\imath}^{*} (\overline{\mathfrak{j}}^{\mathtt{(m)}}_{\mathtt{EH}} - \iota_{\xi}(\jmath^{*}\iota_{Z + q/\gamma}\volg)) - \iota_{\xi}(\jmath^{*}\iota_{Z + q/\gamma}\volg) \\
    &= \mathbb{Q}^{\mathtt{(m)}}_{\mathtt{EH}} +  \int_{\Sigma}\imath^{*}\iota_{\xi}\big(\tensor{Q}{^\sigma_\sigma_\lambda}\tensor{Q}{^\lambda^\beta_\beta} - \tensor{Q}{^\sigma^\beta_\lambda}\tensor{Q}{^\lambda_\sigma_\beta})(\volg) -\frac{1}{\gamma} (\volg)^{\alpha\beta\mu\nu}\tensor{Q}{_\alpha_\mu_\sigma}\tensor{Q}{^\sigma_\nu_\beta}\volg \big)\\
    &\quad + \int_{\partial \Sigma}\overline{\imath}^{*} ( \iota_{\xi}(\jmath^{*}\iota_{Z + q/\gamma}\volg)),
\end{align*}
which indeed vanish over solutions. 

\section{Tetrad Variables}
The tetrad HMS action is defined as the Palatini action in terms of the usual cotetrad $1$-forms $\e_{\alpha}^{I}$, the dual tetrad vector fields  $E^{\alpha}_{I}$, the internal Lorentz metric $\eta_{IJ}$, the generic $1$-form connection $\widetilde{\omega}^{IJ}$ and the Levi-Civita $1$-form connection $\omega^{IJ}$ associated with the metric $g_{\mu\nu} = \eta_{IJ}\e^{I}_{\mu}\e^{J}_{\nu}$. As we did in the Palatini case, the best strategy for solving the field equations is to split the spin-connection's internal indices $\widetilde{\omega}_{IJ}$ into its antisymmetric $\widehat{\omega}_{IJ}$ and symmetric $S_{IJ}$ parts, 
\begin{equation*}
    \widetilde{\omega}_{IJ} = \widehat{\omega}_{IJ} + S_{IJ},
\end{equation*}
so that the action principle is given in terms of these independent variables, 
\begin{equation}
    \mathbb{S}^{\mathtt{(t)}}_{\mathtt{HMS}}(\e, \widehat{\omega}, S) = \int_{\mathcal{M}}L_{\mathtt{HMS}}(\e, \widehat{\omega}, S) - \int_{\partial\Sigma} \overline{\ell}_{\mathtt{HMS}}(\e, \widehat{\omega}).
\end{equation}
The Lagrangian pair is given by the same change of variables used for the Palatini action (\ref{mapPalatini}), but from the metric HMS action,
\begin{align*}
    &L_{\mathtt{HMS}}(\e, \hat{\omega}, S) := L_{\mathtt{HMS}}\big(g = \mathrm{Tr}(\e \otimes \e), Q = \varphi(e, \widetilde{\omega})\big) \\
&\overline{\ell}_{\mathtt{HMS}}(\e, \hat{\omega}) := \overline{\ell}_{\mathtt{HMS}}\big(g = \mathrm{Tr}(\e \otimes \e), Q = \varphi(e, \widetilde{\omega})\big),
\end{align*}
where $g_{\alpha\beta} = \eta_{IJ}\e^{I}_{\alpha}\e^{J}_{\beta}$, $E^{\alpha}_{I}\e{I}_{\beta} = g^{\alpha}_{\beta}$ and $E^{\alpha}_{I}\e^{J}_{\alpha} = \eta^{J}_{I}$ (see Appendix \ref{appendix} for more details on this). 
Thus, the Lagrangian pair is,
\begin{align*}
    & L_{\mathtt{HMS}}(\e, \widehat{\omega}, S) = \frac{1}{2}H_{IJKL}\big(\widehat{F}^{IJ} - \frac{\Lambda}{6}\e^{I}\wedge\e^{J} + \tensor{S}{^I_M}\wedge\tensor{S}{^M^J}\big)\wedge \e^{K}\wedge \e^{L},\\
    & \overline{\ell}_{\mathtt{HMS}}(\e, \widehat{\omega}) = -\frac{1}{2}H_{IJKL}\big(2\N^{I}\mathrm{d}\N^{J} - (\jmath^{*}\widehat{\omega})^{IJ}\big) \wedge \overline{\e}^{K}\wedge\overline{\e}^{L} - \frac{1}{\gamma}\mathrm{d}\overline{\e}_{I}\wedge \overline{\e}^{I},
\end{align*}
where $\overline{\e}^{I}:=\jmath^{*}\e^{I}$, $\N^{I} = \nu^{\alpha}\e_{\alpha}^{I}$ and,
\begin{align*}
    &\tensor{H}{_I_J^K^L} := \tensor{\epsilon}{_I_J^K^L} + \frac{1}{\gamma}(\delta^{K}_{I}\delta^{L}_{J} - \delta^{L}_{I}\delta^{K}_{J}),\\
    & \widehat{F}_{IJ} := \mathrm{d}\widehat{\omega} + \widehat{\omega}_{IK}\wedge\tensor{\widehat{\omega}}{^K_J}.
\end{align*}
In fact, the Lagrangian at the bulk is just the Holst Lagrangian plus a dual term,
\begin{align*}
     &L_{\mathtt{HMS}}(\e, \widehat{\omega}, S) =  L_{\mathtt{Holst}}(\e, \widehat{\omega}) + L_{\mathtt{CH}}(\e, \widehat{\omega}, S),\\
     &L_{\mathtt{Holst}}(\e, \widehat{\omega}) := \frac{1}{2}H_{IJKL}\big(\widehat{F}^{IJ} - \frac{\Lambda}{6}\e^{I}\wedge\e^{J}\big)\wedge \e^{K}\wedge \e^{L},\\
     &L_{\mathtt{CH}}(\e, \widehat{\omega}, S) := \frac{1}{2}H_{IJKL}\tensor{S}{^I_M}\wedge\tensor{S}{^M^J} \wedge\e^{K} \wedge \e^{L},
\end{align*}
which means that the HMS action may be interpreted as the generalization of the Holst action for which the connection is constructed in a frame bundle with structure group $GL($4$, \mathbb{R})$. By varying the action one obtains the Euler forms and the presymplectic potentials,
\begin{align*}
    &\delta L_{\mathtt{HMS}}(\e, \widehat{\omega}, S) = (E^{(1)}_{\mathtt{HMS}})_{L}\wedge \delta \e^{L} + (E^{(2)}_{\mathtt{HMS}})_{KL}\wedge \delta \widehat{\omega}^{KL} + (E^{(3)}_{\mathtt{HMS}})_{JM}\wedge \delta S^{JM} + \mathrm{d}\Theta^{\mathtt{(t)}}_{\mathtt{HMS}},\\
    & \delta \overline{\ell}_{\mathtt{HMS}}(\e, \widehat{\omega}, S) - \jmath^{*}\Theta^{\mathtt{(t)}}_{\mathtt{HMS}} = (\overline{\mathfrak{b}}_{\mathtt{HMS}})_{I}\wedge \delta \overline{\e}^{I} - \mathrm{d}\overline{\theta}^{\mathtt{(t)}}_{\mathtt{HMS}},
\end{align*}
with
\begin{align*}
    & (E^{(1)}_{\mathtt{HMS}})_{L}: = H_{IJKL}\big(\widehat{F}^{IJ} + \tensor{S}{^I_M}\wedge S^{MJ} - \frac{\Lambda}{3}\e^{I}\wedge \e^{J}\big)\wedge \e^{K}, \\
    & (E^{(2)}_{\mathtt{HMS}})_{KL}: = -\frac{1}{2}H_{IJKL}\widehat{D}(\e^{I}\wedge \e^{J}) = H_{IJKL}\e^{I}\wedge \widehat{D}\e^{J}, \\
    & (E^{(3)}_{\mathtt{HMS}})_{JM} := \frac{1}{2}(H_{IKLJ}\delta^{R}_{M} + H_{IKLM}\delta^{R}_{J})\tensor{S}{_R^I}\wedge \e^{K}\wedge \e^{L},\\
    & (\overline{\mathfrak{b}}_{\mathtt{HMS}})_{I}:= \epsilon_{IJKL}(2\N^{K}\mathrm{d}\N^{L} - (\jmath^{*}\widehat{\omega})^{KL})\wedge \overline{\e}^{J} + 2 \epsilon_{MJKL}\N^{L}\overline{E}^{J}_{\alpha}\mathrm{d}\overline{e}^{K}_{\alpha} \wedge \overline{e}^{M}\N_{I} - \frac{2}{\gamma}\widehat{D}\overline{e}_{I},
\end{align*}
and $\widehat{D}\e_{I} := \mathrm{d}\e_{I}+\tensor{\widehat{\omega}}{_I^J}\wedge \e_{J}$. The presymplectic potentials are given by 
\begin{align*}
    & \Theta^{\mathtt{(t)}}_{\mathtt{HMS}}:=\! \frac{1}{2}H_{IJKL}\e^{I}\!\wedge \!\e^{J}\wedge\! \delta \widehat{\omega}^{KL} \!= \Theta^{\mathtt{(t)}}_{\mathtt{PT}} \!+ \frac{1}{\gamma}\e_{I}\wedge\e_{J}\wedge \delta \mathcal{C}^{IJ} \!+ \frac{1}{\gamma} \e_{I}\wedge \e_{J}\wedge \delta \omega^{IJ},\\
    & \overline{\theta}^{\mathtt{(t)}}_{\mathtt{HMS}}: = \epsilon_{IJKL}\overline{\e}\wedge \overline{\e}^{J}\wedge \N^{K} \wedge \delta \N^{L} = \overline{\theta}^{\mathtt{(t)}}_{\mathtt{PT}} - \frac{1}{\gamma}\overline{\e}^{I}\wedge \delta \overline{\e}_{I}.
\end{align*}
Here we have defined the contorsion $\mathcal{C}^{IJ}:= \widehat{\omega}^{IJ} - \omega^{IJ}$. 
To solve the equations of motion, let us start from the third equation and write $S_{IJ} = S_{MIJ}\e^{M}$ with $S_{MIJ} = S_{MJI}$, and substitute this to get
\begin{align*}
    (E^{(3)}_{\mathtt{HMS}})_{JM} = 0  \quad \longrightarrow \quad -2\tensor{S}{_I_{(M}^I}\tensor{\delta}{_{J)}^P} + 2\tensor{S}{_{(MJ)}^P} - \frac{1}{\gamma}\tensor{S}{_N_L_{(M}}\tensor{\epsilon}{_{J)}^N^L^P} = 0.
\end{align*}
We will solve this equation by writing $S_{IJK}$ in terms of its irreducible components,
\begin{equation}\label{Scomponent}
        S_{IJK} = ^{(1)}\!S_{IJK} + ^{(2)}\!S_{IJK} + ^{(3)}\!S_{IJK} + ^{(4)}\!S_{IJK},
\end{equation}
where
\begin{align*}
    & ^{(1)}S_{IJK} : = \frac{1}{4}\tensor{S}{_I_K^K},\\
    & ^{(2)}S_{IJK} := \frac{1}{36}(\eta_{JK}\delta^{L}_{I} - 2\eta_{IJ}\delta^{L}_{K} - 2\eta_{IK}\delta^{L}_{J})(\tensor{S}{_I_K^K} - 4 \tensor{S}{^K_I_K}),\\
    & ^{(3)}S_{IJK} := S_{(IJK)} - \frac{1}{18}(\eta_{IJ}\delta^{L}_{K} + \eta_{KI}\delta^{L}_{J} + \eta_{JK}\delta^{L}_{I})(\tensor{S}{_L_K^K} + 2 \tensor{S}{^K_L_K}),\\
    &^{(4)}S_{IJK}:=  S_{IJK} - ^{(1)}S_{IJK}  - ^{(2)}\!S_{IJK} - ^{(3)}\!S_{IJK}.
\end{align*}
All of these components are symmetric in the last two indices as the original tensor. Furthermore, $^{(3)}S_{(IJK)} = ^{(3)}S_{IJK}$ with $^{(3)}S_{IJK}$ and $^{(4)}S_{IJK}$ both being traceless. Moreover, $^{(4)}S_{IJK}$ satisfies the cyclic property $^{(4)}S_{IJK} + ^{(4)}S_{JKI} + ^{(4)}S_{KIJ} = 0$.\\
\noindent We solve (\ref{Scomponent}) by contracting with $\delta^{J}_{P}$, 
\begin{align*}
    \tensor{S}{_M_I^I} - 4 \tensor{S}{_I_M^I} = 0.
\end{align*}
Symmetrising (\ref{Scomponent}) in the indices $MJP$ results in, 
\begin{align*}
    S_{(JMP)} - \tensor{S}{_I_{(M}^I}\eta_{JP)} = 0 \quad \longleftrightarrow \quad  ^{(3)}\!S_{JMP} + \frac{1}{6}(\tensor{S}{_{(J|}_K^K} - 4 \tensor{S}{^K_{(J|}_K})\eta_{MP)} = 0,
\end{align*}
which implies that $^{(2)}\!S_{IJK} = 0$ and also $^{(3)}\!S_{IJK} = 0$. Plugging these solutions into the original equation gives,
\begin{align*}
    ^{(4)}\!S_{PJM} = \frac{1}{2\kappa}(\delta^{U}_{M}\tensor{\epsilon}{^T_J^V_P} + \delta^{U}_{J}\tensor{\epsilon}{^T_M^V_P})^{(4)}\!S_{TUV},
\end{align*}
where $\kappa \in \mathbb{R}$ is a constant. Solving the previous equation recursively gives us the fourth component of $S_{IJK}$, 
\begin{align*}
     ^{(4)}\!S_{PJM} = -\frac{3}{\kappa^2}^{(4)}\!S_{PJM} \quad \longrightarrow \quad  ^{(4)}\!S_{PJM} = 0.
\end{align*}
Thus, the general solution for $S_{IJ}$ is,
\begin{align*}
    S_{IJ} = \eta_{IJ}U_{K}\e^{K},
\end{align*}
for an arbitrary internal vector field $U^{K}$ in the space of fields. By plugging this solution into the remaining equations of motion, the dependence on $S_{IJ}$ is removed and the system becomes the usual Holst action in tetrad variables.

\noindent The presymplectic form defined by the action is,
\begin{align*}
    \Omega^{\mathtt{(t)}}_{\mathtt{HMS}} = \delta \big(\int_{\Sigma}\imath^{*}\Theta^{\mathtt{(t)}}_{\mathtt{HMS}} - \int_{\partial \Sigma} \overline{\imath}^{*}\overline{\theta}^{\mathtt{(t)}}_{\mathtt{HMS}} \big),
\end{align*}
where
\begin{align*}
    & \delta \Theta^{\mathtt{(t)}}_{\mathtt{HMS}}:= \delta \Theta^{\mathtt{(t)}}_{\mathtt{PT}} + \frac{2}{\gamma}\delta \e_{I}\wedge\e_{J}\wedge \delta \mathcal{C}^{IJ}  - \frac{1}{\gamma}\delta \mathrm{d}(\e_{I}\wedge \delta \e^{I}),\\
    &\delta \overline{\theta}^{\mathtt{(t)}}_{\mathtt{HMS}}:= \delta \overline{\theta}^{\mathtt{(t)}}_{\mathtt{PT}} - \frac{1}{\gamma}\delta \overline{\e}^{I}\wedge \delta \overline{\e}_{I}.
\end{align*}
The presymplectic form is thus given by,
\begin{align*}
    \Omega^{\mathtt{(t)}}_{\mathtt{HMS}} \!&= \!\int_{\Sigma}\imath^{*}\big(\delta \Theta^{\mathtt{(t)}}_{\mathtt{PT}} + \frac{2}{\gamma}\delta \e_{I}\wedge\e_{J}\wedge \delta \mathcal{C}^{IJ} - \frac{1}{\gamma}\delta \mathrm{d}(\delta \e_{I}\wedge \delta \e^{I}) \big) - \int_{\partial \Sigma} \overline{\imath}^{*}\big(\delta \overline{\theta}^{\mathtt{(t)}}_{\mathtt{PT}}- \frac{1}{\gamma}\delta \overline{\e}^{I}\wedge \delta \overline{\e}_{I}\big)\\
    &= \Omega^{\mathtt{(t)}}_{\mathtt{PT}} + \frac{1}{\gamma}\int_{\Sigma} \imath^{*}\big(2\delta \e_{I}\wedge \e_{J} \wedge \delta \mathcal{C}^{IJ} - \delta \mathrm{d}(\e_{I}\wedge \delta \e^{I}) \big) + \frac{1}{\gamma}\int_{\partial \Sigma} \overline{\imath}^{*}(\delta \overline{\e}^{I}\wedge \delta \overline{\e}_{I})\\
    &= \Omega^{\mathtt{(t)}}_{\mathtt{PT}} + \frac{2}{\gamma}\int_{\Sigma} \imath^{*}\big(\delta \e_{I}\wedge \e_{J} \wedge \delta \mathcal{C}^{IJ} \big).
\end{align*}

Since the HMS and Palatini Lagrangians are not cohomologically equivalent, their presymplectic structures are not guaranteed to be equivalent either, and this can be seen from the extra term involving the contorsion and the parameter $\gamma$ in the previous equation. It is only \emph{on-shell}, when $\widehat{\omega} = \omega$, that we have equivalence between the space of solutions of HMS and Palatini actions.
Before computing the currents, it is possible to rewrite the HMS Lagrangian pair by splitting the connection in terms of the contorsion, 
\begin{align*}
    \widetilde{\omega}^{IJ} = \omega^{IJ} + \mathcal{C}^{IJ} + S^{IJ}.
\end{align*}
The Lagrangian pair is then expressed as, 
\begin{align*}
    &L_{\mathtt{HMS}}(\e, \widetilde{\omega}) = L_{\mathtt{PT}}(\e, \widetilde{\omega}) + \frac{1}{\gamma}\Big( \e^{I}\wedge \e^{J}\wedge(\mathcal{C}_{IM}\wedge \tensor{\mathcal{C}}{^M_J} + S_{IM}\wedge \tensor{S}{^M_J} )+ \mathrm{d}(\e^{I}\wedge \e^{J}\wedge \mathcal{C}_{IJ})\Big),\\
    &\overline{\ell}_{\mathtt{HMS}}(\e, \widetilde{\omega}) = \overline{\ell}_{\mathtt{PT}}(\e, \widetilde{\omega}) + \frac{1}{\gamma}\overline{\e}^{I}\wedge \overline{\e}^{J}\wedge \overline{\mathcal{C}}_{IJ}.
\end{align*}
where $\overline{\mathcal{C}}:= \jmath^{*}\mathcal{C}$.
\noindent We can use this decomposition to easily compute the current defined by the HMS action in terms of the Palatini current in the bulk and an additional current  $\mathcal{J}_{B}$, 
\begin{align*}
    \mathcal{J}^{\mathtt{(t)}}_{\mathtt{HMS}} &= \iota_{\xi}L_{\mathtt{HMS}}(\e, \widehat{\omega}, S) - \iota_{\mathrm{X}_{\xi}} \Theta^{\mathtt{(t)}}_{\mathtt{HMS}}\\
    & = \iota_{\xi}\Big(L_{\mathtt{PT}}(\e, \widetilde{\omega}) + \frac{1}{\gamma}\big( \e^{I}\wedge \e^{J}\wedge(\mathcal{C}_{IM}\wedge \tensor{\mathcal{C}}{^M_J} + S_{IM}\wedge \tensor{S}{^M_J} )+ \mathrm{d}(\e^{I}\wedge \e^{J}\wedge \mathcal{C}_{IJ})\big)\Big)\\
    & \quad - \iota_{\mathrm{X}_{\xi}}\Big(\Theta^{\mathtt{(t)}}_{\mathtt{PT}} \!+ \frac{1}{\gamma}\e_{I}\wedge\e_{J}\wedge \delta \mathcal{C}^{IJ} \!+ \frac{1}{\gamma} \e_{I}\wedge \e_{J}\wedge \delta \omega^{IJ}\Big) \\
    & = \mathcal{J}^{\mathtt{(t)}}_{\mathtt{PT}} + \frac{1}{\gamma} \mathcal{J}^{\mathtt{(t)}}_{\mathtt{B}}
\end{align*}
where
\begin{align*}
    & \mathcal{J}^{\mathtt{(t)}}_{\mathtt{B}} :=  \imath_{\xi}(\e^{I}\wedge \e^{J}\wedge(\mathcal{C}_{IM}\wedge \tensor{\mathcal{C}}{^M_J} + S_{IM}\wedge \tensor{S}{^M_J}) + \imath_{\xi}\mathrm{d}(\e^{I}\wedge \e^{J}\wedge \mathcal{C}_{IJ}) \\
    & \quad\quad\quad - \iota_{\mathrm{X}_{\xi}}(\e_{I}\wedge\e_{J}\wedge (\delta \mathcal{C}^{IJ} + \delta \omega^{IJ})).
\end{align*}
Similarly, the current at the boundary contains two parts, one associated with Palatini at the boundary and another one $\overline{j}^{\mathtt{(t)}}_{B}$ proportional to the parameter $\gamma$, 
\begin{align*}
\overline{\mathfrak{j}}^{\mathtt{(t)}}_{\mathtt{HMS}} &= - \iota_{\xi}\overline{\ell}_{\mathtt{HMS}}(\e,\widetilde{\omega}) - \iota_{\mathtt{X}_{\xi}}\overline{\theta}^{\mathtt{(t)}}_{\mathtt{HMS}} \\
&= \iota_{\xi}\big(\overline{\ell}_{\mathtt{PT}}(\e, \widetilde{\omega}) + \frac{1}{\gamma}\overline{\e}^{I}\wedge \overline{\e}^{J}\wedge \overline{C}_{IJ}\big) - \iota_{\mathtt{X}_{\xi}}\big(\overline{\theta}^{\mathtt{(t)}}_{\mathtt{PT}} - \frac{1}{\gamma}\overline{e}^{I}\wedge \delta \overline{\e}_{I}\big) \\
&= \overline{\mathfrak{j}}^{\mathtt{(t)}}_{\mathtt{PT}} + \frac{1}{\gamma}\;\overline{\mathfrak{j}}^{\mathtt{(t)}}_{\mathtt{B}},
\end{align*}
where 
\begin{align*}
\overline{\mathfrak{j}}^{\mathtt{(t)}}_{\mathtt{B}} &:= \imath_{\xi}(\overline{\e}^{I}\wedge \overline{\e}^{J}\wedge \overline{C}_{IJ}) - \imath_{\mathrm{X}_{\xi}}(\overline{\e}^{I}\wedge \delta \overline{\e}_{I}).
\end{align*}
The charges are computed from the currents by following the CPS algorithm, 
\begin{align*}
    \mathbb{Q}^{\mathtt{(t)}}_{\mathtt{HMS}} &= \int_{\Sigma}\imath^{*}\mathcal{J}^{\mathtt{(t)}}_{\mathtt{HMS}} - \int_{\partial \Sigma}\overline{\imath}^{*}\overline{\mathfrak{j}}^{\mathtt{(t)}}_{\mathtt{HMS}} \\
    &= \int_{\Sigma}\imath^{*}\big(\mathcal{J}^{\mathtt{(t)}}_{\mathtt{PT}}+ \frac{1}{\gamma} \mathcal{J}^{\mathtt{(t)}}_{\mathtt{B}} \big) - \int_{\partial \Sigma}\overline{\imath}^{*} \big( \overline{\mathfrak{j}}^{\mathtt{(t)}}_{\mathtt{PT}}+ \frac{1}{\gamma}\;\overline{\mathfrak{j}}^{\mathtt{(t)}}_{\mathtt{B}} \big)\\
    &= \mathbb{Q}^{\mathtt{(t)}}_{\mathtt{PT}}+ \frac{1}{\gamma}\mathbb{Q}^{\mathtt{(t)}}_{\mathtt{B}},
\end{align*}
which vanish identically over solutions. 

%% file: equivalence.tex
\chapter{Symplectic Equivalence}\label{equivalence}

The metric and the tetrad formalisms of the Einstein-Hilbert action in the absence of boundaries are known to be equivalent \emph{on-shell} \cite{frauendiener1992symplectic}. Nevertheless, there seemed to be discrepancies about their equivalence in the presence of boundaries \cite{OliveriBoundaryeffects, de2018gauge, freidel2020edge}. These discrepancies can be attributed to the choice of bulk and boundary Lagrangians, which describe different dynamics and thus the non-equivalence is expected. When considering appropriate Lagrangian pairs such that their actions represent the same dynamics, it is possible to show that the solution spaces of both the metric and tetrad formulations, their presymplectic potentials and presymplectic structures are equivalent. This will be the first objective of this chapter.

The \emph{on-shell} equivalence of Palatini theories with the Einstein-Hilbert action was studied in the absence of boundaries, torsion and assuming metricity for the tetrad and metric formalisms in \cite{dadhich2012equivalence,frauendiener1992symplectic}. Traditionally, the Palatini formalism in metric and tetrad variables considered the Dirichlet functional space, which removed surface terms. In \cite{obukhov1987palatini}, Obukhov introduced the appropriate surface terms for Palatini gravity, a necessary step towards the comparison of the metric and tetrad formalisms. In chapter \ref{Palatiniaction}, we computed the $\mathtt{CPS(\mathbb{S}}$ associated with the Palatini action in a manifold with boundaries, where the connection had torsion and non-metricity. The second goal of this chapter will be to study this Palatini action's cohomological equivalence \emph{off-shell} and \emph{on-shell} with the Einstein-Hilbert action.

Lastly, we will study the equivalence of the HMS action (\ref{HMSaction}) and the Palatini action (\ref{Palatiniactionmetric}). The equivalence between the metric and tetrad formalisms of the Holst action has been under debate in previous studies, which asserted that charges associated with each formalism were not equivalent \cite{SpezialeNote}. This apparent disparity is \emph{removed} in these papers by \emph{dressing} the presymplectic potentials. Their underlying idea is presumably based on the \emph{dressing method} described in \cite{franccois2021bundle}, a cohomologically systematic method to fully or partially remove the gauge dependence of a field theory defined on a manifold without boundaries. The dressing field is interpreted as a change of variables in the action, which in turn modifies the presymplectic potentials. Although following a different approach, the dressing method appears to be equivalent to the relative bicomplex if one does not consider boundaries. 

In this chapter, we will study the equivalence between three actions expressed in both the metric and tetrad formalisms. To do so, we shall start by proving the equivalence of the metric and tetrad formalism of the Einstein-Hilbert action studied in chapter \ref{EHaction}. Then, we will continue to prove in section \ref{PalatiniEH} the equivalence between the Palatini action and the Einstein-Hilbert action, and also the independence of the presymplectic structures from working with metrics or tetrad variables. Lastly, we will study the equivalence of the HMS action and the Palatini action in section \ref{HMSPalatini} for both the metric and tetrad formalisms.  

\section{Between Metric and Tetradic Einstein-Hilbert}

Let the map $\widetilde{\Phi}: M \times \mathcal{F}^{\mathtt{(t)}} \xrightarrow{}  M \times \mathcal{F}^{\mathtt{(m)}}$ be such that $\widetilde{\Phi} = (\text{Id}, \Phi)$ where  $\Phi: \mathcal{F}^{\mathtt{(t)}} \xrightarrow{}  \mathcal{F}^{\mathtt{(m)}}$ is the map from the space of fields for the tetrad formulation to the one for the metric formulation. When working with relative pairs, this map is $\underline{\widetilde{\Phi}}: (M, \partial M) \times \mathcal{F}^{\mathtt{(t)}} \xrightarrow{} (M, \partial M) \times \mathcal{F}^{\mathtt{(m)}}$. This map is surjective but not injective. In fact, 
\begin{align*}
    \Phi(\e) = \Phi(\Tilde{\e}) \quad\quad \Longleftrightarrow \quad\quad \Tilde{\e} = \tensor{\Lambda}{_I^J}\e_{J}, \quad \tensor{\Lambda}{_I^J}\in SO(1,3).
\end{align*}
It allows us to go from an action written in terms of tetrads to one in terms of metrics, 
\begin{equation*}
    \mathbb{S}^{\mathtt{(t)}}_{\mathtt{EH}} = \mathbb{S}^{\mathtt{(m)}}_{\mathtt{EH}}  \circ \Phi,
\end{equation*}
from which it is possible to compute its variation, 
\begin{equation*}
    \delta \mathbb{S}^{\mathtt{(t)}} = \delta(\mathbb{S}^{\mathtt{(m)}}\circ \Phi) = \delta\mathbb{S}^{\mathtt{(m)}}\circ \delta \Phi.
\end{equation*}
Since $\delta\Phi$ is surjective, the relation between the respective solution spaces is, 
\begin{align*}
 \e\in\mathtt{Sol(\mathbb{S}^{\mathtt{(t)}})} &\Longrightarrow \Phi(\e) \in \mathtt{Sol\big(\mathbb{S}^{\mathtt{(m)}}\big)},\\
 g \in \mathtt{Sol\big(\mathbb{S}^{\mathtt{(m)}}\big)} &\Longrightarrow \Phi^{-1}(\{g\}) \subset \mathtt{Sol\big(\mathbb{S}^{\mathtt{(t)}}\big)}.
\end{align*}
In this section, we will prove that the presymplectic structures for both the tetrad and metric approaches,
\begin{align*}
    &\Theta^{\mathtt{(m)}}_{\mathtt{EH}} = \iota_{W}\volg,  && \Theta^{\mathtt{(t)}}_{\mathtt{EH}} = \frac{1}{2}\epsilon_{IJKL}\e^{I}\wedge \e^{J} \wedge \delta \omega^{KL}, \\
    & \overline{\theta}^{\mathtt{(m)}}_{\mathtt{EH}} = \iota_{\overline{V}}\volgg, &&
    \overline{\theta}^{\mathtt{(t)}}_{\mathtt{EH}} = \epsilon_{IJKL}\overline{\e}^{K}\wedge \overline{\e}^{L}\wedge \N \wedge \delta \N^{J},\\
    &  W^{\alpha} := (g^{\lambda\beta}g^{\alpha\kappa} - g^{\lambda\alpha}g^{\kappa\beta})\nabla_{\lambda}\delta g_{\kappa\beta}, && \overline{V}^{\overline{\alpha}}:=- \jmath^{\beta}_{\overline{\beta}}(\overline{g}^{\overline{\alpha}\overline{\beta}}\nu^{\lambda}\delta g_{\lambda\beta}),
\end{align*}
are, in fact, equivalent. Let us start by attempting to prove  the following equality in the bulk,
\begin{equation*}
    \Theta_{\mathtt{EH}}^{\mathtt{(t)}} \overset{?}{=} \widetilde{\Phi}^{*}\Theta_{\mathtt{EH}}^{\mathtt{(m)}}.
\end{equation*}
First, we notice that instead of working directly with the presymplectic potentials, it is easier to work with their duals, 
\begin{align*}
    &(\star_{\eta} \Theta_{\mathtt{EH}}^{\mathtt{(t)}})^{\sigma} = 2 \e^{\sigma}_{K}\e^{\mu}_{L}\delta \omega^{KL}_{\mu},\\
    & (\star_{g} \Theta^{\mathtt{(m)}}_{\mathtt{EH}})^{\mu} = (\star_{g}\iota_{W}\volg)^{\mu} = W^{\mu}.
\end{align*}
The variation of $\omega^{IJ}_{\mu} := \e^{I}_{\beta}\nabla_{\mu}\e^{I\beta}$  is given by, 
\begin{align*}
    & \delta \omega^{IJ}_{\mu} = - \delta g_{\alpha\beta}\e^{K\alpha}\nabla_{\mu}\e^{L\beta} + \delta \e^{K}_{\alpha}\nabla_{\mu}\e^{L\alpha} + \e^{K\beta}\nabla_{\mu}\delta \e^{L}_{\beta} - \e^{K\beta}\e^{L}_{\kappa}\delta \Gamma^{\kappa}_{\mu\beta},
\end{align*}
explicitly, using the expression for $\delta \Gamma^{\kappa}_{\mu\beta}$ given in appendix \ref{appendix_variations}, 
\begin{align*}
    (\star_{\eta} \Theta^{\mathtt{(t)}}_{\mathtt{EH}})^{\sigma} &=  2 \delta(\e^{K}_{\alpha}\e_{K\beta})g^{\sigma\alpha}e^{L\beta}\nabla_{\mu}e^{\mu}_{L} + 2 \e_{\alpha}^{K}\delta \e^{\sigma}_{K}\e^{L\alpha}\nabla_{\mu}\e^{\mu}_{L} + \nabla_{\mu}(2\e^{\mu}_{L}g^{\sigma\beta}\delta \e^{L}_{\beta}) \\
    & \quad - 2 g^{\sigma\beta} \delta \e^{L}_{\beta} \nabla_{\mu}\e^{\mu}_{L} - g^{\sigma\beta}g^{\mu\alpha}(\nabla_{\mu} \delta g_{\alpha\beta } + \nabla_{\beta}\delta  g_{\mu\alpha} - \nabla_{\alpha}\delta g_{\mu\beta}) \\
    &= 2(g^{\sigma\alpha}\e^{L\beta}\e_{\beta K}\delta \e^{K}_{\alpha} + g^{\sigma\alpha}e^{L\beta}\e^{K}_{\alpha}\delta\e_{\beta K} + \delta \e^{\sigma L} - g^{\sigma\beta}\delta \e^{L}_{\beta}) \nabla_{\mu}\e^{\mu}_{L} \\
    & \quad + \nabla_{\mu}(2 \e^{\mu}_{L}g^{\sigma\beta}\delta e^{L}_{\beta}) - \nabla^{\sigma}\delta g \\
    &= W^{\sigma} + \nabla_{\mu}(g^{\sigma\beta}\e^{\mu}_{L}\delta \e^{L}_{\beta} - g^{\alpha\mu}\e^{L \sigma}\delta \e_{L \alpha})\\
    &= (\star_{g} \Theta^{\mathtt{(m)}}_{\mathtt{EH}})^{\sigma} + \nabla_{\mu}(\mathcal{U}^{\mu\sigma}),
\end{align*}
where $\mathcal{U}^{\sigma} := (g^{\sigma\beta} \e^{\mu}_{L} - g^{\alpha\mu}\e^{\sigma}_{L})\delta \e_{L\alpha}$. Reversing the Hodge-star operation gives us the relation between the presymplectic potentials at the bulk from the metric and tetrad formalisms, using the transition function $\widetilde{\Phi}$ between their field spaces, 
\begin{align}
    \Theta_{\mathtt{EH}}^{\mathtt{(t)}} = \widetilde{\Phi}^{*}\Theta_{\mathtt{EH}}^{\mathtt{(m)}} + \mathrm{d}(\star_{g}\;\mathcal{U}).
\end{align}
Analogously for the boundary, the presymplectic potential of the tetrads transforms to metric variables as 
\begin{align*}
    (\overline{\theta}_{\mathtt{EH}}^{\mathtt{(t)}})_{\overline{\alpha}\overline{\beta}} &= \tensor{\epsilon}{^I^J_K_L}(\overline{\e}^{K}\wedge \overline{\e}^{L})_{\overline{\alpha}\overline{\beta}} \N_{I}\delta \N_{J}\\
    &=  \tensor{(\volg)}{^\kappa^\tau_\gamma_\delta}\e^{I}_{\kappa}e^{J}_{\tau}E^{\gamma}_{K}E^{\delta}_{L}(\overline{\e}^{K}_{\alpha}\overline{\e}^{L}_{\beta} - \overline{\e}^{K}_{\beta}\overline{\e}^{L}_{\alpha})\nu_{\sigma}E^{\sigma}_{I}N_{R}\overline{E}^{\rho}\delta\overline{\e}^{R}_{\rho} \\
    &= - 2N_{R}\jmath^{\alpha}_{\overline{\alpha}}\jmath^{\beta}_{\overline{\beta}}\jmath^{\sigma}_{\overline{\sigma}}(\iota_{\nu}\volg)_{\sigma \alpha\beta}\overline{g}^{\overline{\sigma\mu}}\delta \overline{e}^{R}_{\overline{\mu}} \\
    &= (\iota_{\overline{Y}}\volgg)_{\overline{\alpha}\overline{\beta}},
\end{align*}
where $\N_{I} := \nu_{\alpha}E^{\alpha}_{I}$ and $\overline{Y}^{\overline{\mu}} := - 2N_{R}\overline{g}^{\overline{\mu}\overline{\alpha}}\delta \overline{\e}^{R}_{\overline{\alpha}}$. Therefore, the presymplectic potentials are
\begin{align*}
    (\overline{\theta}^{\mathtt{(t)}}_{\mathtt{EH}})_{\overline{\alpha}\overline{\beta}}&= (\iota_{\overline{Y}}\volgg)_{\overline{\alpha}\overline{\beta}},
    & \overline{Y}^{\overline{\alpha}} &:= - 2\nu_{\sigma}E^{\sigma}_{R}\overline{g}^{\overline{\alpha}\overline{\mu}}\delta \overline{\e}^{R}_{\overline{\mu}},\\
    (\overline{\theta}^{\mathtt{(m)}}_{\mathtt{EH}})_{\overline{\alpha}\overline{\beta}}  &= (\iota_{\overline{V}}\volgg)_{\overline{\alpha}\overline{\beta}},
     & \overline{V}^{\overline{\alpha}} &:= - \jmath^{\beta}_{\overline{\beta}}(\overline{g}^{\overline{\alpha}\overline{\beta}}\nu^{\lambda}\delta g_{\lambda\beta}).
\end{align*}
By computing their difference now that they are in the space of fields, we can check their equivalence, 
\begin{align*}
    (\overline{\theta}^{\mathtt{(t)}}_{\mathtt{EH}})_{\overline{\alpha}\overline{\beta}} - (\overline{\theta}^{\mathtt{(m)}}_{\mathtt{EH}})_{\overline{\alpha}\overline{\beta}}&= (\iota_{\overline{Y}}\volgg)_{\overline{\alpha}\overline{\beta}} - (\iota_{\overline{V}}\volgg)_{\overline{\alpha}\overline{\beta}}  = (\iota_{\overline{Y}-\overline{V}}\volgg)_{\overline{\alpha}\overline{\beta}}.
\end{align*}
To this end we compute, 
\begin{align*}
    \overline{V}^{\overline{\alpha}}   - \overline{Y}^{\overline{\alpha}} &= - \jmath^{\beta}_{\overline{\beta}}\overline{g}^{\overline{\alpha}\overline{\beta}}\nu^{\lambda}\delta g_{\lambda\beta} + 2\nu_{\sigma}E^{\sigma}_{R}\overline{g}^{\overline{\alpha}\overline{\mu}}\delta \overline{\e}^{R}_{\overline{\mu}}\\
    &= - \jmath^{\beta}_{\overline{\beta}}\overline{g}^{\overline{\alpha}\overline{\beta}}\nu^{\lambda}\delta (\eta_{IJ}\e^{I}_{\lambda}\e^{J}_{\beta}) + 2\nu_{\sigma}E^{\sigma}_{R}\overline{g}^{\overline{\alpha}\overline{\mu}}\delta \overline{\e}^{R}_{\overline{\mu}}\\
    &= -\jmath^{\beta}_{\overline{\beta}}\overline{g}^{\overline{\alpha}\overline{\beta}}\nu^{\lambda}\eta_{IJ} \e^{J}_{\beta} \delta \e^{I}_{\lambda} - \overline{g}^{\overline{\alpha}\overline{\beta}}\nu_{\lambda}E^{\lambda}_{J}\delta \overline{\e}^{J}_{\overline{\beta}} + 2 \nu_{\sigma}E^{\sigma}_{R}\overline{g}^{\overline{\alpha}\overline{\mu}} \delta \overline{\e}^{R}_{\overline{\mu}}\\ 
    &= -\overline{g}^{\overline{\alpha}\overline{\beta}}\nu^{\lambda}\eta_{IJ} \overline{\e}^{J}_{\overline{\beta}} \delta \e^{I}_{\lambda} + \overline{g}^{\overline{\alpha}\overline{\beta}}\nu_{\lambda}E^{\lambda}_{J}\delta \overline{\e}^{J}_{\overline{\beta}} \\
    &=-\nu^{\lambda}E^{\overline{\alpha}}_{J}\delta \e^{I}_{\lambda} + \overline{g}^{\overline{\alpha}\overline{\beta}}N_{I}\delta \e^{I}_{\beta}\\
   &= (\overline{g}^{\overline{\alpha}\overline{\beta}} \nu^{\sigma}\e^{I}_{\sigma}  - \nu^{\beta}\overline{E}^{\overline{\alpha} I}) \delta \e_{I\beta}\\
    &= \jmath^{*}(\nu^{\sigma}\mathcal{U}^{\alpha}_{\sigma}),
\end{align*}
where $\N^{I}:= \iota_{\nu}\e^{I}$ and $E^{\sigma}_{R}:= \eta_{RL}g^{\sigma\alpha}\e^{L}_{\alpha}$. The presymplectic potentials at the boundary in both cases are thus related by
\begin{align}
    \overline{\theta}^{\mathtt{(t)}}_{\mathtt{EH}} =  \widetilde{\Phi}^{*}\overline{\theta}^{\mathtt{(m)}}_{\mathtt{EH}} + \jmath^{*}(\star_{g}\mathcal{U}).
\end{align}
Expressed in terms of the relative bicomplex, the presymplectic potentials are given by pairs and the equivalence between metric and tetrad formalisms of such structures can be concisely expressed in the form, 
\begin{equation}\label{tetradmetricEHpotentials}
    \boxed{(\Theta_{\mathtt{EH}}^{\mathtt{(t)}},  \overline{\theta}^{\mathtt{(t)}}_{\mathtt{EH}}) = \underline{\widetilde{\Phi}}^{*}\big( \Theta_{\mathtt{EH}}^{\mathtt{(m)}}, \overline{\theta}^{\mathtt{(m)}}_{\mathtt{EH}}\big)  + \underline{\mathrm{d}}\big(\star_{g}\mathcal{U}, 0\big)}
\end{equation}
Once the presymplectic potentials have been compared, we will look at the presymplectic structures constructed from them. By performing the same change of variables as before with the help of the transition function $\widetilde{\Phi}$, we can check that in fact the presymplectic structure transforms from the tetrad to the metric formalism explicitly as
\begin{align*}
    \Omega^{\mathtt{(t)}}_{\mathtt{EH}}&=  \delta \int_{(\Sigma, \partial \Sigma)} \underline{\imath}^{*}(\Theta^{\mathtt{(t)}}_{\mathtt{EH}} ,\overline{\theta}^{\mathtt{(t)}}_{\mathtt{EH}})= \delta \int_{(\Sigma, \partial \Sigma)} \underline{\imath}^{*}\Big( \underline{\widetilde{\Phi}}\big( \Theta_{\mathtt{EH}}^{\mathtt{(m)}}, \overline{\theta}^{\mathtt{(m)}}_{\mathtt{EH}}\big)  + \underline{\mathrm{d}}\big(\star_{g}\mathcal{U}, 0\big)\Big)\\
    &= \Phi^{*} \delta  \int_{(\Sigma, \partial \Sigma)} \underline{\imath}^{*}(\Theta^{\mathtt{(m)}}_{\mathtt{EH}} ,\overline{\theta}^{\mathtt{(m)}}_{\mathtt{EH}}) = \Phi^{*}  \Omega^{\mathtt{(m)}}_{\mathtt{EH}},
\end{align*}
where we have used the definition of (\ref{underlineimath}), (\ref{tetradmetricEHpotentials}), the relative version of Stokes' theorem and the fact that the pullback of $\Phi$ commutes with the $\delta$ derivative. This leads to the first equivalence relation between the presymplectic form of the tetrad and metric formulations of the Einstein-Hilbert action in the presence of boundaries, 
\begin{equation}\label{tetradmetricEHsymplectic}
    \boxed{\Omega^{\mathtt{(t)}}_{\mathtt{EH}} = \Phi^{*}\Omega^{\mathtt{(m)}}_{\mathtt{EH}}}
\end{equation}
Due to the invariance of the tetrads under the action of the structure group $SO(1,3)$, the presymplectic structure of the tetrad formalism $\Omega^{\mathtt{(t)}}_{\mathtt{EH}}$ has more degenerate directions than the one from the metric formalism $\Omega^{\mathtt{(m)}}_{\mathtt{EH}}$. Hence, the previous formula shows that both presymplectic structures are equivalent up to this gauge freedom.  

\noindent In the bulk, the associated charges of both formalisms are related as
\begin{align*}
    \mathcal{J}^{\mathtt{(t)}}_{\mathtt{EH}} &= \imath_{\xi}L^{\mathtt{(t)}}_{\mathtt{EH}} - \imath_{\mathrm{X}_{\xi}}\Theta^{\mathtt{(t)}}_{\mathtt{EH}} = \imath_{\xi} \widetilde{\Phi}^{*}L^{\mathtt{(m)}}_{\mathtt{EH}} - \imath_{\mathrm{X}_{\xi}}\big( \widetilde{\Phi}^{*}\Theta^{\mathtt{(m)}}_{\mathtt{EH}} + \mathrm{d}(\star_{g}\;\mathcal{U})\big)\\ 
    &= \widetilde{\Phi}^{*}\mathcal{J}^{\mathtt{(m)}}_{\mathtt{EH}} -   \imath_{\mathrm{X}_{\xi}}\mathrm{d}(\star_{g}\;\mathcal{U}),
\end{align*}
and at the boundary, 
\begin{align*}
    \overline{j}^{\mathtt{(t)}}_{\mathtt{EH}} &=  -\imath_{\xi}\overline{\ell}^{\mathtt{(t)}}_{\mathtt{EH}} - \imath_{\mathrm{X}_{\overline{\xi}}}\overline{\theta}^{\mathtt{(t)}}_{\mathtt{EH}} = -\imath_{\xi} \widetilde{\Phi}^{*}\overline{\ell}^{\mathtt{(m)}}_{\mathtt{EH}} -  \imath_{\mathrm{X}_{\overline{\xi}}}(\widetilde{\Phi}^{*}\overline{\theta}^{\mathtt{(m)}}_{\mathtt{EH}} + \jmath^{*}(\star_{g}\mathcal{U}))\\
    &= \widetilde{\Phi}^{*}\overline{j}^{\mathtt{(m)}}_{\mathtt{EH}} - \imath_{\mathrm{X}_{\overline{\xi}}}\jmath^{*}(\star_{g}\mathcal{U}).
\end{align*}
Therefore, in the relative bicomplex their relation may be rewritten as, 
\begin{equation}\label{currentsEHfinal}
    \boxed{\big(\mathcal{J}^{\mathtt{(t)}}_{\mathtt{EH}}, \overline{j}^{\mathtt{(t)}}_{\mathtt{EH}}\big) =  \underline{\widetilde{\Phi}}^{*}\big(\mathcal{J}^{\mathtt{(m)}}_{\mathtt{EH}}, \overline{j}^{\mathtt{(m)}}_{\mathtt{EH}} \big) - \underline{\mathrm{d}}\big(\star_{g} \mathcal{U}, 0 \big)}
\end{equation}
which by (\ref{currentsequiv}) means they are in the same equivalence class. The charges for both formalisms are transformed in the same way under the map $\Phi$, 
\begin{align*}
    \mathbb{Q}^{\mathtt{(t)}}_{\mathtt{EH}} &=  \int_{(\Sigma, \partial \Sigma)}\underline{\imath}^{*}\Big(\mathcal{J}_{\mathtt{(EH)}}^{\mathtt{(t)}}, \overline{\jmath}_{\mathtt{(EH)}}^{\mathtt{(t)}}\Big) = \int_{(\Sigma, \partial \Sigma)}\underline{\imath}^{*} \underline{\widetilde{\Phi}}^{*}\Big(\mathcal{J}_{\mathtt{(EH)}}^{\mathtt{(m)}}, \overline{\jmath}_{\mathtt{(EH)}}^{\mathtt{(m)}}\Big) = \Phi^{*} \int_{(\Sigma, \partial \Sigma)}  \Big(\mathcal{J}_{\mathtt{(EH)}}^{\mathtt{(m)}}, \overline{\jmath}_{\mathtt{(EH)}}^{\mathtt{(m)}}\Big) 
\end{align*}
and are shown to be equivalent since the currents and presymplectic structures are the same
\begin{equation}\label{chargesEH}
     \boxed{\mathbb{Q}^{\mathtt{(t)}}_{\mathtt{EH}} = \Phi^{*}\mathbb{Q}^{\mathtt{(m)}}_{\mathtt{EH}}}
\end{equation}

\section{Between Einstein-Hilbert and Palatini}\label{PalatiniEH}
In this section, we will show that the Palatini presymplectic structures are equivalent to those of the Einstein-Hilbert action, first in metric variables and later in tetrad variables. 
Let $ \widetilde{\pi}: M \times \mathcal{F}(g, Q) \rightarrow M \times \mathcal{F}(g)$ so that  $ \widetilde{\pi} = ( \mathrm{Id}, \pi)$ and $\pi$ is the projection that maps the space of fields of a first order Palatini action with independent metric variables $(g,Q)$ to a second order EH action with the metric being the only dynamical field. Similarly, this map over a relative pair is defined as $ \underline{\widetilde{\pi}}: (M, \partial M) \times \mathcal{F}(g, Q) \rightarrow (M, \partial M) \times \mathcal{F}(g)$.

The presymplectic structures of the Palatini action in metric variables consisted of the Einstein-Hilbert term and another one that included the coupling, 
\begin{equation}\label{palatinipotentials}
    (\Theta^{\mathtt{(m)}}_{\mathtt{PT}}, \overline{\theta}^{\mathtt{(m)}}_{\mathtt{PT}}) = \underline{\widetilde{\pi}}^{*}(\Theta^{\mathtt{(m)}}_{\mathtt{EH}}, \overline{\theta}^{\mathtt{(m)}}_{\mathtt{EH}}) +  (\Theta^{\mathtt{(m)}}_{\mathtt{CP}}, 0).
\end{equation}
The presymplectic form was then given by
\begin{align*}
\Omega^{\mathtt{(m)}}_{\mathtt{PT}} &= \delta \Big( \int_{\Sigma} \imath^{*}\Theta^{\mathtt{(m)}}_{\mathtt{PT}}- \int_{\partial \Sigma} \overline{\imath}^{*}\overline{\theta}^{\mathtt{(m)}}_{\mathtt{PT}} \Big),
\end{align*}
where the variation of the presymplectic potential in the bulk is 
\begin{align*}
    \delta \Theta^{\mathtt{(m)}}_{\mathtt{PT}} = \delta \Theta^{\mathtt{(m)}}_{\mathtt{EH}} + \delta \Theta_{\mathtt{CP}}(g,Q) = \delta \Theta^{\mathtt{(m)}}_{\mathtt{EH}},
\end{align*}
since $\delta^2 = 0$. At the boundary, we already have that $\overline{\theta}^{\mathtt{(m)}}_{\mathtt{PT}} =  \widetilde{\pi}^{*}\overline{\theta}^{\mathtt{(m)}}_{\mathtt{EH}}$. Therefore, the presymplectic form is simply the one obtained for the EH action
\begin{equation}\label{palatinisymplecticform}
    \boxed{\Omega^{\mathtt{(m)}}_{\mathtt{PT}}= \pi^{*}\Omega^{\mathtt{(m)}}_{\mathtt{EH}}}
\end{equation}
The current at the bulk consists of two terms associated with each Lagrangian term of the Palatini action 
\begin{align*}
    \mathcal{J}^{\mathtt{(m)}}_{\mathtt{PT}}  &= \imath_{\xi}L^{\mathtt{(m)}}_{\mathtt{EH}} - \imath_{\mathrm{X}_{\xi}}\Theta^{\mathtt{(m)}}_{\mathtt{EH}} + \imath_{\xi}L^{\mathtt{(m)}}_{\mathtt{PT}} - \imath_{\mathrm{X}_{\xi}}\mathrm{d}\Theta^{\mathtt{(m)}}_{\mathtt{CP}}\\
    &=  \mathcal{J}^{\mathtt{(m)}}_{\mathtt{EH}} + \imath_{\xi}(\imath_{A}\volg)  + \mathcal{L}_{\xi}(\imath_{Z}\volg) - \mathrm{d}(\imath_{\xi}\imath_{Z}\volg),
\end{align*}
where we remind the reader that $Z^{\alpha}:= \tensor{Q}{^\alpha^\beta_\beta} - \tensor{Q}{^\beta_\beta^\alpha}$ and  $A^{\alpha}:=\tensor{Q}{^\alpha_\alpha_\lambda}\tensor{Q}{^\lambda^\beta_\beta} - \tensor{Q}{^\alpha^\beta_\lambda}\tensor{Q}{^\lambda_\alpha_\beta}$. 
\noindent At the boundary, the current is given by
\begin{align*}
    \overline{j}^{\mathtt{(m)}}_{\mathtt{PT}} &= - \imath_{\xi}\big(\overline{\ell}^{\mathtt{(m)}}_{\mathtt{GHY}}+ \jmath^{*}(\imath_{Z}\volg)\big) - \imath_{\mathrm{X}_{\xi}}(\overline{\theta}^{\mathtt{(m)}}_{\mathtt{EH}}) \\
    &= \overline{j}^{\mathtt{(m)}}_{\mathtt{EH}} - \jmath^{*}(\imath_{\xi}\imath_{Z}\volg).
\end{align*}
These currents \emph{off-shell} are not in the same equivalence class, 
\begin{equation}\label{currentspalatini}
    \big( \mathcal{J}^{\mathtt{(m)}}_{\mathtt{PT}} , \overline{j}^{\mathtt{(m)}}_{\mathtt{PT}}\big) = \widetilde{\pi}^{*}\big( \mathcal{J}^{\mathtt{(m)}}_{\mathtt{EH}}, \overline{j}^{\mathtt{(m)}}_{\mathtt{EH}}\big)  - \underline{\mathrm{d}}\big( \imath_{\xi}\imath_{Z}\volg, 0 \big) + \big( \mathcal{L}_{\xi}(\imath_{Z}\volg), 0 \big).\\
\end{equation}
Although the currents \emph{off-shell} are not in the same equivalence class, \emph{on-shell},
\begin{equation}\label{currentspalatinionshell}
    \boxed{ \big( \mathcal{J}^{\mathtt{(m)}}_{\mathtt{PT}} , \overline{j}^{\mathtt{(m)}}_{\mathtt{PT}}\big) \overset{\mathtt{Sol}}{=} \widetilde{\pi}^{*} \big( \mathcal{J}^{\mathtt{(m)}}_{\mathtt{EH}}, \overline{j}^{\mathtt{(m)}}_{\mathtt{EH}}\big) - \underline{\mathrm{d}}(\imath_{\xi}\iota_{Z}\volg, 0)}
\end{equation}
which means that the currents are in the same equivalence class and they are therefore cohomologically equivalent. 
Once this has been determined, the charges are computed as usual using (\ref{chargesformula}). They are also trivially equivalent \emph{on-shell} to those given by the Einstein-Hilbert action, 
\begin{equation}\label{chargespalatini}
    \boxed{\mathbb{Q}^{\mathtt{(m)}}_{\mathtt{PT}} = \pi^{*}\mathbb{Q}^{\mathtt{(m)}}_{\mathtt{EH}}}
\end{equation}
Now, we look at the tetrad formulation of Palatini. Let $\Psi: \mathcal{F}(\e, \widetilde{\omega}) \xrightarrow{}\mathcal{F}(g, Q)$ and \\ $\tau: \mathcal{F}(\e, \widetilde{\omega}) \rightarrow \mathcal{F}(\e)$. The presymplectic potentials for the tetrad Einstein-Hilbert and tetrad Palatini actions were, in fact, already equivalent, 
\begin{align*}
    &\Theta^{\mathtt{(t)}}_{\mathtt{PT}} = \frac{1}{2}\epsilon_{IJKL}\e^{I}\wedge \e^{J} \wedge \delta \widehat{\omega}^{KL} = \tau^{*}\Theta^{\mathtt{(t)}}_{\mathtt{EH}},\\
    &\overline{\theta}^{\mathtt{(t)}}_{\mathtt{PT}} = \epsilon_{IJKL}\overline{\e}^{I}\wedge \overline{\e}^{J}\wedge N^{K}\wedge \delta N^{L} = \tau^{*}\overline{\theta}^{\mathtt{(t)}}_{\mathtt{EH}},
\end{align*}
so the presymplectic structures can be trivially written as,
\begin{equation}
    \boxed{\Omega^{\mathtt{(t)}}_{\mathtt{PT}}= \tau^{*}\Omega^{\mathtt{(t)}}_{\mathtt{EH}}}
\end{equation}
The vectors of the form $(0, \mathrm{W})\in T \mathtt{Sol}\big(\mathbb{S}^{\mathtt{(t)}}_{\mathtt{PT}}\big)$ correspond to degenerate directions of $\Omega^{\mathtt{(t)}}_{\mathtt{PT}}$. So, even in the presence of boundaries, all the structures are equivalent to each other, with a projective symmetry given by $Q$, in both the metric and tetrad formulations of Palatini theories. 

\begin{EvalBox}{}
\begin{center}
    \begin{tikzpicture}
  \matrix (m) [matrix of math nodes,row sep=3em,column sep=4em,minimum width=2em]
  {
     \Omega^{\mathtt{(t)}}_{\mathtt{PT}}& \Omega^{\mathtt{(m)}}_{\mathtt{PT}}\\
     \Omega^{\mathtt{(t)}}_{\mathtt{EH}}& \Omega^{\mathtt{(m)}}_{\mathtt{EH}} \\};
  \path[-stealth]
    (m-1-1) edge node [above] {$\Psi$} (m-1-2)
    (m-1-1) edge node [left] {$\tau$} (m-2-1)
    (m-2-1) edge node [below] {$\Phi$} (m-2-2)
    (m-1-2) edge node [right] {$\pi$} (m-2-2);
\end{tikzpicture}
\end{center}
\end{EvalBox}
\newpage
\section{Between Palatini and Hojman-Mukku-Sayed}\label{HMSPalatini}
In this section, we will show that the HMS presymplectic structures are equivalent to those of Palatini, first in metric variables and later in tetrad variables. 
An advantage with respect to the comparison with the Einstein-Hilbert action is that the HMS and Palatini actions are defined on the same space of fields, so we can use again the transition function $\Psi$ between the metric and the tetrad formalisms. 

The presymplectic potentials for the HMS and Palatini actions in metric variables are,
\begin{align*}
    &\Theta^{\mathtt{(m)}}_{\mathtt{HMS}} = \Theta^{\mathtt{(m)}}_{\mathtt{PT}} + \frac{1}{\gamma}\delta(\iota_{q}\volg),\\
    & \overline{\theta}^{\mathtt{(m)}}_{\mathtt{HMS}} = \overline{\theta}^{\mathtt{(m)}}_{\mathtt{PT}} = \overline{\theta}^{\mathtt{(m)}}_{\mathtt{EH}},
\end{align*}
where $q^{\mu}:=(\volg)^{\mu\alpha\beta\nu}Q_{\alpha\beta\nu}$.
So, \emph{off-shell}, the presymplectic potentials are not in the same cohomology class, but are equivalent due to  (\ref{presymplecticpotentialequivalence}), 
\begin{equation}\label{HMSpotentialsoffshell}
    \big( \Theta^{\mathtt{(m)}}_{\mathtt{HMS}}, \overline{\theta}^{\mathtt{(m)}}_{\mathtt{HMS}} \big) = \big( \Theta^{\mathtt{(m)}}_{\mathtt{PT}}, \overline{\theta}^{\mathtt{(m)}}_{\mathtt{PT}}\big) + \frac{1}{\gamma}\underline{\delta }\big(\imath_{q}\volg, 0 \big).
\end{equation}
The relative cohomology class of the HMS presymplectic potentials depends on the parameter $\gamma$, unlike the Palatini action, where this term does not appear. If we compute them \emph{on-shell}, they are in the same equivalence class since $q^{\mu} = 0$, 
\begin{equation}
    \boxed{\big( \Theta^{\mathtt{(m)}}_{\mathtt{HMS}}, \overline{\theta}^{\mathtt{(m)}}_{\mathtt{HMS}} \big) \overset{\mathtt{Sol}}{=}\big( \Theta^{\mathtt{(m)}}_{\mathtt{PT}}, \overline{\theta}^{\mathtt{(m)}}_{\mathtt{PT}}\big)}
\end{equation}

\noindent Let us look now at the tetrad HMS and Palatini actions. The presymplectic potentials of the HMS action in tetrad variables are,
\begin{align*}
    &\Theta^{\mathtt{(t)}}_{\mathtt{HMS}} = \Theta^{\mathtt{(t)}}_{\mathtt{PT}} + \frac{1}{\gamma}\e_{I}\wedge \e_{J}\wedge \delta \mathcal{C}^{IJ} - \frac{1}{\gamma}\mathrm{d}(\e_{I}\wedge \delta \e^{I}),\\
    & \overline{\theta}^{\mathtt{(t)}}_{\mathtt{HMS}} = \overline{\theta}^{\mathtt{(t)}}_{\mathtt{PT}} - \frac{1}{\gamma}\overline{\e}^{I}\wedge \delta \overline{\e}_{I}.
\end{align*}
\emph{Off-shell}, the HMS and Palatini presymplectic potentials are not equal in the relative cohomology due to the term involving the contorsion $\mathcal{C}^{IJ}$ and the presence of the parameter $\gamma$,
\begin{equation}\label{HMSpotentialstetradoffshell}
    \big( \Theta^{\mathtt{(t)}}_{\mathtt{HMS}}, \overline{\theta}^{\mathtt{(t)}}_{\mathtt{HMS}} \big) = 
     \big( \Theta^{\mathtt{(t)}}_{\mathtt{PT}}, \overline{\theta}^{\mathtt{(t)}}_{\mathtt{PT}} \big) - \frac{1}{\gamma}\underline{\mathrm{d}}\big( \e_{I} \wedge \delta \e^{I}, 0 \big)  + \frac{1}{\gamma}\big( \e_{I}\wedge \e_{J} \wedge \delta \mathcal{C}^{IJ}, 0 \big).
\end{equation}
\emph{On-shell}, the presymplectic potentials are cohomologically equivalent because the last term vanishes over solutions and we are left with a relation between the two, given by (\ref{presymplecticpotentialequivalence}) as
\begin{equation}\label{HMSpotentialstetradonshell}
    \boxed{\big( \Theta^{\mathtt{(t)}}_{\mathtt{HMS}}, \overline{\theta}^{\mathtt{(t)}}_{\mathtt{HMS}} \big) \overset{\mathtt{Sol}}{=} 
     \big( \Theta^{\mathtt{(t)}}_{\mathtt{PT}}, \overline{\theta}^{\mathtt{(t)}}_{\mathtt{PT}}\big)  - \frac{1}{\gamma}\underline{\mathrm{d}}\big( \e_{I} \wedge \delta \e^{I}, 0 \big)}
\end{equation}

\noindent Another way of proving that the spaces of solutions of HMS and Palatini are in fact the same in either metric or tetrad variables is by showing that the map  $\delta \Psi$ is surjective. This condition is sufficient since
\begin{equation*}
    \delta_{(\e, \widetilde{\omega})}\mathbb{S}^{\mathtt{(t)}}_{\mathtt{HMS}}= \delta_{(\e, \widetilde{\omega})}(\mathbb{S}^{\mathtt{(m)}}_{\mathtt{HMS}} \circ \Psi)=\delta_{\Psi(\e, \widetilde{\omega})}\mathbb{S}^{\mathtt{(m)}}_{\mathtt{HMS}} \circ \delta_{(\e, \widetilde{\omega})}\Psi, 
\end{equation*}
which means that,
\begin{align*}
\mathtt{Sol}(\mathbb{S}^{\mathtt{(t)}}_{\mathtt{HMS}}) = \Psi^{-1}\mathtt{Sol}(\mathbb{S}^{\mathtt{(m)}}_{\mathtt{HMS}}) = \Psi^{-1}\mathtt{Sol}(\mathbb{S}^{\mathtt{(m)}}_{\mathtt{PT}}) = \mathtt{Sol}(\mathbb{S}^{\mathtt{(t)}}_{\mathtt{PT}}).
\end{align*}
This is a straight-forward approach to show the equivalence between solution spaces, alternative to our explicit computations of their symplectic structures, which might be useful for other field theories.

%% file: conclusions.tex
\chapter{Conclusions}\label{conclusions}

In this thesis, we have studied the covariant phase space (\ref{cpspair}) of three gravitational actions in the presence of boundaries with the help of the relative bicomplex framework. This approach unequivocally specifies how to compute the covariant phase space of a field theory defined on a manifold with boundaries. In this way, the relative bicomplex surpasses traditional covariant phase space methods, which are only suited for boundary-free field theories. The relative bicomplex classifies action principles in terms of equivalence classes of Lagrangian pairs and thus provides us with a technique to compare the spaces of solutions, presymplectic potentials, presymplectic forms, currents and charges associated with two action principles. In the hope of displaying its practicality, we exemplified its use with two background dependent field theories: the scalar and Yang-Mills fields. Despite being very well-known field theories, the sophistication of the relative bicomplex in dealing with boundaries proved itself superior even with these familiar examples. 

The first result of this thesis is the equivalence between the metric and tetrad formulations of the Einstein-Hilbert action (\ref{EHaction}). We were able to show that, without considering the compatible boundary terms (as previously done in the literature), it is incorrect to compare the presymplectic structures of both formalisms. Nevertheless, by using the CPS algorithm (\ref{cpsalgorithm}), we were able to provide the correct relation between them in the presence of boundaries (\ref{tetradmetricEHpotentials}),
\begin{equation}
    (\Theta_{\mathtt{EH}}^{\mathtt{(t)}},  \overline{\theta}^{\mathtt{(t)}}_{\mathtt{EH}}) = \underline{\widetilde{\Phi}}^{*}\big( \Theta_{\mathtt{EH}}^{\mathtt{(m)}}, \overline{\theta}^{\mathtt{(m)}}_{\mathtt{EH}}\big)  + \underline{\mathrm{d}}\big(\star_{g}\mathcal{U}, 0\big).
\end{equation}
We further showed that their presymplectic structures coincide (\ref{tetradmetricEHsymplectic}), 
\begin{equation}
    \Omega^{\mathtt{(t)}}_{\mathtt{EH}} = \Phi^{*}\Omega^{\mathtt{(m)}}_{\mathtt{EH}},
\end{equation}
that the currents are equivalent since they are in the same equivalence class (\ref{currentsEHfinal}) and that the generalised Noether charges are also equivalent (\ref{chargesEH}).
Introducing the connection as an independent variable led us to a first-order formalism or Palatini theories, particularly useful for coupling fermionic matter to gravity. We considered the Palatini action in vacuum on a manifold with boundaries, assuming torsion and nonmetricity. Thus, the second result of this thesis is the equivalence of the space of solutions of the Palatini action and the space of solutions of the previously studied Einstein-Hilbert action, that is, the equivalence between first order and second order actions in pure gravity. We confirmed the projective nature of the connection and showed that the metric variables satisfied Einstein's equations also in the presence of boundaries. 

We then showed the equivalence between the metric and tetrad formulations of the Palatini action with boundaries, torsion and nonmetricity. Using the CPS algorithm, we arrived at the relation between the presymplectic potentials (\ref{palatinipotentials}) and saw that their equivalence is only satisfied \emph{on-shell},
\begin{equation}
    (\Theta^{\mathtt{(m)}}_{\mathtt{PT}}, \overline{\theta}^{\mathtt{(m)}}_{\mathtt{PT}}) \overset{\mathtt{Sol}}{=} \underline{\widetilde{\pi}}^{*}(\Theta^{\mathtt{(m)}}_{\mathtt{EH}}, \overline{\theta}^{\mathtt{(m)}}_{\mathtt{EH}}).
\end{equation}

However, because the coupling term is a vertical total-derivative, the presymplectic structures identically coincide (\ref{palatinisymplecticform}). We also studied the generalised Noether currents and showed that, although not in the same equivalence class, they coincide \emph{on-shell} (\ref{currentspalatini}), (\ref{currentspalatinionshell}), as do the Noether charges (\ref{chargespalatini}), 
\begin{equation}
    \mathbb{Q}^{\mathtt{(m)}}_{\mathtt{PT}} = \pi^{*}\mathbb{Q}^{\mathtt{(m)}}_{\mathtt{EH}}.
\end{equation}
In the last chapter, we presented a new action referred to as the Hojman-Mukku-Sayed (HMS) action, which is a generalisation of the Holst action with boundaries, torsion and nonmetricity (\ref{HMSaction}). Specifically, we derived a new boundary Lagrangian recovering the Gibbons-Hawking-York Lagrangian at the boundary. The HMS action is built from the Palatini action by adding a dual-term, proportional to the Barbero-Immirzi parameter $\gamma$. In the metric formalism, its equivalence to the metric Palatini action was only demonstrated \emph{on-shell}, since the presymplectic potentials associated with each Lagrangian pair are not in the same equivalence class (\ref{HMSpotentialsoffshell}) due to a term proportional to $\gamma$ which is not a total derivative in the relative bicomplex,
\begin{equation}
    \big( \Theta^{\mathtt{(m)}}_{\mathtt{HMS}}, \overline{\theta}^{\mathtt{(m)}}_{\mathtt{HMS}} \big) = \big( \Theta^{\mathtt{(m)}}_{\mathtt{PT}}, \overline{\theta}^{\mathtt{(m)}}_{\mathtt{PT}}\big) + \frac{1}{\gamma}\underline{\delta}\big( \imath_{q}\volg, 0 \big).
\end{equation}
In the tetrad formalism, the change of variables involves derivatives, and the associated presymplectic potentials (\ref{HMSpotentialstetradoffshell}) are not equivalent \emph{off-shell},
\begin{equation}
    \big(\Theta_{\mathtt{HMS}}^{\mathtt{(t)}}, \overline{\theta}_{\mathtt{HMS}}^{\mathtt{(t)}}\big) = \big(\Theta_{\mathtt{PT}}^{\mathtt{(t)}}, \overline{\theta}_{\mathtt{PT}}^{\mathtt{(t)}}\big) + \frac{1}{\gamma}\big( \e_{I} \wedge \e_{J} \wedge \delta \mathcal{C}^{IJ}, 0\big) - \frac{1}{\gamma}\underline{\mathrm{d}}\big( \e_{I} \wedge \delta \e^{I}, 0\big),
\end{equation}
but are equivalent \emph{on-shell} (\ref{HMSpotentialstetradonshell}) to those of Palatini,
\begin{equation}
    \big( \Theta^{\mathtt{(t)}}_{\mathtt{HMS}}, \overline{\theta}^{\mathtt{(t)}}_{\mathtt{HMS}} \big) \overset{\mathtt{Sol}}{=} 
     \big( \Theta^{\mathtt{(t)}}_{\mathtt{PT}}, \overline{\theta}^{\mathtt{(t)}}_{\mathtt{PT}}\big) - \frac{1}{\gamma}\underline{\mathrm{d}}\big( \e_{I} \wedge \delta \e^{I}, 0\big).
\end{equation}
Therefore, the non-equivalence off-shell of the HMS action with the Palatini action is due to the term proportional to the parameter $\gamma$. This term might lead to non-equivalent quantisations, but it does not play a role in the classical regime since it vanishes over solutions. Notably, the role of the parameter $\gamma$ seems to drastically differ from the parameter $\theta$ in QCD, which is only topological in nature even when the gauge field is coupled to bosons or fermions. This is not the case for the parameter $\gamma$. Although for bosons, the coupling terms are independent of the connection and do not affect the equations of motion. If one considers fermions, the space of solutions differs from the one corresponding to the Palatini action.

The natural continuation of this thesis would be to compare the presymplectic structures obtained for each action principle with the ones obtained in the canonical Hamiltonian formalism. In particular, the ADM formalism in the presence of boundaries. Another line of research that spurs from our results includes coupling the action principles we studied with torsion and nonmetricity to fermionic matter. A more ambitious future goal would be to extend the relative bicomplex framework to deal with null and spacelike boundaries and incorporate the treatment of null-horizons (isolated horizons and null-infinity) and dynamical-horizons. The inclusion of null and spacelike boundaries is an active area of research that has been explored with traditional methods in \cite{chandrasekaran2018symmetries, Ashtekarnull1, Ashtekarnull2}. Nevertheless, the development of an unambiguous procedure, provided by the relative bicomplex, is still lacking and is left for future research.

%% file: appendix.tex
\appendix
\chapter{Geometric Structures}\label{appendix}
This appendix gathers some selected definitions repeatedly used in the thesis and attempts to consider with some care the relationship between them. These relationships are sometimes overlooked in other treatments of gravitational theories but are of major importance when comparing the tetrad and metric formalisms of any field theory. For a complete treatment of the material presented, the reader is referred to \cite{husemoller1966fibre}. 
\section{Bundles}
\vspace{-0.25cm}
Given a fibre bundle $(E, \pi, B, F)$, the differential manifold $B$ is called the base space, the manifold $E$ is the total space, and the map $\pi: E \xrightarrow{} B $ is called the projection of the bundle. For each $b\in B$, the space $\pi^{-1}(b)$ is called the fibre of the bundle at $b\in B$ and is homeomorphic to $F$. 
\vspace{-0.3cm}
\begin{definition}[Vector Bundle]
An $n$-dimensional vector bundle is a fiber bundle $(E, \pi, B, F)$ whose fibre $F$ is homeomorphic to a vector space $V$ over a field $k = \mathbb{R}\; \text{or} \; \mathbb{C} $ \;  (or even the division ring of quaternions $\mathbb{H}$), and whose structure group is $GL(V) = GL(n, k)$.  
\end{definition}

\begin{definition}[Right $G$-action]
Let $(G, \circ)$ be a Lie group and let $M$ be a smooth manifold. A right $G$-action on $M$ is a smooth map
\begin{align*}
    \triangleleft: \;\; M \times G &\longrightarrow M \\
     (m, g) & \longmapsto \triangleleft_{g}(m) =  m \triangleleft g
\end{align*}
that satisfies, 
\begin{itemize}
        \item $\forall m \in M, \; m \triangleleft e = m$,  where $e$ is the identity element of $G$. 
    \item $\forall g_{1}, g_{2} \in G, \; \forall m \in M: \; m \triangleleft (g_{1} \circ g_{2}) = (m \triangleleft g_{1})\triangleleft g_{2}$.
\end{itemize}
\end{definition}

\begin{definition}[Principal G-bundle]
Let $G$ be a Lie group. A principal $G$-bundle is a fibre bundle $(E, \pi, B, F)$, together with a right $G$-action such that $\forall b \in B$, 
\begin{itemize}
    \item $\forall y \in \pi^{-1}(b), y \triangleleft_{g} = y $,  implies that g is the identity of $G$ (the action is free).
    \item $\forall y, y' \in \pi^{-1}(b), \; \exists g \in G$ such that $y' = y \circ g$ (the action is transitive).
    \item $\forall y \in \pi^{-1}(b)\; \Longrightarrow y \triangleleft_{g} \in \pi^{-1}(b), \; \forall g \in G$ (the action preserves the fibres).
\end{itemize}
Hence, a principal $G$-bundle  $(E, \pi, B, F)$ is a bundle whose fibre is homeomorphic to the group $G$, and so the action of $G$ upon $F$ simply permutes the elements within each fibre.
\end{definition}
\begin{definition}[Vertical Subspace] Let $(E, \pi, B, F)$ be a fibre bundle and let $y \in E$. Let $T_{y}E$ be the tangent space at $y$ of the space $E$. The vertical subspace at $y$ is the vector subspace of $V_{y}E \subset T_{y}E$ given by
\begin{align*}
    V_{y}E:= \text{ker}(\pi_{*})_{y} = \{ X_{y} \in T_{y}E \; | \; (\pi_{*})_{y}(X_{y}) = 0\},
\end{align*}
where $(\pi_{*})$ is the push-forward of the projection map. When $(E, \pi, B, F)$ is a principal $G$-bundle, the natural vertical subspace generated by the vector fields arising from the Lie-algebra $\mathfrak{g}$ of the structure group is given by the fundamental vector field mapping for the right action, defined as 
\begin{align*}
    i : E \times \mathfrak{g} &\longrightarrow VE \\
    (y, A) & \longmapsto \xi^{A}_{y},
\end{align*}
which means that $\mathfrak{g}$ gives rise to only vertical vectors in the fibres of $E$ and hence it gives a canonical trivialization of the vertical subspace $VE$.
\end{definition}

Any complementary vector subspace to $V_{y}E$ is called a horizontal subspace which is not uniquely defined. Making a particular choice on the horizontal subspace is one way to define a connection on $E$.
\begin{definition}[Connection] Let $(E, \pi, B, F)$ be a principal $G$-bundle. Let $\sigma_{y}: T_{b}B \xrightarrow{}T_{y}E$ be the horizontal lift such that $b = \pi(y)$ with $(\pi)_{*}\circ \sigma_{y} = \text{id}_{T_{b}B}$, and $\sigma_{y \triangleleft g} = (\triangleleft_{g})_{*} \sigma_{y}$. A connection on a principal $G$-bundle $(E, \pi, B, F)$ is a collection of horizontal subspaces $H_{y}E \subset T_{y}E$ such that,
\begin{align*}
    & T_{y}E = H_{y}E \oplus V_{y}E \\
    & X_{y} = \text{Hor}(X_{y}) + \text{Ver}(X_{y})
\end{align*}
where $\text{Hor}(X_{y}) := \sigma_{y} \pi_{*}  X_{y}$ and $\text{Ver}(X_{y}) := X -  \sigma_{y} \pi_{*}  X_{y}$. Notice that $\text{Ver}(X_{y}) \in V_{y}E$ crucially depends on the choice of connection.  
\end{definition}

Another equivalent way of defining a connection is as a $\mathfrak{g}$-valued one-form.
\begin{definition}[Adapted Connection Form]
An adapted connection form $\omega$ is a collection of $\mathfrak{g}$-valued $1$-forms, adapted to a decomposition $TE = HE \oplus VE$, such that
\begin{align*}
    \omega = \{  \omega_{y} : T_{y}E \xrightarrow{} T_{e}G  \; | \; y \in E\}
\end{align*}
so that each $\omega_{y}$ sends 
and element of $T_{y}E$ to an element of $\mathfrak{g}$,
\begin{align*}
     \omega_{y} : T_{y}E & \longrightarrow \mathfrak{g} \\
      \xi_{y} &\longmapsto i^{-1}(\text{Ver}(\xi_{y})) = i^{-1}(\xi^{A}_{y}) = A.
\end{align*}
Thus the connection form annihilates the horizontal vectors and takes the vertical ones into the Lie algebra.
\end{definition} 
The adapted connection form generalizes the notion of the Maurer-Cartan form on a Lie group $G$ to a principal $G$-bundle. Therefore, the horizontal space is completely determined by its kernel, $HE =\text{ker}(\omega)$. 

In fact, there always exists a principal connection $\Phi\in \Omega^{1}(E ; VE)$ on a principal $G$-bundle, such that $\ker(\Phi) = HE$ and $(\triangleleft_{g})^{*}\Phi = \Phi$ (i.e. the action of the group is equivariant). Then, we can define a connection form prior to a choice of horizontal space as follows. 
\begin{definition}[Connection Form]
Let $(E, \pi, B, F)$ be a principal G-bundle. A $\mathfrak{g}$-valued $1$-form $\omega \in \Omega^{1}(E, \mathfrak{g})$ is called a connection form if it satisfies the following properties:
\begin{enumerate}
    \item $\omega(i(y, A)) = A$ for all $A \in \mathfrak{g}$.
    \item $(\triangleleft_{g})^{*}(\omega(\xi_{y})) = \omega (T_{y}(\triangleleft_{g}\xi_{y})) = \mathrm{Ad}(g^{-1}) \omega(\xi_{y})$ for all $g \in G$ and $\xi_{y} \in T_{y}E$. 
    \item $\mathcal{L}_{\xi^{A}}\;\omega = -\mathrm{ad}(A)\omega$ where  $\xi^{A}:= i(\cdot, A) $ and $\mathrm{ad}$ is the tangent map of the adjoint map. 
\end{enumerate}
%Conversely, each $\omega \in \Omega^{1}(E, \mathfrak{g})$ that satisfies $(1)$ defines a principal connection $\Phi$ in $E$ if and only if $(2)$ is satisfied. 
\end{definition}

\begin{definition}[Covariant derivative of a Lie-algebra valued form] 
Let $(E, \pi, B, F)$ be a principal $G$-bundle. Let $\varphi$ be a  Lie-algebra valued $r$-form on $E$. We define its covariant derivative by
    \begin{align*}
        D\varphi (v_{1}, ..., v_{r+1}) = \mathrm{d}\varphi (\text{Hor}(v_{1}), ... , \text{Hor}(v_{r+1}))
    \end{align*}
The \emph{curvature form} $\Omega$ of the \emph{connection form} $\omega$ is defined as $\Omega = D \omega$ and satisfies the Maurer-Cartan structure equations.
\end{definition}
The tangent bundle of an $n$-dimensional manifold $(TM, \tau, M, F)$ is a typical example of a vector bundle where its fibre is $F = \mathbb{R}^{n}$. 

\begin{definition}[Frame Bundle]
Let $(TM, \tau, M, \mathbb{R}^{n})$ be the tangent bundle. Then the frame bundle is the principal $G$-bundle $(\mathrm{F}(TM), \pi, M, \mathrm{GL}(n, \mathbb{R}))$ associated with the tangent bundle whose elements $u \in F(TM)$ are linear isomorphisms $u: \mathbb{R}^{n} \longrightarrow T_{x}M$ for some $x \in M$. Explicitly,
\begin{equation*}
   \begin{split}
        u: & \mathbb{R}^{n} \longrightarrow T_{\pi(u)}M \\
      & \vec{v} \longmapsto u(v^{I}\vec{e}_{I}) = v^{I}\e^{\alpha}_{I}\frac{\partial}{\partial x^{\alpha}} 
\end{split} 
\end{equation*}

where $\vec{e}_{I} \in \mathbb{R}^{k}$, $I = 1, \ldots, n$ is the canonical coordinate basis and $\frac{\partial}{\partial x^{\alpha}}$ is the coordinate basis of $T_{\pi(u)}M$, with $x=\pi(u)$. 
\end{definition}
The inverse of a frame at $\pi(u) \in M$  in $\mathrm{F}(TM)$ is a map so that, 
\begin{align*}
    u^{-1} : T_{\pi(u)}M &\longrightarrow \mathbb{R}^{n} \\
    \xi &\longmapsto u^{-1}\big(\xi^{\alpha}\frac{\partial}{\partial x^{\alpha}}\big) = \xi^{\alpha}\e^{I}_{\alpha}\vec{e}_{I}
\end{align*}

The tangent bundle to the frame bundle is $( T\mathrm{F}(TM), \sigma, \mathrm{F}(TM), F'' = \mathbb{R}^{k})$, where $k = n + n^2$ . An element of this bundle at $u\in \mathrm{F}(TM)$ is explicitly given in local coordinates by 
\begin{align*}
    \vec{\xi}_{u} = \xi^{\alpha}\frac{\partial}{\partial x^{\alpha}} + \xi^{\alpha}_{I}\frac{\partial}{\partial e^{\alpha}_{I}}
\end{align*}
This bundle is $T\mathrm{F}(TM) = \mathrm{F}(TM) \times \mathbb{R}^{k}$ so its fibres are isomorphic to $\mathbb{R}^{k}$. 
\begin{definition}[Soldering Form]
Let $(TM, \tau, M, \mathbb{R}^{n})$ be the tangent bundle and let its associated tangent frame bundle be $(\mathrm{F}(TM), \pi, M, F' = GL(n, \mathbb{R}))$. A soldering form on $\mathrm{F}(TM)$ is a vector valued $1$-form such that  for $u\in \mathrm{F}(T_{\pi(u)}M)$ with $\pi(u) \in M$, 
\begin{align*}
    \theta: T\mathrm{F}(TM) &\longrightarrow \mathbb{R}^{n} \\
    \vec{\xi}_{u} &\longmapsto \theta(\vec{\xi}_{u}) =  u^{-1}(\pi_{*}\vec{\xi}_{u}) = \xi^{\alpha}\e^{I}_{\alpha}\vec{e}_{I}, 
\end{align*}
where the push-forward map $(\pi_{*})$ is given by, 
\begin{align*}
    \pi_{*}: T\mathrm{F}(TM) &\longrightarrow TM \\
    \vec{\xi}_{u} &\longmapsto (\pi_{*})_{u}(\vec{\xi}_{u}) = \xi^{\alpha}\frac{\partial}{\partial x^{\alpha}}.
\end{align*}
Therefore, the soldering form  is a linear isomorphism that relates or ``solders'' a vector on $\mathrm{F}(TM)$ to an element of the tangent bundle, $\theta(\xi) \in \mathbb{R}^{n}$. 
\end{definition}

\begin{EvalBox}{}
\begin{center}
    \begin{tikzpicture}
  \matrix (m) [matrix of math nodes,row sep=3em,column sep=4em,minimum width=2em]
  {     F = \mathbb{R}^{n} & T\mathrm{F}(TM) & F'' = \mathbb{R}^{k} \\
  TM & \mathrm{F}(TM) & F' = GL(n, \mathbb{R}) \\
  M & & 
  \\};
  \path[-stealth]
    (m-1-3) edge node [above] {} (m-1-2)
    (m-1-1) edge node [above] {} (m-2-1)
    (m-1-1) edge node [left] {$u$} (m-2-1)
    (m-2-1) edge node [left] {$\tau$} (m-3-1)
    (m-2-2) edge node [above] {$\pi$} (m-3-1)
    (m-1-2) edge node [above] {$\theta$} (m-1-1)
    (m-1-2) edge node [above] {$\pi_{*}$} (m-2-1)
    (m-2-3) edge node [above] {} (m-2-2)
    (m-1-2) edge node [right] {$\sigma$} (m-2-2);
\end{tikzpicture}
\end{center}
\end{EvalBox}

We add structure by considering a metric $g$ in the base manifold $M$,
\begin{align*}
    g(x): T_{x}M \times T_{x}M &\longrightarrow \mathbb{R} \\
    (v^{\alpha}, w^{\beta}) & \longmapsto g_{\alpha\beta}v^{\alpha}w^{\beta}.
\end{align*}
We further equip the fibre of the tangent bundle $F = \mathbb{R}^{n}$ with the Minkowski metric $\eta$,
\begin{align*}
    \eta : F \times F &\longrightarrow \mathbb{R} \\
    (y^{I}, z^{J}) & \longmapsto \eta_{IJ}y^{I}z^{J}.
\end{align*}
Using an element of the frame bundle $u$, along with  $v, w \in \mathbb{R}^{n}$, we can write 
\begin{align*}
    g(u(\vec{v})u(\vec{w})) &= g(v^{I}\e^{\alpha}_{I}\;\vec{\partial}_{\alpha}, w^{I}\e^{\beta}_{J}\;\vec{\partial}_{\beta}) = v^{I}w^{I}\e^{\alpha}_{I}\e^{\beta}_{J}g_{\alpha\beta} .
\end{align*}
If we specify the frame to be a linear isometry, we have the following relation 
\begin{align*}
    \eta_{IJ} = \e^{\alpha}_{I}\e^{\beta}_{J}g_{\alpha\beta}.
\end{align*}

By the fundamental theorem of Riemannian geometry, there is a unique connection on the tangent bundle $\nabla$ which is torsionless and metric compatible, namely the Levi-Civita connection. This affine connection gives rise to a choice of horizontal subspace $H_{u}F(TM)$ in $T_{u}F(TM)$ at each point $u \in F(TM)$, which defines a connection form on $F(TM)$, usually referred to as the \emph{spin-connection}. 

\begin{proposition}
There exists a unique smooth $\omega \in \Omega^{1}(F(TM), \mathfrak{g})$ connection form in $F(TM)$ called the spin-connection such that, 
\begin{enumerate}
    \item $\omega(\xi_{u}) = 0 \Longleftrightarrow \xi_{u} \in H_{u}\mathrm{F}(TM)$.
    \item $\forall A \in \mathfrak{g} \Longrightarrow \omega(\xi^{A}) = A$ where $\xi^{A}:= i(\cdot, A)$. 
    \item $(\triangleleft_{g}^{\;*}\;\omega)(\xi_{u}) = \mathrm{Ad}(g^{-1})\;\omega(\xi_{u})$ for all $g\in G$ and $\xi_{u} \in T_{u}\mathrm{F}(TM)$.
\end{enumerate}
\end{proposition}
\begin{proof}

(1)\quad The general coordinate expression of $\xi \in T\mathrm{F}(TM)$ is given by, 
\begin{align*}
    \xi = \xi^{\alpha} \frac{\partial}{\partial x^{\alpha}} + \xi^{\alpha}_{I}\frac{\partial}{\partial \e^{\alpha}_{I}}.
\end{align*}
Requiring this vector to be horizontal is equivalent to demanding $\xi^{\alpha}_{I} = - \Gamma^{\alpha}_{\beta\gamma}\xi^{\beta}e^{\gamma}_{I}$, where the Christoffel symbols are given by the Levi-Civita connection, $\nabla$ on $M$. The general coordinate expression for $\omega \in \Omega^{1}(F(TM), \mathfrak{g})$ is, 
\begin{equation*}
     \tensor{\omega}{^J_K} = \tensor{\omega}{_\alpha^J_K}\mathrm{d}x^{\alpha} + \tensor{(\omega^{I}_{\alpha})}{^J_K}\mathrm{d}\e^{\alpha}_{I}.
\end{equation*}
Now, let $\xi_{H}$ be a horizontal vector field. Requiring that the first property in the definition holds is explicitly demanding that
\begin{align*}
     \tensor{\omega}{^J_K}(\xi_{H}) = \tensor{\omega}{_\alpha^J_K}\xi^{\alpha} - \tensor{(\omega^{I}_{\alpha})}{^J_K}\Gamma^{\alpha}_{\beta\gamma}\xi^{\beta} \e^{\gamma}_{I} = 0 \Longleftrightarrow \tensor{\omega}{_\alpha^J_K} = \tensor{(\omega^{I}_{\beta})}{^J_K}\Gamma^{\beta}_{\alpha\gamma} \e^{\gamma}_{I}.
\end{align*}
Hence, asking for $(1)$ is equivalent to asking the spin-connection to have the coordinate expression,
\begin{equation}\label{omega1}
    \tensor{\omega}{^J_K}
 = \tensor{(\omega^{I}_{\beta})}{^J_K}\Gamma^{\beta}_{\alpha\gamma}\e^{\gamma}_{I}\mathrm{d}x^{\alpha} + \tensor{(\omega^{I}_{\beta})}{^J_K}\mathrm{d}\e^{\beta}_{I}
 \end{equation}
(2) \quad An element of the Lie-algebra is given by a matrix, $A = \tensor{A}{^J_K}$. Let $\xi^{A}_{u}:= i(u, A)$ be a vertical vector field. Then, using the coordinate expression for $\omega$ we have 
\begin{align*}
\begin{rcases}
    &\xi^{A}_{u}:= \tensor{A}{^L_I}\e^{\alpha}_{L}\frac{\partial}{\partial \e^{\alpha}_I} \\
    &\omega(\xi^{A}_{u}) = A 
\end{rcases}
\Longrightarrow \tensor{A}{^J_K} = \tensor{(\omega^{I}_{\alpha})}{^J_K}\tensor{A}{^L_I}\e^{\alpha}_{I}.
\end{align*}
The solution to this equation is $\tensor{(\omega^{I}_{\alpha})}{^J_K} = \e^{J}_{\alpha}\delta^{I}_{K}$. Therefore,
\begin{align*}
    (1)+ (2)\Longrightarrow \tensor{\omega}{^J_K} = \Gamma^{\beta}_{\alpha\gamma}\e^{J}_{\beta}\e^{\gamma}_{K}\;\mathrm{d}x^{\alpha} + \e^{J}_{\beta}\;\mathrm{d}\e^{\beta}_{K}.
\end{align*}
(3) \quad Now, let us consider how the spin-connection acts on a general $ \xi_{u} \in T_{u}\mathrm{F}(TM)$, 
\begin{align}
    \tensor{\omega}{^J_K}(\xi_{u}) = \Gamma^{\beta}_{\alpha\gamma}\e^{J}_{\beta}\e^{\gamma}_{K}\xi_{u}^{\alpha} + \e^{J}_{\beta}\;\xi_{u\;K}^{\beta}.
\end{align}
The right action on a frame $u \in \mathrm{F}(TM)$ is given by, 
\begin{align*}
    &u(\vec{\e}_{I}) = \e^{\alpha}_{I}\frac{\partial}{\partial x^{\alpha}}, & (u \triangleleft g)(\vec{\e}_{I}) = \tensor{G}{^I_J}\;\e^{\alpha}_{I}\frac{\partial}{\partial x^{\alpha}},\\
    & u^{-1}(\vec{\partial_{x}}) = \e^{I}_{\alpha}\vec{\e}_{I}, & (u^{-1}\triangleleft g)(\vec{\partial_{x}}) = \tensor{(G^{-1})}{^I_J}\;\e^{J}_{\alpha}\;\vec{e}_{I},
\end{align*}
hence explicitly we have that, 
\begin{align*}
    \tensor{(\omega\triangleleft g)}{^J_K} &= \Gamma^{\beta}_{\alpha\gamma} \tensor{(G^{-1})}{^{J}_{S}}\; \e^{S}_{\beta} \;\tensor{G}{^{P}_{K}}\;\e^{\alpha}_{P} \;\mathrm{d}x^{\alpha} + \tensor{(G^{-1})}{^{J}_{S}} \;\e^{S}_{\beta} \;\tensor{G}{^{P}_{K}}\; \mathrm{d} \e^{\beta}_{P} \\
    &= \tensor{(G^{-1})}{^{J}_{S}} \big( \Gamma^{\beta}_{\alpha\gamma} \;\e^{S}_{\beta} \;\e^{\gamma}_{P}\;\mathrm{d}x^{\alpha} + \e^{S}_{\beta}\; \mathrm{d}\e^{\beta}_{P} \big) \tensor{G}{^P_K} \\
    &= \tensor{(g^{-1}\omega \;g)}{^J_K}.
\end{align*}
\end{proof}
The spin-connection defined on the principal bundle when pull-backed through a (local) section $s: U \subset M \xrightarrow{}\mathrm{F}(TM)$ such that $s(x) = (x^{\alpha}, \e^{\alpha}_{I}(x))$ to the base manifold is also called the spin-connection. Explicitly, 
\begin{align*}
\tensor{(s^{*}\omega)}{^J_K} = \Gamma^{\beta}_{\alpha\gamma} \e^{J}_{\beta}(x)\e^{\gamma}_{K}(x) \mathrm{d}x^{\alpha} + \e^{J}_{\beta}(x) \frac{\partial \e^{\beta}_{K}}{\partial x^{\alpha}} \mathrm{d}x^{\alpha} = \Big( \e^{J}_{\beta} \frac{\partial \e^{\beta}_{K}}{\partial x^{\alpha}} + \Gamma^{\beta}_{\alpha\gamma} \e^{J}_{\beta} \e^{\gamma}_{K}\Big) \mathrm{d}x^{\alpha} 
\end{align*}
and when evaluated on a vector $V \in \mathfrak{X}(M)$,  $\tensor{(s^{*}\omega)}{^J_K}(V) = \e^{J}\nabla_{\vec{V}}\E_{K}$ where $e^{J}:= e^{J}_{\alpha}\mathrm{d} x^{\alpha}$ and $E_{I}:= \e^{\alpha}_{I}\frac{\partial}{\partial x^{\alpha}}$, making the usual abuse of notation of denoting with the same letter the elements that live in the frame bundle and in the base manifold. 
These Lie algebra-valued forms $\omega$ induce an affine connection $\nabla$ on their associated vector bundles (\cite{kolar2013natural} p. 109). When this $\omega$ connection form is the spin-connection, the induced affine connection is the Levi-Civita one $\nabla^{\small{\mathrm{LC}}}$. Crucially, the difference between two affine connections is a tensor, schematically $\nabla - \nabla^{\small{\mathrm{LC}}} = Q$, and this is the object that is used as the dynamical variable in first order field theories. 

\section{Variations}\label{appendix_variations}
Some variations of the basic objects used in the metric and tetrad formalisms  follow. Let us denote the determinant of the metric by $g$, the Riemann tensor by $\tensor{R}{^\rho_\sigma_\mu_\nu}$, and the trace of the extrinsic curvature by $K$, 
\begin{align*}
    & \tensor{R}{^\rho_\sigma_\mu_\nu}:= \partial_{\mu} \Gamma^{\rho}_{\nu\sigma} - \partial_{\nu}\Gamma^{\rho}_{\mu\sigma} + \Gamma^{\rho}_{\mu\lambda}\Gamma^{\lambda}_{\nu\sigma} - \Gamma^{\rho}_{\nu\lambda}\Gamma^{\lambda}_{\mu\sigma}, \\
    & \nabla_{\alpha}(\delta \Gamma^{\alpha}_{\mu\nu}) = \partial_{\alpha} \delta \Gamma^{\alpha}_{\mu\nu} + \Gamma^{\alpha}_{\alpha\gamma}\delta \Gamma^{\gamma}_{\mu\nu} - \Gamma^{\gamma}_{\alpha\mu}\delta \Gamma^{\alpha}_{\gamma\nu} - \Gamma^{\gamma}_{\alpha\nu} \delta \Gamma^{\alpha}_{\gamma\mu}, \\
 &\delta \mathrm{R}= - \mathrm{R}^{\alpha\beta}\delta g_{\alpha\beta} + \nabla^{\alpha}\nabla^{\beta}\delta g_{\alpha\beta} - \nabla^{\alpha}\nabla_{\alpha}\delta g,\\
 & \delta (\mathrm{R}_{\beta\gamma})=\nabla_{\alpha}\;\delta(\tensor{\Gamma}{^\alpha_\beta_\gamma})-\nabla_{\beta}\;\delta(\tensor{\Gamma}{^\alpha_\alpha_\gamma}),\\
 & \delta g^{\alpha\beta} = - g^{\alpha\lambda}g^{\beta\gamma}\delta g_{\lambda \gamma}, \\
&\tensor{\delta\Gamma}{^\alpha_\beta_\gamma}=\frac{1}{2}g^{\alpha\mu}\Big(\nabla_{\!\beta}\delta g_{\mu\gamma}+\nabla_{\!\gamma}\delta g_{\beta\mu}-\nabla_{\!\mu}\delta g_{\beta\gamma}\Big),\\
& \delta\volg=\frac{\delta g}{2}\volg,\\
& \delta \nu_{\alpha}=\frac{1}{2}\nu_\alpha\nu^\beta\nu^\gamma\delta g_{\beta\gamma},\\
& \delta \K_{\alpha\beta}=\frac{1}{2}\jmath^{*}(\K_{\alpha\beta}\nu^\sigma\nu^\gamma\delta g_{\sigma\gamma} + \nu^{\mu}\nabla_{\mu}\delta g_{\alpha\beta}-\nu^\mu\nabla_{\!\alpha}\delta g_{\mu\beta}-\nu^\mu\nabla_{\!\beta}\delta g_{\mu\alpha}),\\
& \delta \K=\frac{1}{2}\jmath^{*}\Big(\nu^{\mu}\nabla_{\mu}\delta g-\nu^\alpha\nabla^\beta\delta g_{\alpha\beta}-K^{\alpha\beta}\delta g_{\alpha\beta}- \nabla^{\beta}\nu^{\alpha}\delta g_{\alpha\beta}\Big),\\
 &\delta \N^{I} = - \N_{J} \overline{E}^{I}_{\alpha}\delta\overline{\e}^{J\alpha},\\
 & \delta g_{\alpha\beta} = \eta_{IJ} ( \e^{I}_{\alpha}\delta \e^{J}_{\beta} + \e^{I}_{\beta}\delta \e^{J}_{\alpha}).
\end{align*}

\section{Irreducible Decompositions}\label{appendix_irreducible}
Let $g_{\alpha\beta}$ be a metric on $M$, where dim$(M) = n$.  Let $\tensor{T}{^\alpha_\beta_\sigma}$ be a $3$-tensor in $M$, antisymmetric in its last two indices. Its irreducible decomposition against the $SO(1,3)$ (pseudo-orthogonal group) is given by \cite{mccrea1992irreducible, itin2022decomposition, jimenez2022metric}
\begin{align*}
 &\tensor{T}{^\alpha_\beta_\sigma} = ^{(1)}\tensor{T}{^\alpha_\beta_\sigma} + ^{(2)}\tensor{T}{^\alpha_\beta_\sigma} + ^{(3)}\tensor{T}{^\alpha_\beta_\sigma}, \\ 
    & ^{(1)}\tensor{T}{^\alpha_\beta_\sigma} = \frac{1}{n-1}\big( \tensor{T}{^\lambda_\beta_\lambda}\delta^{\alpha}_{\sigma} - \tensor{T}{^\lambda_\lambda_\sigma}\delta^{\alpha}_{\beta}\big), \\
    & ^{(2)}\tensor{T}{^\alpha_\beta_\sigma} = g^{\alpha\lambda}T_{[\lambda\beta\sigma]} = \frac{1}{3!}g^{\alpha\lambda} \big( T_{\alpha\beta\sigma} + T_{\beta\sigma\alpha} + T_{\sigma\alpha\beta} \big),\\
    & ^{(3)}\tensor{T}{^\alpha_\beta_\sigma} = \tensor{T}{^\alpha_\beta_\sigma} - ^{(1)}\tensor{T}{^\alpha_\beta_\sigma} - ^{(2)}\tensor{T}{^\alpha_\beta_\sigma}.
\end{align*}
The number of independent components of the tensor is $\frac{n^2(n-1)}{2}$. The sum of components of each of the terms of its irreducible decomposition is given in order by,
\begin{align*}
    \frac{n^2(n-1)}{2} = n \; + \; \frac{n(n-1)(n-2)}{3!} \; + \;  \frac{n(n+2)(n-2)}{3}.
\end{align*}
The non-trivial properties of the components are 
\begin{align*}
& ^{(1)}\tensor{T}{^\alpha_\beta_\sigma} = \tensor{T}{^\lambda_\beta_\lambda},\\
&  ^{(2)}\tensor{T}{^\alpha_\beta_\alpha} = 0,    \\
& ^{(3)}\tensor{T}{^\alpha_\alpha_\sigma} = 0,  \\
   &  ^{(3)}T_{\alpha\beta\sigma} + ^{(3)}T_{\beta\sigma\alpha}  + ^{(3)}T_{\sigma\alpha\beta}= 0.
\end{align*}
In a similar way, let $\tensor{M}{^\alpha_\beta_\sigma}$ be a $3$-tensor in $M$, symmetric in its last two indices. Its irreducible decomposition is given by
\begin{align*}
    &M_{\alpha\beta\sigma} = ^{(1)}M_{\alpha\beta\sigma} +^{(2)}M_{\alpha\beta\sigma} + ^{(3)}M_{\alpha\beta\sigma} + ^{(4)}M_{\alpha\beta\sigma}, \\
    &^{(1)}M_{\alpha\beta\sigma} = \frac{1}{n}\tensor{M}{_\alpha_\lambda^\lambda}g_{\beta\sigma},\\
    &^{(2)}M_{\alpha\beta\sigma} = \frac{2}{n(n-1)(n+2)}(g_{\beta\sigma}\delta^{\mu}_{\alpha} - 2 g_{\alpha\beta}\delta^{\mu}_{\sigma} - 2g_{\alpha\sigma}\delta^{\mu}_{\beta})(\tensor{M}{_\mu_\lambda^\lambda} - 4 \tensor{M}{_\lambda_\mu^\lambda}),\\
    &^{(3)}M_{\alpha\beta\sigma} = M_{(\alpha\beta\sigma)} - \frac{1}{3(n+2)}(g_{\alpha\beta} \delta^{\mu}_{\sigma} + g_{\sigma\alpha}\delta^{\mu}_{\beta} + g_{\beta\sigma}\delta^{\mu}_{\alpha})(\tensor{M}{_\mu_\lambda^\lambda} + 2 \tensor{M}{_\lambda_\mu^\lambda}),\\
    &^{(4)}M_{\alpha\beta\sigma} =  M_{\alpha\beta\sigma} - ^{(1)}M_{\alpha\beta\sigma} -  ^{(2)}M_{\alpha\beta\sigma} - ^{(3)}M_{\alpha\beta\sigma}.
\end{align*}
\newpage
The number of independent components of the tensor is $\frac{n^2(n+1)}{2}$. The sum of components of each of the terms of its irreducible decomposition is given in order by,
\begin{align*}
    \frac{n^2(n+1)}{2} = n + n + \frac{n(n-1)(n+4)}{3!} + \frac{n(n^2 - 4)}{3}.
\end{align*}
The non-trivial properties of these components are
\begin{align*}
    & ^{(1)}\tensor{M}{_\alpha_\beta^\alpha} = \frac{1}{n}\tensor{M}{_\beta_\lambda^\lambda}  && ^{(3)}\tensor{M}{_\alpha_\beta^\beta} =0,\\
    & ^{(2)}\tensor{M}{_\alpha_\beta^\beta} = 0 &&  ^{(3)}\tensor{M}{^\alpha_\alpha_\sigma} = 0,\\
    & ^{(2)}\tensor{M}{_\alpha_\beta^\alpha} = \tensor{M}{_\lambda_\beta^\lambda} - \frac{1}{n}\tensor{M}{_\beta_\lambda^\lambda} && ^{(4)}\tensor{M}{_\alpha_\beta^\beta} = 0 ,\\
    & ^{(4)}\tensor{M}{^\alpha_\beta_\alpha} = 0 && ^{(4)}M_{\alpha\beta\sigma} + ^{(4)}M_{\sigma\alpha\beta} + ^{(4)}M_{\beta\sigma\alpha} = 0 ,\\
    & ^{(1)}\tensor{M}{_\alpha_\beta^\beta} = \tensor{M}{_\beta_\lambda^\lambda}.
\end{align*}